\documentclass[11pt,twoside]{article}
\usepackage[utf8]{inputenc}
\usepackage{times,fancyhdr}
\usepackage[dvips]{graphicx}
\usepackage{latexsym}
\usepackage[affil-it]{authblk}
\usepackage{amsmath}
\usepackage{amssymb}
\usepackage{mathtools}
\usepackage[title, titletoc]{appendix}

\newcommand{\KB}{K_{\mathrm{B}}}
\newcommand{\Ktwo}{\mathcal{K}_2}
\newcommand{\baralpha}{\bar{\alpha}}
\newcommand{\barbeta}{{\bar{\beta}}}
\newcommand{\coup}{g_m}
\newcommand{\X}{X}
\newcommand{\Rgal}{R_p}
\newcommand{\dotRgal}{\dot{R}_p}

\usepackage[dvipsnames]{xcolor}
\usepackage[colorlinks,allcolors=Blue]{hyperref}

\usepackage[backend=biber,style=phys,biblabel=brackets,pageranges=false,eprint=true]{biblatex}
\DeclareSourcemap{
 \maps[datatype=bibtex,overwrite=true]{
  \map{
    \step[fieldsource=Collaboration, final=true]
    \step[fieldset=usera, origfieldval, final=true]
  }
 }
}

\renewbibmacro*{author}{%
  \printnames{author}
  \iffieldundef{usera}{
  }{
    (\printfield{usera} Collaboration)
  }
}

\DeclareFieldFormat{eprint:arxiv}{%
    arXiv\addcolon
    \ifhyperref
    {\href{https://arxiv.org/abs/#1}{%
        #1%
    }}
    {%
        #1%
    }}

\AtEveryBibitem{\clearfield{eid}}

\usepackage[top=4cm, bottom=4cm, left=4cm, right=4cm]{geometry}

\begin{document}

\begin{titlepage}

\title{A novel Cherenkov radiation constraint for hybrid MOND dark matter models}
\author{Tobias Mistele\thanks{\href{mailto:mistele@fias.uni-frankfurt.de}{mistele@fias.uni-frankfurt.de}}}
\affil{\small Frankfurt Institute for Advanced Studies\\
Ruth-Moufang-Str. 1,
D-60438 Frankfurt am Main, Germany
}
\date{}
\maketitle

\begin{abstract}
Modified gravity models often contain modes
that couple to normal matter
and propagate with slightly less than the speed of light.
High-energy cosmic rays then lose energy due to Cherenkov radiation,
which constrains such models.
This is also true for some {\sc MOND} (Modified Newtonian Dynamics) models.
However, these constraints are difficult to make precise
because {\sc MOND} is inherently non-linear
and because the results may depend on the specific high-acceleration behavior of these models, i.e. the behavior outside the {\sc MOND} regime.
Recently, various hybrid {\sc MOND} dark matter models were proposed,
where cold dark matter ({\sc CDM}) phenomenology on cosmological scales and {\sc MOND} phenomenology on galactic scales share a common origin.
Such models typically contain a mode that is directly coupled to matter (for {\sc MOND}), but with non-relativistic sound speed (for {\sc CDM}).
Thus, even non-relativistic objects like stars can emit gravitational Cherenkov radiation.
We calculate a lower bound on the associated energy loss.
We use a controlled approximation that depends only on the {\sc MOND} regime of these models.
We apply our results to three concrete models:
For the original superfluid dark matter model ({\sc SFDM}), we rule out a part of the parameter space, including the most commonly used parameters.
For two-field {\sc SFDM}, we find no constraint since the matter coupling of the relevant mode is suppressed by mixing.
For the recently-proposed model by Skordis and Z\l o\'{s}nik, we find no constraint since the matter coupling is suppressed in non-static situations.
\end{abstract}

\end{titlepage}

\label{sec:cherenkov}

\section{Introduction}

A simple explanation for the problem of missing non-baryonic mass on cosmological scales is the existence of a collisional fluid.
On galactic scales, a simple explanation is a modified force law, specifically Modified Newtonian Dynamics ({\sc MOND}) \cite{Milgrom1983a,Milgrom1983b,Milgrom1983c,Bekenstein1984}.
A natural idea is to combine the phenomenology of a collisionless fluid on cosmological scales and that of {\sc MOND} on galactic scales in a single model.
Recently, various such models were proposed.
We refer to them as hybrid MOND dark matter models.

All such hybrid models produce both MOND-like and CDM-like phenomena.
Some hybrid models simply introduce two independent sectors,
    one of which provides a collisionless fluid on cosmological scales,
    while the other provides a MOND-like force on galactic scales.
Examples are the $\nu$HDM model \cite{Angus2009, Haslbauer2020} or the two-field model from Ref.~\cite{Khoury2015}.
In such models,
    the MOND and CDM phenomena exist independently of each other.
In contrast, other models seek a common origin for the cosmological and galactic phenomena.
Examples are the original superfluid dark matter (SFDM) model \cite{Berezhiani2015, Berezhiani2018} and the recently-proposed model by Skordis and Z\l o\'{s}nik ({\sc SZ} model) \cite{Skordis2020, Skordis2021}.

Such models with a common origin for the cosmological and galactic phenomena
    typically contain a component that plays a double role.
That is, a component that is involved both
    in providing a significant energy density (for the CDM phenomenology)
    and in providing a modified force law (for the MOND phenomenology).
Such a double role may lead to tensions between these two different roles.
This happens for example in SFDM \cite{Mistele2020}.
This is why Ref.~\cite{Mistele2020} proposed two-field SFDM,
    which weakens the link between the cosmological and galactic phenomena.
Similarly, for the SZ model,
    a MOND-like force on galactic scales tends to induce a too large pressure of the DM-like fluid in the early universe if one is not careful \cite{Skordis2020}.
Here,
    we will discuss another consequence of components that play such a double role,
    namely a novel type of Cherenkov radiation.

\begin{figure}
 \centering
 \includegraphics[width=.4\textwidth]{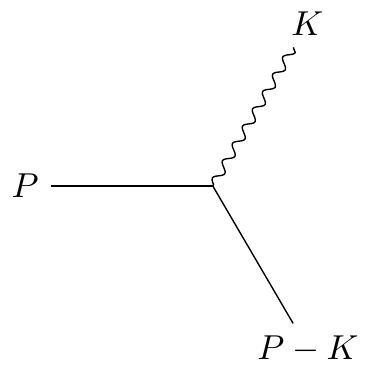}
 \caption{
        The Feynman diagram for Cherenkov radiation.
        The straight lines denote matter, the wiggly line denotes the radiation mode, e.g. a massless modified gravity mode.
        The incoming matter particle has four-momentum $P$ and radiates away energy and momentum $K$.
        This process is kinematically allowed only if the matter particle moves faster than the massless mode propagates.
    }
 \label{fig:cherenkov-feynman}
\end{figure}

Modified gravity theories often contain a massless mode that is directly coupled to matter
    in order to provide a modified force law.
If this massless mode propagates with a speed $c_s$ that is smaller than the speed of light,
    matter can have a velocity $V$ that is larger than $c_s$.
In this case, matter loses energy by emitting radiation in the form of the massless modified gravity mode.
This radiation is called Cherenkov radiation.
In general,
    Cherenkov radiation is emitted whenever a massless mode is directly coupled to matter and the matter velocity $V$ is larger than $c_s$.
For brevity, we will refer to the speed $c_s$ as the speed of sound,
    but it is not in general necessary that there is a hydrodynamical description.
In particle physics language, Cherenkov radiation corresponds to the Feynman diagram shown in Fig.~\ref{fig:cherenkov-feynman}.
A direct coupling to matter implies that the vertex in this Feynman diagram exists.
When $V$ is larger than $c_s$, the external legs can go on-shell.
For smaller $V$, this process is kinematically forbidden by energy-momentum conservation.

Usually, gravitational Cherenkov radiation is studied for highly relativistic matter \cite{Caves1980, Moore2001, Elliott2005, Bruneton2007b, Milgrom2011, Chesler2017}.
Indeed, if the sound speed is close to the speed of light, $c_s \approx 1$, only relativistic objects can emit Cherenkov radiation,
    because only relativistic objects can satisfy $V > c_s$.
But if $c_s$ is much smaller than the speed of light, $c_s \ll 1$,
    even non-relativistic matter can emit Cherenkov radiation.
An example of this phenomenon is dynamical friction in collisional fluids \cite{Ostriker1999, Berezhiani2019b, Berezhiani2020}:
If such a fluid constitutes dark matter in galaxies,
    then stars or globular clusters may experience dynamical friction.
That is, if normal matter moves faster than the dark matter sound speed,
    it loses energy through Cherenkov radiation,
    i.e. it radiates away sound waves.\footnote{
    The standard Chandrasekhar dynamical friction for collisionless fluids \cite{Chandrasekhar1943} is related but different.
    E.g. no sound waves are emitted and the friction is due to the (instantaneous) Newtonian gravitational force.
    Also, this friction force does not drop off sharply below a critical velocity.
    Still, the order of magnitude of the friction force can be the same in the collisional and the collisionless case \cite{Ostriker1999}.
}

The timescale on which a non-relativistic object loses a significant fraction of its energy due to this dynamical friction is \cite{Ostriker1999}
\begin{align}
 \tau_{\mathrm{df}} = |\dot{E}|/E_{\mathrm{kin}} = \frac{V^3}{M G^2 \rho_{\mathrm{DM}}} \,,
\end{align}
where we neglected $\mathcal{O}(1)$ factors, $E_{\mathrm{kin}}$ is the object's kinetic energy, $M$ is its mass, and $\rho_{\mathrm{DM}}$ is the dark matter density.
This timescale is dynamically relevant for heavy objects since $\tau_{\mathrm{df}} \propto 1/M$.
For example, globular clusters in the Fornax dwarf spheroidal may have a timescale $\tau_{\mathrm{df}}$ smaller than the age of the Universe \cite{Tremaine1976, Sanchez-Salcedo2006}.
In comparison, individual stars have much smaller masses
    so that the timescale $\tau_{\mathrm{df}}$ is much larger and not phenomenologically relevant.
For example, $\tau_{\mathrm{df}} \sim 10^{20}\,\mathrm{yr}$ for the Sun in a dark matter fluid with $\rho_{\mathrm{DM}} = 0.3\,\mathrm{GeV}/\mathrm{cm}^3$.

A standard dark matter fluid is not directly coupled to matter,
    only indirectly through Newtonian gravity.
This is why $\tau_{\mathrm{df}}$ scales as $1/G^2$.
If, instead, matter were directly gravitationally coupled to the dark matter fluid,
    this timescale would scale as $1/G$.
This is typical for gravitational Cherenkov radiation in modified gravity models.
In this case,
    the relevant timescale may be much shorter
    so that even comparably light objects can lose energy on a dynamically relevant timescale.

Hybrid MOND-dark-matter models with a common origin for MOND and CDM phenomena typically allow for Cherenkov radiation.
The reason is as follows.
These models have a massless mode that couples to normal matter,
    namely the mode that carries the MOND-like force.
If there is a common origin of this MOND-like force and the cosmological DM-like fluid,
    this massless mode naturally propagates with a non-relativistic speed.
A non-relativistic propagation speed is natural because the dark matter fluid must be pressureless on cosmological scales.

For example, in SFDM, the collisionless fluid on cosmological scales is a non-relativistic superfluid, which also forms a cored halo around galaxies.
The MOND-like force is carried by the phonons of this superfluid.
Thus, these phonons are non-relativistic and directly coupled to normal matter.
This allows for Cherenkov radiation.
Since the sound speed is non-relativistic, this allows Cherenkov radiation even from non-relativistic objects.
That is, even stars may lose energy through Cherenkov radiation, i.e. to phonons.

This is a novel type of Cherenkov radiation.
It comes from a direct coupling to matter as in standard modified gravity models.
But it also allows for non-relativistic objects to emit Cherenkov radiation like dynamical friction in collisional fluids.

We can constrain models allowing for this type of Cherenkov radiation by requiring that stars do not lose a significant amount of their energy on galactic timescales.
Specifically, either the sound speed $c_s$ must be large enough so that most stars are subsonic and Cherenkov radiation is avoided kinematically.
Or the timescale $\tau_E$ on which stars lose a significant amount of energy due to Cherenkov radiation must be much larger than galactic timescales.
We will show that the timescale $\tau_E$ scales as
\begin{align}
 \tau_E \propto \frac{1}{c_s^2 \, g_m^2 G} \,,
\end{align}
where $c_s$ is the non-relativistic sound speed of the massless mode and $g_m \sqrt{G}$ its coupling to matter.
Thus, to avoid Cherenkov radiation by having a large energy loss timescale $\tau_E$ one can make either $c_s$ or $g_m$ small.
This is different from the relativistic limit considered in standard Cherenkov radiation constraints,
    where $\tau_E$ cannot be made large by making $c_s$ small.

Our results for $\tau_E$ will be a conservative lower bound,
    because we assume strict cuts in the phase space integrals in our calculation.
We do this for two reasons.
First,
    to avoid technical difficulties due to the non-linearities inherent in any MOND model.
Second, to rely only on the MOND regime of each model,
    i.e. to be independent of the behavior for accelerations larger than the MOND acceleration scale $a_0$.
This makes our calculation simple and robust.
For example, we do not need to consider potential higher-derivative terms \cite{Babichev2011} which may become important in the solar system,
    as has been considered in both SFDM and the SZ model \cite{Berezhiani2018, Skordis2020}.

These strict cuts also imply that there is no significant recoil.
In a quasi-particle picture:
    Any single emission carries very little energy and momentum.
    A macroscopic change happens only after a large number of these soft emissions.
In contrast, for standard gravitational Cherenkov radiation, the emissions are not soft.
    Supersonic objects lose a significant amount of energy and momentum with each emission, i.e. recoil is important \cite{Chesler2017}.

We first demonstrate the general idea and calculations behind our Cherenkov constraints for a simple prototype model in Sec.~\ref{sec:general}.
In Secs.~\ref{sec:standardsfdm}, \ref{sec:twofieldsfdm}, and \ref{sec:relmond} we apply these results, with appropriate modifications, to the original SFDM model, two-field SFDM, and the SZ model, respectively.
For the original SFDM model, we can rule out a MOND limit in the Milky Way for a certain part of the parameter space.
In contrast, two-field SFDM and the SZ model evade our constraints due to two different mechanisms that we explain in detail.
After a short discussion in Sec.~\ref{sec:discussion}, we summarize in Sec.~\ref{sec:conclusion}.
The present paper supplements and extends the shorter work Ref.~\cite{Mistele2021}.
We employ units with $ c = \hbar = 1 $ and the metric signature $ (+, -, -, -) $, unless otherwise stated.
Small Greek indices run from $ 0 $ to $ 3 $ and denote spacetime dimensions.

\section{Cherenkov radiation in a prototype hybrid model}
\label{sec:general}

In this section, we will introduce the general idea behind our Cherenkov radiation constraint using a prototype model.
Specifically, we consider a prototype model for a massless mode with non-relativistic sound speed $c_s \ll 1$ that directly couples to matter with a standard gravitational coupling constant.
As discussed above, this is a natural setup for hybrid MOND dark matter models with a common origin for the cosmological and galactic scale phenomena.
The three models we explicitly discuss below in Secs.~\ref{sec:standardsfdm}, \ref{sec:twofieldsfdm}, and \ref{sec:relmond} are not exactly of the form discussed here.
However, we prefer to keep the discussion simple for now and later adapt our results as needed.

We consider a matter object, for example a star,
    with mass $M$ and velocity $V$ at a distance $\Rgal$ from the center of its host galaxy.
We denote the host galaxy's baryonic mass inside a radius $\Rgal$ by $M_{\mathrm{gal}}$.
In general, we use a capital $R$ for distances to the center of the host galaxy
    and a lower case $r$ for distances to the matter object, also referred to as the perturber.

Our main result will be an estimate for the timescale $\tau_E$ on which non-relativistic objects like stars lose a significant fraction of their energy,
    if their velocity $V$ is larger than a critical velocity $V_{\mathrm{crit}} = \mathcal{O}(\bar{c})$,
    i.e. if they are supersonic,
\begin{align}
 \tau_E = \frac{2\cdot10^8\,\mathrm{yr}}{f_a f_p^2 \, \coup^2} \cdot \left(\frac{V/\bar{c}}{2}\right)^2 \cdot \left(\frac{a_0}{a_b^{\mathrm{gal}}}\right) \cdot \left(\frac{V}{200\,\mathrm{km}/\mathrm{s}}\right) \cdot \left(\frac{1.2 \cdot 10^{-10}\,\mathrm{m}/\mathrm{s}^2}{a_0}\right) \,.
\end{align}
Here,
    $\bar{c}$ sets the scale of the sound speed $c_s$,
    $\coup$ is a model-dependent coupling constant,
    $a_0$ is the MOND acceleration scale,
    and $a_b^{\mathrm{gal}}$ is the Newtonian acceleration due to baryons in the host galaxy at the position of the matter object.
The factor $f_a$ depends on the relative orientation of $\vec{V}$ and the MOND force of the background galaxy.
The factor $f_p$ is determined by the radius where the perturber's field becomes smaller than the galaxy's background field.
Both $f_a$ and $f_p$ are typically of order $1$.
For a standard gravitational coupling, $\coup$ is also of order $1$.
Details are explained below.

Thus, stars with $V > V_{\mathrm{crit}}$ may lose a significant fraction of their energy on galactic timescales.
This constrains hybrid MOND-DM models.
Such models
    either need a sound speed large enough such that most stars are subsonic,
    or a timescale $\tau_E$ that is much larger than galactic timescales.

\subsection{Prototype Lagrangian}

Consider a real scalar field $\varphi$ that satisfies a MOND-type equation on galactic scales in the static limit.
Then, roughly, $\varphi \propto \sqrt{G M_{\mathrm{gal}} a_0} \ln(R)$.
Typically, the Lagrangian for perturbations $\delta$ around a galaxy's static background field $\varphi_0$ can be written as
\begin{align}
 \label{eq:prototype}
 \mathcal{L} = \frac12 \frac{1}{\bar{c}^2} (\partial_t \delta)^2 - \frac12 \left( (\vec{\nabla} \delta)^2 + (\hat{a} \vec{\nabla} \delta)^2 \right) - \frac{\coup}{\sqrt{2} M_{\mathrm{Pl}}} \delta \, \delta_b \,,
\end{align}
after an appropriate definition of $\delta$.
We will discuss concrete examples of this later.
Here,
    $\coup$ and $\bar{c}$ are parameters that may depend on the background field $\varphi_0$,
    and $\delta_b$ is a perturbation of the baryonic density $\rho_b$.
Further, $\hat{a}$ is a unit vector that points into the direction of the background $\vec{\nabla} \varphi_0$,
    i.e. into the direction of the background MOND force or in the opposite direction, depending on the signs.
The dispersion relation is $\omega = c_s |\vec{k}|$ with the sound speed
\begin{align}
  c_s^2 = \bar{c}^2(1 + \gamma^2) \,,
\end{align}
where $\gamma$ is the cosine of the angle between the perturbation's wavevector $\vec{k}$ and $\hat{a}$.

A prototypical MOND Lagrangian is \cite{Bekenstein1984, Bruneton2007b}
\begin{align}
 \mathcal{L} = \frac{2 M_{\mathrm{Pl}}^2}{3a_0} \sqrt{|K_\varphi|} K_\varphi - \rho_b \, \varphi \,,
\end{align}
where $K_\varphi = \nabla_\alpha \varphi \nabla^\alpha \varphi$ is a standard kinetic term and $a_0$ is the MOND acceleration scale.
After rescaling, the second-order Lagrangian for perturbations is that of Eq.~\eqref{eq:prototype} with
\begin{align}
 \bar{c} = 1 \,, \quad
  \coup = \sqrt{\frac{a_0}{|\vec{\nabla} \varphi_0|}} \,.
\end{align}
Thus, the sound speed is relativistic (even superluminal \cite{Bruneton2007, Bruneton2007b, Babichev2008}) and $\coup$ is roughly of order 1 on galactic scales since $|\vec{\nabla} \varphi_0|$ is roughly of order $a_0$ there.

In many hybrid MOND-DM models, we expect this to be still qualitatively true, except that $\bar{c}$ gives a non-relativistic sound speed
\begin{align}
\bar{c} \ll 1 \,.
\end{align}
For example, in standard SFDM, $\delta$ corresponds to the phonons of a non-relativistic superfluid.
These have a non-relativistic sound speed so that $\bar{c} \ll 1$.
We will discuss this and other examples in more detail below.

As already mentioned above, the models we discuss below do not exactly have the form of our prototype Lagrangian.
For example, standard SFDM has an additional term that mixes spatial and time derivatives of $\delta$.
Two-field SFDM has a more complicated coupling $g_m$ and no $\hat{a}$ term for the relevant mode.
The SZ model also has a more complicated coupling $g_m$.
We will discuss how the calculation needs to be adjusted in each case in Secs.~\ref{sec:standardsfdm}, \ref{sec:twofieldsfdm}, and \ref{sec:relmond}, respectively.
Nevertheless, the general considerations for our prototype Lagrangian give a useful qualitative picture.

\subsection{Cherenkov radiation}
\label{sec:general:cherenkov}

Consider a non-relativistic perturber, e.g. a star, that moves in a galaxy.
Following Refs.~\cite{Berezhiani2019b, Berezhiani2020}, we model the perturber as a real scalar field $\chi$ with mass $M$.
In the non-relativistic limit, this scalar field's energy density is
\begin{align}
\rho_\chi \approx \frac12 (\dot{\chi}^2 + M^2 \chi^2) \approx M^2 \chi^2 \,.
\end{align}
We take this as the baryon density's perturbation $\delta_b$ in our prototype Lagrangian from Eq.~\eqref{eq:prototype}.
We can then calculate the perturber's energy loss rate
\begin{align}
 \label{eq:dotEgeneral}
 \dot{E} = - \int \omega d\Gamma \,,
\end{align}
where $d\Gamma$ is the differential decay rate and $\omega = c_s |\vec{k}|$ is the energy of the quasi-particle that is radiated away.
It is important to impose cutoffs in the integral in Eq.~\eqref{eq:dotEgeneral} that reflect the regime in which our prototype Lagrangian is valid.
For now, we keep the calculation general and assume cutoffs $k_{\mathrm{min}}$ and $k_{\mathrm{max}}$ with $k_{\mathrm{min}} < k_{\mathrm{max}}$ in the momentum integral.
We will discuss the numerical values of these cutoffs below.
We find (see Appendix~\ref{sec:Edotcalculation}),
\begin{align}
 \label{eq:dotE}
 |\dot{E}| = f_a \frac{\bar{c}^2}{16 \pi V} \frac{\coup^2 M^2}{M_{\mathrm{Pl}}^2} (k_{\mathrm{max}}^2 - k_{\mathrm{min}}^2) \cdot \Theta(V - V_{\mathrm{crit}})\,.
\end{align}
Here, $f_a$ is a factor that depends on the direction of $\hat{a}$ relative to $\vec{V}$.
The critical velocity $V_{\mathrm{crit}}$ also depends on the direction of $\vec{V}$ and denotes the velocity below which no Cherenkov radiation is emitted.
For $\vec{V} \parallel \hat{a}$:
\begin{subequations}
\begin{align}
 f^\parallel_a &= \frac{1}{1 - (\bar{c}/V)^2}\,,\\
 V^\parallel_{\mathrm{crit}} &= \sqrt{2} \bar{c}\,.
\end{align}
\end{subequations}
And for $\vec{V} \perp \hat{a}$:
\begin{subequations}
\begin{align}
 f^\perp_a &= \frac{1}{\sqrt{1 + (\bar{c}/V)^2}}\,,\\
 V^\perp_{\mathrm{crit}} &= \bar{c}\,.
\end{align}
\end{subequations}
The factor $f_a$ does not change the order of magnitude of the energy loss.
Numerically, for $V > V_{\mathrm{crit}}$, it varies between $1$ and $2$ for $\vec{V} \parallel \hat{a}$ and between $1$ and $1/\sqrt{2}$ for $\vec{V} \perp \hat{a}$.

Note that $\dot{E}$ scales as $\bar{c}^2$.
This is because the correct normalization for using the standard QFT formalism with the standard Feynman rules is not the one shown in Eq.~\eqref{eq:prototype}.
Instead, the correct normalization is to scale the $1/\bar{c}^2$ in front of the time derivatives away \cite{Berezhiani2019b}.
This scaling makes the coupling proportional to $\bar{c}$ and the amplitude squared proportional to $\bar{c}^2$.

The critical velocity $V_{\mathrm{crit}}$ reflects the fact that Cherenkov radiation is allowed only for supersonic perturbers, $V > c_s$.
Here, the sound speed $c_s$ depends on the direction of the wave vector $\vec{k}$ of the perturbation relative to that of the background field $\hat{a}$, i.e. $c_s = \bar{c} \sqrt{1 + \gamma^2}$ depends on $\gamma$.
This is why $f_a$ and $V_{\mathrm{crit}}$ depend on the relative orientation of $\vec{V}$ and $\hat{a}$.
Without the $\hat{a}$-term in our prototype Lagrangian, this would just be $f_a = 1$ and $V_{\mathrm{crit}} = \bar{c}$, independently of the direction of $\vec{V}$.

Our result for the energy loss $|\dot{E}|$ scales differently from what Refs.~\cite{Caves1980, Moore2001, Elliott2005, Kimura2012} have found for standard gravitational Cherenkov radiation.
This is because these consider the limit of a relativistic perturber and sound speed, $V \approx 1$ and $c_s \approx 1$, while we consider a non-relativistic perturber and sound speed, $V \ll 1$ and $c_s \ll 1$.

The timescale on which perturbers like stars lose a significant amount of their energy $E$ is roughly $E/|\dot{E}|$.
Thus, assuming $V > V_{\mathrm{crit}}$, we define the timescale $\tau_E$ as
\begin{align}
 \label{eq:taukin1}
 \tau_E \equiv \frac{E_{\mathrm{kin}}}{|\dot{E}|} \equiv \frac{\frac12 M V^2}{|\dot{E}|} = \frac{8 \pi V^3 M_{\mathrm{Pl}}^2}{f_a \bar{c}^2 \coup^2 M k_{\mathrm{max}}^2} \, \frac{1}{1 - (k_{\mathrm{min}}/k_{\mathrm{max}})^2} \,.
\end{align}
As alluded to in the introduction, this scales as $1/G \propto M_{\mathrm{Pl}}^2$, which is different from the standard dynamical friction timescale that scales as $1/G^2 \propto M_{\mathrm{Pl}}^4$ \cite{Chandrasekhar1943, Ostriker1999, Berezhiani2019b}.
Stars have both kinetic and potential energy.
For simplicity, our definition of $\tau_E$ includes only the kinetic energy.
Including the gravitational energy does not change the order of magnitude.
Indeed, we will now discuss a concrete example to see that $\tau_E$ is a useful quantity.

For a star in a galaxy, we have
\begin{align}
 \partial_t \left(E_{\mathrm{kin}} + E_{\mathrm{grav}} \right) = \dot{E} \,,
\end{align}
where $\dot{E}$ is the energy loss through Cherenkov radiation, and $E_{\mathrm{grav}}$ is the gravitational energy.
If $E_{\mathrm{grav}}$ depends only on the star's position $\Rgal$ in the galaxy, we have
\begin{align}
 \partial_t \left( \frac12 M V^2 \right) + M \dotRgal \, a_{\mathrm{grav}}  = \dot{E} \,,
\end{align}
where $a_{\mathrm{grav}}$ is the gravitational acceleration produced by the host galaxy at the star's position.
In the MOND-dominated part of the galaxy with a flat rotation curve, we have $a_{\mathrm{grav}} = \sqrt{G M_{\mathrm{gal}} a_0}/\Rgal$.
If we further assume that the star is always approximately on a circular orbit, we have $V^2 = \sqrt{G M_{\mathrm{gal}} a_0}$, which is constant.
This gives
\begin{align}
 \frac{\dotRgal}{\Rgal} = - \frac{1}{2 \tau_E} \,.
\end{align}
As long as $\tau_E$ depends only weakly on $\Rgal$, this means that stars transition to smaller galactic radii as $\exp(-t/2\tau_E)$ due to Cherenkov radiation.

If this calculation is correct, stars in the flat part of the rotation curve lose energy by transitioning to smaller galactic radii, not by losing their velocity.
This is due to the assumption that the star's velocity is always the circular velocity, even as it transitions to smaller radii.
In the next subsection, we show numerically that this assumption is justified.

\subsection{Orbits of stars emitting Cherenkov radiation}
\label{sec:Edotcalculation:orbits}

Here, we numerically investigate how stars orbit around a galaxy under the influence of the non-relativistic Cherenkov radiation derived above.
We model the effect of the Cherenkov radiation as a friction force that reproduces the energy loss $\dot{E}$ calculated above.
Modelling the Cherenkov radiation as a friction force is justified because the energy loss happens through a large number of emissions, each of which carries away only a very small fraction of the star's energy.
This is different from the high-energy Cherenkov radiation emitted from ultra-relativistic protons discussed in Refs.~\cite{Caves1980, Moore2001, Elliott2005, Kimura2012}, where a single emission carries away a significant fraction of the proton's energy and thus produces a significant recoil \cite{Chesler2017}.

For simplicity, we consider the deep-MOND regime of a galaxy and we consider orbits $\vec{X}(t)$ that would be circular without the energy loss due to Cherenkov radiation.
Concretely, we consider initial conditions $Y(t_0) = R_0$, $X(t_0) = Z(t_0) = 0$, $\dot{X}(t_0) = V_0$, and $\dot{Y}(t_0) = \dot{Z}(t_0) = 0$, with equation of motion
\begin{align}
 \label{eq:orbiteom}
 \ddot{\vec{X}} = \vec{a}_{\mathrm{grav}} - \left(R_\eta \cdot \dot{\vec{X}}\right) \eta \,.
\end{align}
Here, $V_0$ is the MOND circular velocity $(G M_{\mathrm{gal}} a_0)^{1/4}$,
    $\vec{a}_{\mathrm{grav}}$ is the MOND gravitational acceleration,
    and $\eta$ is the coefficient of the friction force due to Cherenkov radiation.
The direction of this friction force is that of $R_\eta \dot{\vec{X}}$, where $R_\eta$ is a rotation matrix.

For our prototype Lagrangian,
    the friction force points in the direction of $\vec{V}$ for both $\vec{V} \perp \hat{a}$ and $\vec{V} \parallel \hat{a}$,
    i.e. $R_\eta$ is just the identity matrix.
For a general orientation of $\vec{V}$, the force is rotated within the $X$-$Y$ plane.
This is discussed in Appendix~\ref{sec:Edotcalculation:direction}.
For standard SFDM, the force is rotated in the $X$-$Y$ plane even for the special case $\vec{V} \perp \hat{a}$.
This is because spatial and time derivatives are mixed there, see Sec.~\ref{sec:standardsfdm} and Appendix~\ref{sec:Edotcalculation:standardsfdm:direction}.

We can find the coefficient $\eta$ by contracting Eq.~\eqref{eq:orbiteom} with $\dot{\vec{X}} = \vec{V}$ and then comparing to the energy balance equation
\begin{align}
 \partial_t \left( \frac 12 M V^2 + E_{\mathrm{grav}}\right) = \dot{E}_{\mathrm{Cherenkov}} \,,
\end{align}
where $E_{\mathrm{grav}}$ is the gravitational energy and $\dot{E}_{\mathrm{Cherenkov}}$ is the energy loss rate due to Cherenkov radiation.
This gives
\begin{align}
 \eta = \frac{|\dot{E}_{\mathrm{Cherenkov}}|}{M (\vec{V}^T R_\eta \vec{V})} = \frac{1}{2 \tau_E \cos \theta_\eta} \,,
\end{align}
where $\theta_\eta$ is the angle by which $R_\eta$ rotates the vector $\vec{V}$ in the $X$-$Y$ plane.
Since $\tau_E$ scales as $V^3$, we choose to write
\begin{align}
 \ddot{\vec{X}} = \vec{a}_{\mathrm{grav}} - \frac{R_\eta \cdot \dot{\vec{X}}}{2 \cos \theta_\eta V^3} \cdot \frac{V^3}{\tau_E} \,,
\end{align}
where the factor $V^3/\tau_E$ is independent of $V$ and is time-dependent only through factors like $a_0/a_b^{\mathrm{gal}}$ that occur in $\tau_E$ in some models and depend on the star's position in the galaxy.
For simplicity, we assume such factors to be roughly constant for now.

We first consider the case where $\theta_\eta = 0$ that is relevant for our prototype Lagrangian for circular orbits.
The resulting orbits will not be exactly circular, that is we do not have $\vec{V} \perp \hat{a}$ exactly.
But we expect this to be a reasonable approximation for the numerical values we consider.
For very large $\tau_E$, it is clear that the friction term does not have any effect.
Similarly, it is clear that stars will quickly fall in to the center of the galaxy for very small $\tau_E$.
The most interesting case is when $\tau_E$ is roughly of the order of galactic timescales.
Thus, here we choose as numerical values
\begin{align}
 R_0 = 30\,\mathrm{kpc} \,, \quad
 V_0 = 200\,\mathrm{km}/\mathrm{s}\,, \quad
 \frac{\tau_E}{V^3} = \frac{5 \cdot 10^9\,\mathrm{yr}}{(200\,\mathrm{km}/\mathrm{s})^3} \,.
\end{align}
We do not have to choose a mass for the star since the equations are independent of this mass.
Our choice of $V_0$ and $R_0$ also fixes the host galaxy's mass,
    since we require orbits to be circular in the absence of Cherenkov radiation, i.e. $\sqrt{G M_{\mathrm{gal}} a_0}/\Rgal = V_0^2/\Rgal$.
The resulting equation of motion is similar to that in the case of dynamical friction in galaxies, see e.g. Ref.~\cite{VandenBosch1999}.

\begin{figure}
 \centering
 \includegraphics[width=.65\textwidth]{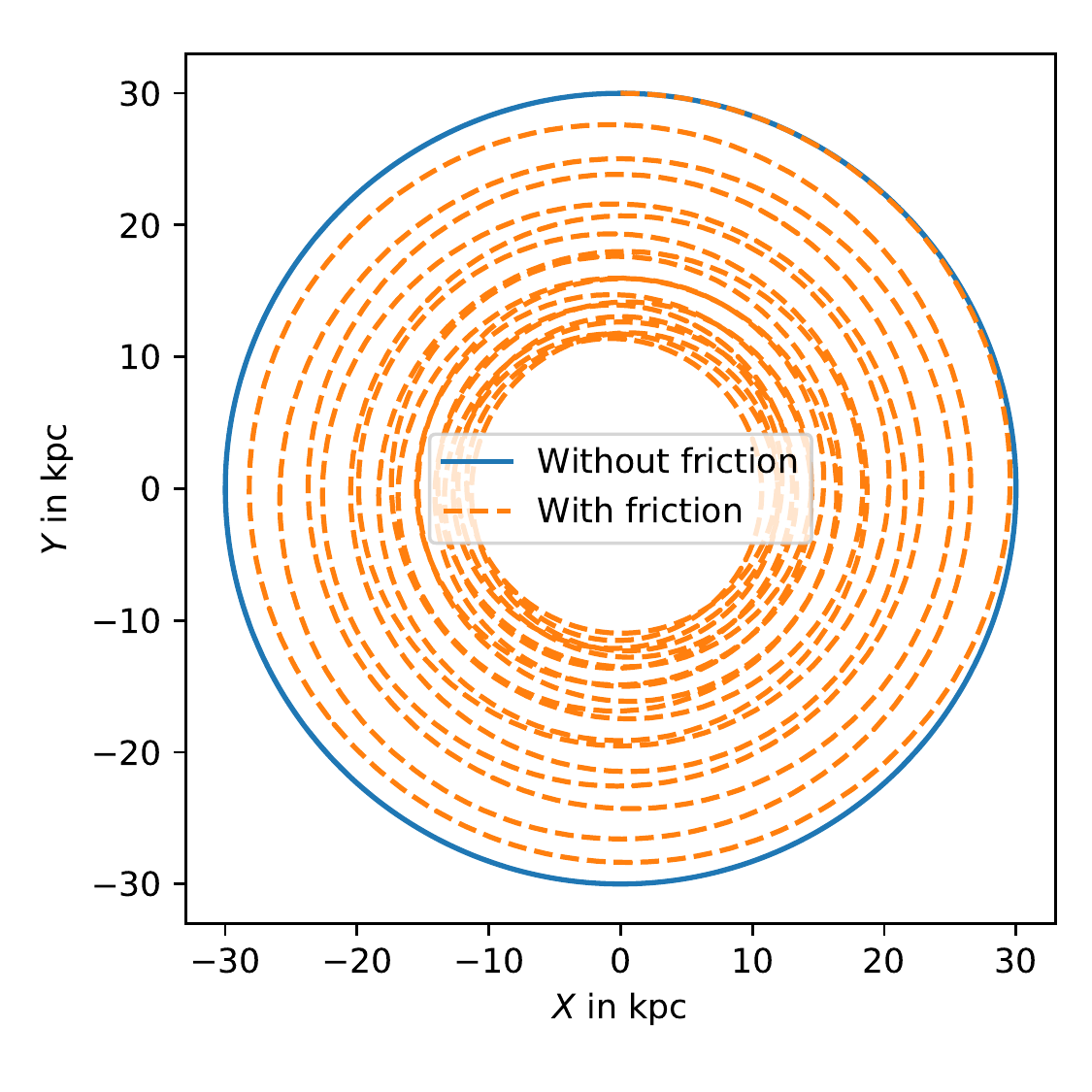}
 \caption{
     Orbit of a perturber in the MOND regime of a galaxy with (dashed orange line) and without (solid blue line) friction due to Cherenkov radiation.
     The friction corresponds to a timescale $\tau_E = 5\cdot10^9\,\mathrm{yr}$ and acts in the direction of the velocity, i.e. $\theta_\eta = 0$.
     The initial conditions are $Y(0) = 30\,\mathrm{kpc}$, $X(0) = 0$, and $\dot{X}(0) = V_0 = 200\,\mathrm{km}/\mathrm{s}$, $\dot{Y}(0) = 0$.
     The galaxy mass is chosen such that these initial conditions give a circular orbit without friction.
     The orbits are integrated for a total time of $10^{10}\,\mathrm{yr}$.
    }
 \label{fig:orbitXY}
\end{figure}

\begin{figure}
 \centering
 \includegraphics[width=.49\textwidth]{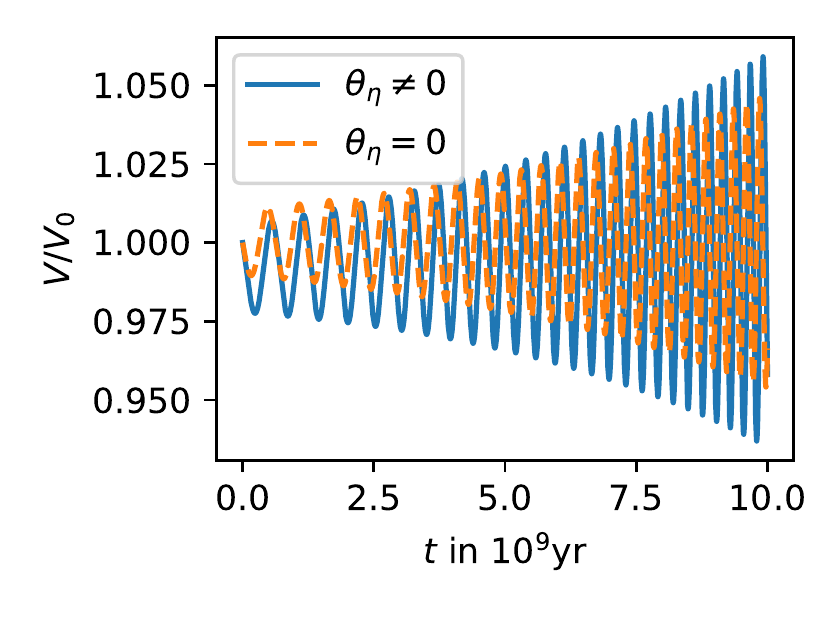}
 \includegraphics[width=.49\textwidth]{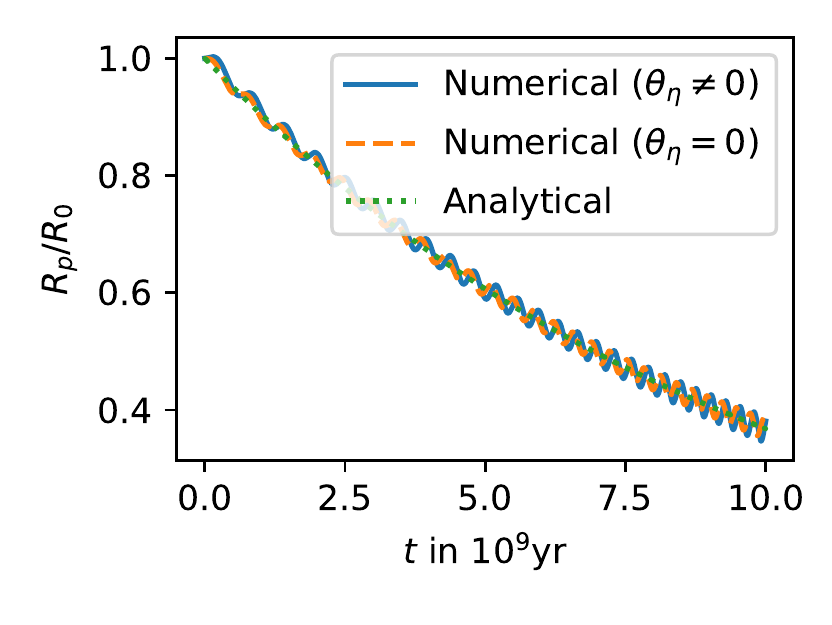}
 \caption{
     Left: Velocity of a perturber losing energy through Cherenkov radiation relative to its initial velocity.
     The dashed orange line corresponds to the orbit with friction from Fig.~\ref{fig:orbitXY} ($\theta_\eta = 0$).
     The solid blue line shows an orbit with the same parameters but with a force that is rotated according to Eq.~\eqref{eq:thetaeta} ($\theta_\eta \neq 0$).
     Right: The perturber's position $\Rgal$ in the host galaxy relative to its initial position for the same orbits as in the left panel (solid blue line for $\theta_\eta = 0$ and dashed orange line for $\theta_\eta \neq 0$), and additionally for the analytical estimate Eq.~\eqref{eq:dotRgalappendix} (dotted green line).
 }
 \label{fig:orbitRV}
\end{figure}

We numerically integrate the equation of motion for $10^{10}\,\mathrm{yr}$ in Mathematica \cite{Mathematica12}.
The resulting orbit in the $X$-$Y$ plane is shown in Fig.~\ref{fig:orbitXY}.
We see that the star transitions to smaller galactic radii due to the friction force.
Fig.~\ref{fig:orbitRV}, left, shows that the mean velocity of the star stays roughly constant during this transition, but there are small oscillations on the percentage level.
Fig~\ref{fig:orbitRV}, right, shows the star's distance to the host galaxy's center $\Rgal = |\vec{X}|$ both for the numerical calculation and for our analytical estimate from Sec.~\ref{sec:general:cherenkov},
\begin{align}
 \label{eq:dotRgalappendix}
 \frac{\dotRgal}{\Rgal} = - \frac{1}{2\tau_E}\,.
\end{align}
Up to small oscillations, the numerical and analytical results agree with each other.
This justifies the analytical estimate from Sec.~\ref{sec:general:cherenkov}.

Consider now the case $\theta_\eta \neq 0$.
For concreteness we take $\theta_\eta$ from standard SFDM,
    where $\theta_\eta = 0$ for $\vec{V} \parallel \hat{a}$, but $\theta_\eta \neq 0$ for $\vec{V} \perp \hat{a}$,
    as shown in Appendix~\ref{sec:Edotcalculation:standardsfdm:direction}.
In both cases, the force stays within the $X$-$Y$ plane, so that we do not need to include a $Z$ component.
For $\vec{V} \perp \hat{a}$, we have
    \begin{align}
    \label{eq:thetaeta}
     \tan \theta_\eta = - \frac12 \frac{f_\barbeta V}{\bar{c}} \,,
    \end{align}
    where $f_\barbeta$ is a constant that depends on the model parameter $\barbeta$, see Sec.~\ref{sec:standardsfdm}.
The orbit we consider is mostly, but not exactly, circular.
Still, we expect that using Eq.~\eqref{eq:thetaeta} is a reasonable approximation.
We further adopt
\begin{align}
 f_\barbeta = 1/\sqrt{3} \,, \quad \bar{c} = 50\,\mathrm{km}/\mathrm{s} \,.
\end{align}
The value of $f_\barbeta$ is the largest possible value for the parameter range we consider, see Sec.~\ref{sec:standardsfdm}.
The value for $\bar{c}$ is unusually small.
This gives an unusually large $\tan \theta_\eta$.
Despite this, the resulting orbit is very similar to that for $\theta_\eta = 0$.
This can be seen from the dashed orange lines in Fig.~\ref{fig:orbitRV}.
The oscillations differ slightly, but the qualitative behavior is the same.

The details of the oscillations also depend on the choice of initial condition.
But we have numerically verified that, at least for initially approximately circular orbits, these oscillations always stay small and the radial decay of the orbit is accurately captured by Eq.~\eqref{eq:dotRgalappendix}.\footnote{
    In particular, the oscillations are completely absent for special initial conditions.
    Such initial conditions are the physically expected ones in many astrophysical situations since gravitational wave emission circularizes orbits \cite{Peters1964}.
    However, the energy loss due to Cherenkov radiation we consider here is physically quite similar to the energy loss from supersonic dynamical friction in a collisional medium, which has the opposite effect \cite{Cardoso2021, Macedo2013}.
    Of course, one still has normal gravitational wave emission in addition, at least in models like SFDM or the SZ model.
    But even if we neglect the opposite effect from Cherenkov radiation,
        we expect the timescale of circularization due to normal gravitational wave emission to be much larger than the age of the universe for the masses and separations we consider.
    For example, in General Relativity, this timescale is about $10^{35}\,\mathrm{yr}$ for a star with mass $1\,M_\odot$ orbiting a galaxy with mass $10^{10}\,M_\odot$ (treated as a point particle) at a distance of $10\,\mathrm{kpc}$ \cite{Amaro-Seoane2007}.
    Indeed, if orbits of stars were to circularize quickly, we wouldn't observe significant asymmetric drift \cite{Martinsson2013}.
    In the models we consider here, this timescale may be modified.
    One possible reason being that not only the usual Newtonian force acts on the star.
    But we don't expect this timescale to be sufficiently reduced to become relevant on galactic scales.
}
Additionally, as mentioned above, there may be $\Rgal$-dependent factors in $\tau_E$ that we have so far neglected.
We have numerically verified that our conclusions hold even with an $\Rgal$-dependent $\tau_E$.
Specifically, we have verified that for $\tau_E \propto \Rgal$ (relevant for standard SFDM, see Sec.~\ref{sec:standardsfdm}) and $\tau_E \propto \Rgal^2$ (relevant for some cases of dynamical friction \cite{VandenBosch1999}),
    Eq.~\eqref{eq:dotRgalappendix} still accurately captures the transition to smaller radii.
We have further verified that the orbital decay still happens on a timescale $\tau_E$ in these cases.

\subsection{Regime of validity}
\label{sec:general:validity}

We now come back to the question of how to choose the cutoffs $k_{\mathrm{min}}$ and $k_{\mathrm{max}}$ in our calculation of the energy loss due to Cherenkov radiation.
We will argue that reasonable values are
\begin{subequations}
\begin{align}
 k_{\mathrm{min}} &\sim 1/\mathrm{kpc} \sim 10^{-26}\,\mathrm{eV} \,, \\
 k_{\mathrm{max}} &\sim f_p \cdot \sqrt{\frac{a_0}{G M}} \sqrt{\frac{a_b^{\mathrm{gal}}}{a_0}} \sim 10^{-22}\,\mathrm{eV} \cdot f_p \cdot \sqrt{\frac{a_b^{\mathrm{gal}}}{a_0}}  \,.
\end{align}
\end{subequations}
Here, $k_{\mathrm{min}}$ is determined by the scale on which the background field $\varphi_0$ varies.
In galaxies, this is typically $\mathcal{O}(\mathrm{kpc})$.
The upper cutoff $k_{\mathrm{max}}$ is determined by the distance from the perturber where its field drops below the galaxy's background field,
    i.e. the distance from where on we can actually treat the perturber's field as a perturbation.
This is related to the perturber's MOND radius $r_{\mathrm{MOND}} = \sqrt{G M/a_0}$ outside of which the MOND regime begins.
It's also related to the galaxy's Newtonian baryonic acceleration $a_b^{\mathrm{gal}} = G M_{\mathrm{gal}}/\Rgal^2$ at the perturber's position.
For $M = M_\odot$, we have $r_{\mathrm{MOND}} \approx 3 \cdot 10^{-5}\,\mathrm{kpc}$,
    corresponding to $k_{\mathrm{max}} \approx 10^{-22}\,\mathrm{eV}$.
Since we assume the galaxy to be in the MOND regime for our calculations, the factor $\sqrt{a_b^{\mathrm{gal}}/a_0}$ is smaller than 1, but large enough such that $k_{\mathrm{max}}$ is still much larger than $k_{\mathrm{min}}$.
The factor $f_p$ is a model-dependent correction factor.

Our method of calculation gives a nonzero energy loss only for $k_{\mathrm{max}} > k_{\mathrm{min}}$.
This is because these are the limits of a momentum integral, see Appendix~\ref{sec:Edotcalculation}.
Thus, our method gives useful results only for $M \lesssim 10^8\,M_\odot$.
This means we can apply our results to stars and also to many globular clusters.
But most dwarf galaxies are too heavy.
In the following, we will mainly consider stars.

We choose the cutoffs $k_{\mathrm{min}}$ and $k_{\mathrm{max}}$ to ensure that our prototype Lagrangian from Eq.~\eqref{eq:prototype} is valid.
The conditions for this are:
\begin{itemize}
 \item Our calculation assumes that $\coup$ and $\bar{c}$ do not vary on the scales under consideration.
       Also, in deriving a Lagrangian like Eq.~\eqref{eq:prototype} for perturbations,
        one usually neglects derivatives of the background fields against derivatives of the perturbations.
       This restricts us to scales smaller than $\sim\mathrm{kpc}$ since the galactic background varies on this scale.
       This means we need to choose
       \begin{align}
        k_{\mathrm{min}} \gtrsim 1/\mathrm{kpc} \,.
       \end{align}
 \item Close to a star, the baryonic acceleration is larger than the MOND acceleration scale $a_0$ and we are in the Newtonian regime.
       Depending on the model, other terms in the Lagrangian may become important there.
       For example, higher derivative terms might become important \cite{Babichev2011}.
       Thus, our prototype Lagrangian may not be valid in this regime.
       Or it may be valid but with different parameters $g_m$ and $\bar{c}$.
       This Newtonian regime ends at a radius of about
       \begin{align}
        r_{\mathrm{MOND}} = \sqrt{G M/a_0} \,.
       \end{align}
       This means we should restrict our calculation to larger scales, i.e. we should choose
       \begin{align}
        k_{\mathrm{max}} \lesssim \frac1{r_{\mathrm{MOND}}} \,.
       \end{align}
 \item Close to a star, the field due to the star dominates compared to the background field of the host galaxy.
       Thus, we are not allowed to expand around the galaxy's background field.
       This regime ends roughly at a radius
       \begin{align}
        \label{eq:rmaxpert}
        r_{\mathrm{pert}} = \Rgal \sqrt{M/M_{\mathrm{gal}}} \,.
       \end{align}
       We can write this as
       \begin{align}
        r_{\mathrm{pert}} = r_{\mathrm{MOND}} \sqrt{\frac{a_0}{a_b^{\mathrm{gal}}}} \,.
       \end{align}
       Thus, we should choose
       \begin{align}
        \label{eq:kmaxpert1}
        k_{\mathrm{max}} \lesssim \frac1{r_{\mathrm{MOND}}} \sqrt{\frac{a_b^{\mathrm{gal}}}{a_0}} \,.
       \end{align}
 \item The estimate Eq.~\eqref{eq:rmaxpert} gives the radius where the static field of a static perturber becomes smaller than the host galaxy's background field.
       However, we are interested in a dynamical situation where Cherenkov radiation is emitted.
       It is possible that a different condition controls when the dynamical field due to the perturber is small compared to the galaxy's static background field.
       In general, this cannot be estimated in a model-independent way.
       Thus, we introduce a factor $f_p \lesssim 1$ to allow for a smaller cutoff
       \begin{align}
        \label{eq:kmaxpert2}
        k_{\mathrm{max}} \lesssim f_p \frac1{r_{\mathrm{MOND}}} \sqrt{\frac{a_b^{\mathrm{gal}}}{a_0}} \,.
       \end{align}
       For standard SFDM, we have explicitly checked which values of $f_p$ are allowed,
            see Appendix~\ref{sec:standasfdm:perturbationssmall}.
       We find that we can set $f_p = 1$ for $\vec{V} \perp \hat{a}$,
        \begin{align}
         f_p^\perp =1\,.
        \end{align}
       For $\vec{V} \parallel \hat{a}$, we may need to decrease $f_p$ a bit.
       The worst case is for parallel (rather than anti-parallel) $\vec{V}$ and $\hat{a}$ with $\barbeta = 3/2$ and a velocity $V$ that is just barely larger than the critical velocity.
       In this case, we estimate
       \begin{align}
        \left(f_p^\parallel\right)^{\mathrm{worst\,case}} \approx 1 / \sqrt{2.8} \,.
       \end{align}
       For larger values of $\barbeta$, larger velocities, or for anti-parallel $\vec{V}$ and $\hat{a}$,
            we can choose $f_p^\parallel$ much closer to $1$.
       As explained in Appendix~\ref{sec:Edotcalculation:standardsfdm:cutoffs},
          introducing cutoffs in the way we do means we cannot take $V \to V_{\mathrm{crit}}$ without perturbations becoming large.
       Thus, here we restrict ourselves to $V$ at least $1\%$ away from $V_{\mathrm{crit}}$.
       This does not significantly affect any of our conclusions.
 \item We need to choose $k_{\mathrm{max}} < 2 M (V - \bar{c} \sqrt{2})$.
       This is the standard cutoff in Cherenkov radiation calculations, see Appendix~\ref{sec:Edotcalculation}.
       However, numerically, this is not relevant in our case because the other cutoffs we considered are much more restricting.
       Concretely, we have $2 M V \sim 10^{63}\,\mathrm{eV} \cdot (M/M_\odot) \cdot (V/200\mathrm{km/s})$.
       This is the reason our energy loss behaves like a friction force without a significant recoil.
       We restrict $k$ to be much smaller than the perturber's momentum $M V$.
\end{itemize}

In principle, there can be additional model-dependent cuts,
    i.e. further restrictions on the validity of our prototype Lagrangian.
Here, we assume that there are no such cuts or, at least, that they are less restricting than the cuts we already considered above.

As long as $k_{\mathrm{min}} \ll k_{\mathrm{max}}$, the result Eq.~\eqref{eq:taukin1} for the timescale $\tau_E$ is dominated by $k_{\mathrm{max}}$.
Then, with our particular choice of $k_{\mathrm{min}}$ and $k_{\mathrm{max}}$,
\begin{align}
 \tau_E = \frac{V^3}{f_a f_p^2 \bar{c}^2 \coup^2 a_0} \cdot \frac{a_0}{a_b^{\mathrm{gal}}} \,.
\end{align}
This is independent of the perturber's mass, but depends on the position of the perturber in the galaxy.
Numerically,
\begin{align}
 \tau_E = \frac{2\cdot10^8\,\mathrm{yr}}{f_a f_p^2 \, \coup^2} \cdot \left(\frac{V/\bar{c}}{2}\right)^2 \cdot \left(\frac{a_0}{a_b^{\mathrm{gal}}}\right) \cdot \left(\frac{V}{200\,\mathrm{km}/\mathrm{s}}\right) \cdot \left(\frac{1.2 \cdot 10^{-10}\,\mathrm{m}/\mathrm{s}^2}{a_0}\right) \,.
\end{align}
Thus, for $\coup$ of order 1 and $V > V_{\mathrm{crit}}$, stars in galaxies lose a significant fraction of their energy on timescales $\tau_E$ which are not much larger than typical galactic timescales.

As already mentioned above, our result for $|\dot{E}|$ is a conservative lower bound.
The actual energy loss may be higher.
In particular, modes with $k > k_{\mathrm{max}}$ and $k < k_{\mathrm{min}}$ may contribute to the energy loss, but are not considered here.

\section{Application to standard SFDM}
\label{sec:standardsfdm}

We will now apply the results of Sec.~\ref{sec:general} to superfluid dark matter \cite{Berezhiani2015}.
We refer to this original model from Ref.~\cite{Berezhiani2015} as standard {\sc SFDM} to differentiate it from the two-field {\sc SFDM} model from Ref.~\cite{Mistele2020} that we discuss in Sec.~\ref{sec:twofieldsfdm}.
The idea of {\sc SFDM} is to introduce a new type of particle which condenses to a superfluid on galactic scales where the phonons then mediate a {\sc MOND}-like force.
On cosmological scales, this particle simply behaves as cold dark matter.

Concretely, Ref.~\cite{Berezhiani2015} proposes the following non-relativistic Lagrangian for the field $\theta$ that carries both the superfluid and the {\sc MOND} force,
\begin{align}
 \label{eq:sfdm}
 \mathcal{L} = \frac{2\Lambda}{3} (2m)^{3/2} \sqrt{|X - \barbeta Y|} X - \frac{\baralpha \Lambda}{M_{\rm{Pl}}} \rho_b \, \theta \,,
\end{align}
with
\begin{align}
 X = \dot{\theta} + \hat{\mu} -  (\vec{\nabla} \theta)^2/(2m) \,, \quad Y = \dot{\theta} + \hat{\mu} \,, \quad \hat{\mu} = \mu_{\rm{nr}} - m \phi_{\rm{N}} \,.
\end{align}
Here, $m$ is the mass of the superfluid constituent's particles, $\mu_{\rm{nr}}$ is the non-relativistic chemical potential, $\Lambda$ is a constant with mass-dimension one, $\baralpha$ is a dimensionless constant, $\barbeta$ parametrizes finite-temperature effects,
    and $\phi_N$ is the Newtonian gravitational potential.
The non-relativistic chemical potential $\mu_{\rm{nr}}$ is related to the relativistic chemical potential $\mu$ by $\mu = m + \mu_{\rm{nr}}$ with $\mu_{\rm{nr}} \ll m$.
In the static limit, the phonon force $\vec{a}_\theta = - (\alpha \Lambda/M_{\mathrm{Pl}}) \vec{\nabla} \theta$ has a MOND-like form for
\begin{align}
 \label{eq:mondlimit}
 (\vec{\nabla} \theta)^2 \gg 2 m \hat{\mu} \,.
\end{align}
We refer to this limit as the MOND limit of SFDM.
In this case,\footnote{This assumes $\barbeta$ to be of order 1. Otherwise, the {\sc MOND} limit is reached for $(\vec{\nabla} \theta)^2 \gg 2 m \hat{\mu} \barbeta$.}
\begin{align}
 X \approx X - \barbeta Y \approx - (\vec{\nabla} \theta)^2/(2m) \,.
\end{align}
The quantity $\barbeta$ must be larger than $3/2$ to fix an instability\footnote{In the {\sc MOND} limit, $\barbeta > 1$ is sufficient, but more generally $\barbeta > 3/2$ is needed.} and it must be smaller than $3$ so that the superfluid energy density is positive \cite{Berezhiani2015}.

In the following, we assume the MOND limit $(\vec{\nabla} \theta)^2 \gg 2 m \hat{\mu}$.
In Refs.~\cite{Mistele2020, Mistele2022} we estimated that many galaxies cannot reach this MOND limit.
Here, we assume the MOND limit anyway for two reasons.
First, this simplifies the equations so that our results for standard SFDM serve as an illustrative example of how one can constrain hybrid models using Cherenkov radiation from non-relativistic objects.
Second, %
    our results provide an independent method to rule out that a given galaxy is in the MOND limit.
That is, compared to our estimates from Refs.~\cite{Mistele2020, Mistele2022}, our Cherenkov radiation constraints can rule out a MOND limit for a different set of galaxies and using a different physical mechanism.
Of course, our constraints can be avoided by simply allowing galaxies to go outside the MOND limit.
But the MOND limit is one of the main motivations behind SFDM.
Knowing under which circumstances such a MOND limit can exist is important.

We will use the observed Milky Way rotation curve to rule out a MOND limit in the Milky Way for certain values of the SFDM model parameters.
Specifically, assuming a MOND limit, we find constraints on the parameter combination $\sqrt{\baralpha}/m$ for each fixed value of the parameter $\barbeta$, which parametrizes finite-temperature corrections.
For the fiducial value $\barbeta = 2$ from Ref.~\cite{Berezhiani2018},
we will rule out roughly the interval
\begin{align}
    0.34\,\mathrm{eV}^{-1} \lesssim \frac{\sqrt{\baralpha}}{m} \lesssim 3.29\,\mathrm{eV}^{-1} \,.
\end{align}
This includes the fiducial value $\sqrt{\baralpha}/m = 2.4\,\mathrm{eV}^{-1}$ from Ref.~\cite{Berezhiani2018}.

\subsection{Perturbations}

We consider perturbations on top of a background galaxy in the MOND limit $(\vec{\nabla} \theta)^2 \ll 2 m \hat{\mu}$.
We further consider the MOND regime $a_b < a_0$,
    so that potential higher-order terms are negligible, as discussed in Sec.~\ref{sec:general:validity}.
Then, the perturbed Lagrangian is (see Appendix~\ref{sec:standardsfdmpert})
\begin{align}
 \mathcal{L} = \frac12 \dot{\delta}^2 \bar{c}^{-2} - \frac12 (\vec{\nabla} \delta)^2 - \frac12 (\hat{a} \vec{\nabla} \delta)^2 + \bar{c}^{-1} (3-\barbeta) \bar{f}_\barbeta \hat{a} \vec{\nabla} \delta \dot{\delta} - \frac{\coup}{\sqrt{2} M_{\mathrm{Pl}}} \delta \, \delta_b\,,
\end{align}
with
\begin{align}
 \bar{c} =  3 \bar{f}_\barbeta \, \frac{|\vec{a}_{\theta_0}|}{a_0} \frac{\baralpha^2 \Lambda}{m}\,, \quad \coup = \sqrt{\frac{a_0}{|a_{\theta_0}|}} \,, \quad \hat{a} = \frac{\vec{\nabla} \theta_0}{|\vec{\nabla} \theta_0|} = - \frac{\vec{a}_{\theta_0}}{|a_{\theta_0}|} \,,
\end{align}
and
\begin{align}
 \bar{f}_\barbeta =  \frac{1}{\sqrt{3 \left(\barbeta-1\right) \left(\barbeta+3\right)}} \,.
\end{align}
The quantities $\theta_0$ and $a_{\theta_0}$ refer to the background values of $\theta$ and $a_\theta$.
This Lagrangian has the form of our prototype Lagrangian from Eq.~\eqref{eq:prototype} up to the term proportional to $\bar{f}_{\barbeta}$ that mixes spatial and time derivatives.

The quantity $\bar{c}$ determines the sound speed up to order 1 corrections, which we discuss below.
This gives a non-relativistic sound speed, as expected for a non-relativistic superfluid.
For the fiducial parameters from Ref.~\cite{Berezhiani2018} ($m = 1\,\mathrm{eV}$, $\baralpha = 5.7$, $\Lambda = 0.05\,\mathrm{meV}$, $\barbeta = 2$), we have
\begin{align}
 \bar{c} = 1.25 \cdot 10^{-3} \frac{|\vec{a}_{\theta_0}|}{a_0} = 375\,\mathrm{km}/\mathrm{s} \cdot \frac{|\vec{a}_{\theta_0}|}{a_0} \,.
\end{align}
Since we consider the MOND regime $a_b < a_0$, we have $|\vec{a}_{\theta_0}|/a_0 \lesssim 1$.
Thus, indeed $\bar{c} \ll 1$.
The full dispersion relation is, see Appendix~\ref{sec:standardsfdmpert},
\begin{subequations}
\begin{align}
 \omega = \bar{c} |\vec{k}| \left( \sqrt{1 + \gamma^2(1 + f_\barbeta^2)} + f_\barbeta \gamma \right) \,,
\end{align}
\end{subequations}
with $f_\barbeta \equiv (3 - \barbeta) \bar{f}_\barbeta$.
This is the standard result from our prototype Lagrangian up to corrections from $f_\barbeta$, i.e. up to corrections from the mixing of spatial and time derivatives.

Unfortunately,
    due to this mixing of spatial and time derivatives,
    our standard calculation of the energy loss $\dot{E}$ from Appendix~\ref{sec:Edotcalculation} based on the standard QFT formalism does not apply here.
Instead of adjusting the QFT formalism for our case at hand,
    we instead choose to do a classical calculation for standard SFDM.
This calculation is done in Appendix~\ref{sec:Edotcalculation:standardsfdm} and follows that of the standard electromagnetic Cherenkov radiation from Ref.~\cite{Jackson1998}.
The result from the classical calculation has the same form as that from our previous QFT calculation, but with slightly adjusted critical velocities $V_{\mathrm{crit}}$ and $\hat{a}$-dependent factors $f_a$ to account for the mixing of spatial and time derivatives.
We also use this classical calculation to explicitly determine the factor $f_p$.
That is, we check when perturbations are small compared to the galaxy's background field.
We find that we can use $f_p = 1$ for our purposes,
    see Appendix~\ref{sec:standasfdm:perturbationssmall} and Sec.~\ref{sec:general:validity}.

For $\vec{V} \parallel \hat{a}$, the critical velocity $V_{\mathrm{crit}}$ is
    \begin{align}
     V_{\mathrm{crit}}^\parallel = \bar{c} \left(\sqrt{2+f_\barbeta^2} \pm f_\barbeta\right) \,.
    \end{align}
    The plus and minus signs are for parallel and antiparallel orientations, respectively.
    For the parallel case, $V_{\mathrm{crit}}^\parallel$ can be significantly larger compared to our prototype Lagrangian where $V_{\mathrm{crit}}^\parallel = \sqrt{2} \bar{c}$.
    For the antiparallel case, $V_{\mathrm{crit}}^\parallel$ can be significantly smaller.
The timescale $\tau_E$ has the same form as for our prototype Lagrangian
    but with a modified factor $f_a$,
    \begin{align}
     f_a^\parallel = \frac{1}{1 - (\bar{c}/V)^2 \mp 2 f_\barbeta (\bar{c}/V)} \,,
    \end{align}
    where the minus sign is for the parallel and the plus sign for the antiparallel case.
Depending on this sign, $f_a^\parallel$ is smaller or larger than for our prototype Lagrangian where $f_a^\parallel = 1/(1-(\bar{c}/V)^2)$.
Both $V_{\mathrm{crit}}^\parallel$ and $f_a^\parallel$ agree with our standard calculation in the $f_\barbeta = 0$ case.

For $\vec{V} \perp \hat{a}$, the critical velocity $V_{\mathrm{crit}}$ is
        \begin{align}
         V_{\mathrm{crit}}^\perp = \bar{c} \, \sqrt{\frac{2}{2 + f_\barbeta^2}} \,.
        \end{align}
    In this case, some angular integrals are difficult to evaluate analytically, see Appendix~\ref{sec:Edotcalculation:standardsfdm:cutoffs}.
    This makes it difficult to analytically obtain the timescale $\tau_E$.
    But we can do a somewhat more conservative estimate and find as a lower limit on $f_a^\perp$,
        \begin{align}
        \label{eq:sfdmstandard:faperp}
         f_a^\perp = \frac{1}{\sqrt{2}} \frac{1}{1 + f_\barbeta^2} \,.
        \end{align}
The critical velocity $V_{\mathrm{crit}}^\perp$ reproduces our standard result for the $f_\barbeta=0$ case.
But $f_a^\perp$ does not due to the more conservative estimate,
        see Appendix~\ref{sec:Edotcalculation:standardsfdm:cutoffs}.

\subsection{Constraints}
\label{sec:standardsfdm:constraints}

\begin{figure}
 \centering
 \includegraphics[width=.7\textwidth]{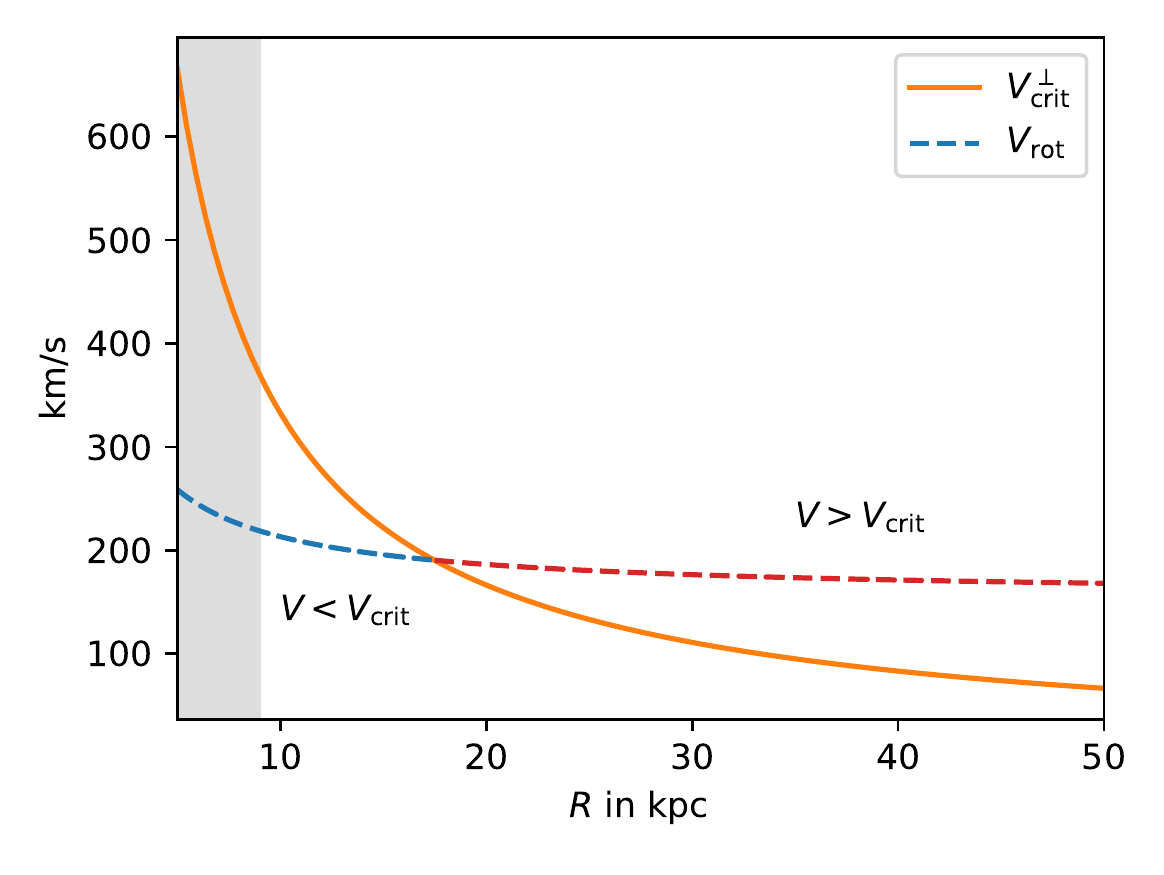}
 \caption{
     Critical velocity $V_{\mathrm{crit}}^\perp$ (solid orange line) and rotation curve $V_{\mathrm{rot}}$ (dashed blue and red line) of a galaxy with mass $M_{\mathrm{gal}} = 5 \cdot 10^{10}\,M_\odot$ concentrated at its center in standard SFDM with the fiducial parameter values from Ref.~\cite{Berezhiani2018}.
     The shaded region is not in the MOND regime since $a_b^{\mathrm{gal}} > a_0$.
     Stars emit Cherenkov radiation when they move faster than the critical velocity.
    }
 \label{fig:sfdm-vrot-vs-soundspeed}
\end{figure}

We can now put quantitative constraints on standard SFDM.
In the MOND limit, a good approximation is the so-called no-curl approximation $|\vec{a}_{\theta_0}| = \sqrt{a_0 a_b^{\mathrm{gal}}}$ \cite{Hossenfelder2020} with $a_b^{\mathrm{gal}} = G M_{\mathrm{gal}}/\Rgal^2$.
This gives
    \begin{align}
    \label{eq:standardsfdm:cbar}
     \bar{c} = \frac{3 \bar{f}_\barbeta}{\sqrt{8\pi}} \frac{\sqrt{\bar{\alpha}}}{m} \sqrt{\frac{M_{\mathrm{gal}}}{M_{\mathrm{Pl}}}} \frac{1}{\Rgal} \,.
    \end{align}
For circular orbits, the critical velocity is $V_{\mathrm{crit}}^\perp = \bar{c} \sqrt{2/(2 + f_\barbeta^2)}$.
This critical velocity scales as $1/\Rgal$ inside a given galaxy.
In contrast, rotation curves are flat at large $\Rgal$.
Thus, there is a galactocentric radius where $V_{\mathrm{crit}}$ drops below the rotation curve velocity $V_{\mathrm{rot}}$.
Beyond this radius,
    stars with velocity $V_{\mathrm{rot}}$ lose energy on timescales $\tau_E$.
This is illustrated in Fig.~\ref{fig:sfdm-vrot-vs-soundspeed}.

For standard SFDM, the timescale $\tau_E$ is
    \begin{align}
     \tau_E = \frac{V^3}{f_a f_p^2 \bar{c}^2  a_0} \cdot \sqrt{\frac{a_0}{a_b^{\mathrm{gal}}}} \,.
    \end{align}
    This depends on parameters of the model mainly through the factor $\bar{c}^{-2}$.
    In principle, $a_0 = \bar{\alpha}^3 \Lambda^2/M_{\mathrm{Pl}}$ also depends these parameters.
    But $a_0$ is mostly fixed by requiring standard MOND-like rotation curves.
    Thus, $\tau_E$ depends on the model parameters as $\tau_E \propto 1/\bar{c}^2$ so that our Cherenkov constraints can be avoided in two ways:
        Either, $\bar{c}$ is sufficiently large so that most stars are subsonic and therefore don't emit Cherenkov radiation since Cherenkov radiation is kinematically forbidden.
        Or $\bar{c}$ is sufficiently small so that stars that are supersonic lose only little energy through Cherenkov radiation.

This means that we need for most stars that either
\begin{align}
    \label{eq:cherenkov-conditions}
    V < V_{\mathrm{crit}} \propto \bar{c} \quad \mathrm{or} \quad \tau_{\mathrm{min}} < \tau_E \propto \frac{1}{\bar{c}^2} \,,
\end{align}
for some reasonable minimum timescale $\tau_{\mathrm{min}}$.
This rules out an interval of $\bar{c}$ values which, through Eq.~\eqref{eq:standardsfdm:cbar}, corresponds to an interval of $\sqrt{\baralpha}/m$ values.
More quantitatively, a rotation curve velocity $V$ at a radius $\Rgal$ rules out the following interval of $\sqrt{\baralpha}/m$ values
\begin{align}
 \label{eq:standardsfdm:rotcurve:interval1}
 \frac{V \Rgal}{\bar{f}_\barbeta} \frac{\sqrt{8\pi}}{3} \sqrt{\frac{M_{\mathrm{Pl}}}{M_{\mathrm{gal}}}} \cdot \left(
     \sqrt{\frac{V}{\tau_{\mathrm{min}} f_a f_p^2} \sqrt{\frac{\Rgal^2}{a_0 G M_{\mathrm{gal}}}} } \,,
     \sqrt{1 + \frac12 f_\barbeta^2}
\right)
\,,
\end{align}
where we used the notation $a \cdot (x_1, x_2) \equiv (a x_1, a x_2)$.
For rotation curves with approximately circular orbits, we can set $f_a = f_a^\perp$ (see Eq.~\eqref{eq:sfdmstandard:faperp}) and $f_p = 1$ as discussed in Sec.~\ref{sec:general:validity} and Appendix~\ref{sec:standasfdm:perturbationssmall}.

For simplicity, we separately consider three different values of $\barbeta$
    \begin{align} \barbeta \in \{3/2, 2, 3\}\,. \end{align}
This covers the range of plausible values for $\barbeta$.
Namely, $\barbeta$ must be larger than $3/2$ to fix an instability\footnote{In the MOND limit, $\barbeta > 1$ is sufficient, but more generally $\barbeta > 3/2$ is needed.} and it must be smaller than $3$ so that the superfluid energy density is positive \cite{Berezhiani2015}.
The fiducial value from Ref.~\cite{Berezhiani2018} is $\barbeta = 2$.
For a fixed value of $\barbeta$, we have $\bar{c}\propto \sqrt{\baralpha}/m$.
This allows us to constrain $\sqrt{\baralpha}/m$ given a value of $\barbeta$.

In standard SFDM, $a_0$ is given by $\bar{\alpha}^3 \Lambda^2/M_{\mathrm{Pl}}$.
Thus, strictly speaking, the lower boundary of the interval in Eq.~\eqref{eq:standardsfdm:rotcurve:interval1} depends on a different combination of model parameters than the upper boundary.
However, in practice this is not important.
    Namely, the lower boundary depends on $a_0$ only very mildly, as $a_0^{-1/4}$.
    For a useful MOND regime, we must choose $a_0$ close to $10^{-10}\,\mathrm{m}/\mathrm{s}^2$ \cite{Lelli2017b, Berezhiani2018}.
    Thus, to get a conservative estimate, we simply take the unusually small but fixed value
        \begin{align}
        \bar{a}_0 \equiv 0.5\cdot10^{-10}\,\mathrm{m}/\mathrm{s}^2 \,,
        \end{align}
    when evaluating the lower boundary of Eq.~\eqref{eq:standardsfdm:rotcurve:interval1} instead of $a_0 = \bar{\alpha}^3 \Lambda^2/M_{\mathrm{Pl}}$.

We can now rule out various intervals of $\sqrt{\bar{\alpha}}/m$ using the observed Milky Way rotation curve.
We assume a Milky Way baryonic mass of $M_{\mathrm{gal}} = 6 \cdot 10^{10}\,M_\odot$ following Refs.~\cite{Mistele2020, Hossenfelder2020},
    and the rotation curve data from Refs.~\cite{McGaugh2019b, Eilers2019}.
We choose $\tau_{\mathrm{min}} = 10^{10}\,\mathrm{yr}$,
    i.e. stars should not lose a significant fraction of their energy in $10^{10}\,\mathrm{yr}$.
For $\barbeta = 3/2$, $2$, and $3$, we list the excluded intervals of $\sqrt{\bar{\alpha}}/m$ from three different galactic radii in Table~\ref{tab:standardsfdm:constraints}.
These radii are, roughly, $\Rgal = 15\,\mathrm{kpc}$, $20\,\mathrm{kpc}$, and $25\,\mathrm{kpc}$.
The ruled out intervals from these three different radii overlap.
Thus, taken together they rule out $\sqrt{\bar{\alpha}}/m$ in a larger interval
    \begin{align}
     \label{eq:standardsfdm:rotcurve:intervalnumeric}
        \sqrt{\frac{6\cdot10^{10}\,M_\odot}{M_{\mathrm{gal}}}} \left(
         \sqrt{\frac{10^{10}\,\mathrm{yr}}{\tau_{\mathrm{min}}}} \left(\frac{6\cdot 10^{10}\,M_\odot}{M_{\mathrm{gal}}}\right)^{1/4} q_l,
         q_h
         \right) \cdot \mathrm{eV}^{-1} \,,
    \end{align}
    for some dimensionless numbers $q_l$ and $q_h$.
We have
    \begin{align}
    \label{eq:standardsfdm:q}
    \begin{split}
     q_l &= 0.25\,, q_h = 2.34\,, \quad \mathrm{for\,} \barbeta =3/2 \,, \\
     q_l &= 0.34\,, q_h = 3.29\,, \quad \mathrm{for\,} \barbeta =2 \,, \\
     q_l &= 0.51\,, q_h = 5.01\,, \quad \mathrm{for\,} \barbeta =3 \,.
    \end{split}
    \end{align}
This also rules out the fiducial value $\sqrt{\baralpha}/m \approx 2.4\,\mathrm{eV}^{-1}$ for $\barbeta=2$ from Ref.~\cite{Berezhiani2018}.

The precise ruled out values of $\sqrt{\baralpha}/m$ depend on our choice for $M_{\mathrm{gal}}$ and $\tau_{\mathrm{min}}$,
    as shown in Eq.~\eqref{eq:standardsfdm:rotcurve:intervalnumeric}.
For example, Ref.~\cite{McGaugh2019b} adopts the somewhat higher baryonic Milky Way mass $M_{\mathrm{gal}} = 7.4\cdot10^{10}\,M_\odot$.
This would decrease both the upper and lower boundaries of the excluded $\sqrt{\baralpha}/m$ interval by about $10\%$.
The value of $\tau_{\mathrm{min}}$
    enters our result as the inverse square root,
    but affects only the lower interval boundary.
Uncertainties in the measured rotation curve affect both the upper and lower boundaries of the excluded $\sqrt{\baralpha}/m$ interval, see Eq.~\eqref{eq:standardsfdm:rotcurve:intervalnumeric}.
However, the formal errors in the rotation curve from Refs.~\cite{Eilers2019, McGaugh2019b} are less than $5\%$ for $\Rgal \lesssim 25\,\mathrm{kpc}$.
Thus, the uncertainties in $M_{\mathrm{gal}}$ and $\tau_{\mathrm{min}}$ are the dominant sources of uncertainty in our constraints.

In addition, we assumed for simplicity that $a_b^{\mathrm{gal}} = G M_{\mathrm{gal}}/\Rgal^2$ which is valid only in spherical symmetry, but not for a disk galaxy like the Milky Way.
We have verified that this underestimates the correct axisymmetric value of $a_b^{\mathrm{gal}}$ by about $10-15\%$ in the range of radii considered here.
Taking this into account would decrease both the upper and lower boundaries of the excluded interval by about $5-7\%$ which is not a large change compared to the other uncertainties.
Thus, we choose to keep things simple for now and keep the spherically-symmetric assumption for $a_b^{\mathrm{gal}}$.

It may be possible to push the upper boundary of our excluded interval of $\sqrt{\baralpha}/m$ even higher by considering hypervelocity stars (HVS) or globular clusters (GCs).
The reason is that HVS and GCs can be at much larger galactic radii and have much higher velocities than the Milky Way stellar rotation curve.
Thus, the quantity $V \Rgal$ can be much larger, which allows to reach much higher values of $\sqrt{\baralpha}/m$.
However, in practice, this is more complicated, because HVS and GCs cannot be assumed to be on circular orbits.
Thus,
    we are forced to make various worst-case assumptions
    if we know only the distance to the host galaxy's center and total velocity
    and nothing more about the orbit.
That is, in general, we are forced to assume the worst-case critical velocity $V_{\mathrm{crit}}$ and direction-dependent factors $f_a$ and $f_p$.
We must also choose a very conservative timescale $\tau_{\mathrm{min}}$
    since we cannot, in general, assume the HVS or GC to have been at its current position for most of the host galaxy's lifetime.

As a result, we cannot get good constraints from HVS or GCs with the simple procedure we use for the Milky Way rotation curve.
Doing better is certainly possible but requires a more detailed modelling of their orbits.
We leave such a more involved analysis for future work.

\begin{table}
 \centering
 \begin{tabular}{c|c|c|c|c}
  $\Rgal$ &
  $V$ &
  $(q_l, q_h)$ for $\barbeta=3/2$ &
  $(q_l, q_h)$ for $\barbeta=2$ &
  $(q_l, q_h)$ for $\barbeta=3$ \\
  $\mathrm{kpc}$ &
  $\mathrm{km}/\mathrm{s}$ &
  &
  &
  \\
  \hline
  ${ 15.2 }$ & ${ 220 }^{ +1 }_{ -1 }$ & $\left( 0.25, 1.56 \right)$ & $\left( 0.34, 2.19 \right)$ & $\left( 0.51, 3.34 \right)$ \\
${ 20.3 }$ & ${ 203 }^{ +3 }_{ -3 }$ & $\left( 0.35, 1.92 \right)$ & $\left( 0.46, 2.70 \right)$ & $\left( 0.69, 4.11 \right)$ \\
${ 24.8 }$ & ${ 202 }^{ +6 }_{ -6 }$ & $\left( 0.47, 2.34 \right)$ & $\left( 0.62, 3.29 \right)$ & $\left( 0.93, 5.01 \right)$
 \end{tabular}
 \caption{
     Assuming a MOND limit in the Milky Way, we rule out $\sqrt{\bar{\alpha}}/m$ intervals $(q_l, q_h) \cdot \mathrm{eV}^{-1}$ from the observed Milky Way rotation curve at different radii.
     We assume $M_{\mathrm{gal}} = 6 \cdot 10^{10}\,M_\odot$ and $\tau_{\mathrm{min}} = 10^{10}\,\mathrm{yr}$.
     See Eq.~\eqref{eq:standardsfdm:rotcurve:intervalnumeric} for how the exclusion interval depends on these quantities.
     The rotation curve data is that of Ref.~\cite{Eilers2019} adopted to the assumptions of Ref.~\cite{McGaugh2019b}, as also used in Ref.~\cite{Hossenfelder2020}.
     The different radii taken together exclude a larger total interval, see Eq.~\eqref{eq:standardsfdm:q}.
 }
 \label{tab:standardsfdm:constraints}
\end{table}

The parameter combination $\sqrt{\baralpha}/m$ that we have constrained is phenomenologically important since the superfluid energy density is directly proportional to powers of it.
For example, in the MOND limit and using the no-curl approximation,
\begin{align}
 \rho_{\mathrm{SF}} = 2 \left(1 - \frac{\barbeta}{3}\right) \left(\frac{m}{\sqrt{\baralpha}}\right)^2 M_{\mathrm{Pl}} \sqrt{a_0 \, a_b^{\mathrm{gal}}} \,,
\end{align}
where we used $\Lambda m^3 = (m/\sqrt{\baralpha})^3 \sqrt{a_0 M_{\mathrm{Pl}}}$.
Thus, our constraints likely have further implications for situations where the superfluid energy density is important, for example for strong lensing.
Investigating this is left for future work.

Our results rule out that the Milky Way is in the MOND limit $(\vec{\nabla} \theta)^2 \gg 2 m \hat{\mu}$ for a certain range of parameters.
As mentioned above, this does not completely rule out this range of parameters.
Being in the MOND limit is one of the main motivations behind SFDM.
Still, we might just accept that the Milky Way is outside the MOND limit\footnote{
    Indeed, the Milky Way model discussed in Ref.~\cite{Hossenfelder2020} goes outside the MOND limit $(\vec{\nabla} \theta)^2 \gg 2 m \hat{\mu}$, see Ref.~\cite{Mistele2020}.
    But we have verified that one can adjust the $\hat{\mu}$ boundary condition in order to obtain other solutions with $\varepsilon_* \lesssim 0.15$ at the radii $15\,\mathrm{kpc} \lesssim R \lesssim 25\,\mathrm{kpc}$ that are relevant here.
    Here, $\varepsilon_* \equiv 2 m \hat{\mu}/(\baralpha M_{\mathrm{Pl}} |a_b^{\mathrm{gal}}|)$ and the condition $\varepsilon_* \ll 1$ is equivalent to the condition $(\vec{\nabla} \theta)^2 \gg 2 m \hat{\mu}$ \cite{Mistele2022}.
    So equilibrium solutions in the MOND limit may, barely, be possible.
    The price to pay for this is that the superfluid core ends shortly after the last measured rotation curve data point.
    The associated dark matter mass is probably too small to be plausible \cite{Hossenfelder2020}.
    Our Cherenkov radiation constraint rules out such solutions without relying on knowing which dark matter masses are or are not plausible in SFDM.
} and hope that this makes it one of only few outliers,
    while most galaxies actually are in the MOND limit.
This should be checked in future work by doing a similar analysis for a comprehensive sample of galaxies with measured rotation curves.
We discuss this in more detail in Sec.~\ref{sec:discussion}.

\section{Application to two-field SFDM}
\label{sec:twofieldsfdm}

In standard {\sc SFDM}, the phonon field $\theta$ plays a double role: It carries the superfluid's energy density and it mediates a {\sc MOND}-like force.
Ref.~\cite{Mistele2020} showed that these two roles are in tension with each other.
As a solution, Ref.~\cite{Mistele2020} proposed a model which is phenomenologically close to standard {\sc SFDM} but splits the two roles of $\theta$ between two fields.
Here, we will refer to this model as two-field {\sc SFDM}.
Its Lagrangian reads
\begin{align}
\mathcal{L} = \mathcal{L}_- + f(K_+ + K_- - m^2) - \frac{\baralpha \Lambda}{M_{\rm{Pl}}} \, \theta_+\, \rho_b \,,
\end{align}
where $K_\pm = \nabla_\alpha \theta_\pm \nabla^\alpha \theta_\pm$ and $\mathcal{L}_-$ is the standard Lagrangian of a complex scalar field $\phi_- = \rho_- e^{-i \theta_-} / \sqrt{2}$ with quartic self-interaction,
\begin{align}
\mathcal{L}_- = (\nabla_\alpha \phi)^*(\nabla^\alpha \phi) - m^2 |\phi|^2 - \lambda_4 |\phi|^4 \,.
\end{align}
The function $f(K) \equiv \sqrt{|K|} K$ is similar to standard SFDM but now contains both $\theta_+$ and $\theta_-$.

The field $\phi_-$ carries the superfluid's energy density, but is not directly coupled to normal matter.
In an equilibrium with chemical potential $\mu$, we have $\dot{\theta}_- = \mu$ and $\rho_-^2 = 2 m \hat{\mu}/\lambda_4$, where $\hat{\mu} = \mu_{\rm{nr}} - m \Phi_{\rm{N}}$ with the non-relativistic chemical potential $\mu_{\rm{nr}}$ as in standard {\sc SFDM}.
In contrast, $\theta_+$ is coupled directly to normal matter and satisfies a {\sc MOND}-like equation in equilibrium.
The four parameters $\baralpha$, $m$, $\lambda_4$, and $\Lambda$ can be expressed through four other parameters that are more closely related to phenomenology,
\begin{equation}
 \begin{alignedat}{2}
 r_0 &= \sqrt{\lambda_4} \, \frac{M_{\rm{Pl}}}{m^2} \,,   \quad & \frac{\sigma}{m} &= \frac{1}{8 \pi} \frac{\lambda_4^2}{m^3} \,, \\
 a_0 &=\frac{\baralpha^3 \Lambda^2}{M_{\rm{Pl}}} \,, \quad &  \bar{a} &= a_0 \left(10^7 \frac{\lambda_4}{\baralpha^2}\right)^2 \,.
 \end{alignedat}
\end{equation}
Here, $r_0$ is the typical length scale of the dark matter halo, $\sigma$ is the self-interaction cross-section calculated from $\mathcal{L}_-$, $a_0$ is the {\sc MOND} acceleration scale, and $\bar{a}$ is an acceleration scale below which equilibrium solutions typically become unstable \cite{Mistele2020}.

Two-field SFDM contains two gapless low-energy modes, roughly corresponding to the two fields $\theta_+$ and $\theta_-$.
Of these, only the mode corresponding to $\theta_-$ has a non-relativistic sound speed
\begin{align}
 c_s = \sqrt{\frac{\hat{\mu}}{m}} \,,
\end{align}
so that only this mode can potentially be radiated away by stars as Cherenkov radiation.
The coupling of this mode to normal matter is suppressed because $\theta_-$ does not directly couple to normal matter.
This is because only $\theta_+$ carries  the MOND-like force in this model, not $\theta_-$.
Cherenkov radiation is possible only through a mixing of $\theta_-$ and $\theta_+$ from the $f(K_+ + K_- - m^2)$ term of the Lagrangian.

We can write down a low-energy effective Lagrangian for this non-relativistic mode.
This gives our prototype Lagrangian Eq.~\eqref{eq:prototype} without the $\hat{a}$ term and with (see Appendix~\ref{sec:twofieldsfdmpert})
\begin{align}
 \bar{c} = \sqrt{\frac{\hat{\mu}}{m}} \,, \quad \coup = -\frac{\sqrt{\lambda_4}}{\baralpha} \frac{\gamma}{1+\gamma^2} \frac{a_0}{|\vec{a}_{\theta_+^0}|} \,.
\end{align}
This is different from our prototype model in two ways.
First, since there is no $\hat{a}$ term, we have $c_s = \bar{c}$ without the $\sqrt{1+\gamma^2}$ factor.
As a result, the critical velocity $V_{\mathrm{crit}}$ is $V_{\mathrm{crit}} = \bar{c}$ independently of the orientation of $\vec{V}$.
Similarly, we have $f_a = 1$ independently of the orientation of $\vec{V}$.
Apart from that, the energy-loss time scale $\tau_E$ is not affected by the missing $\hat{a}$ term.
Second, the coupling $\coup$ depends on $\gamma$.
This requires a more careful evaluation of the integrals in the $\dot{E}$ calculation, as we will discuss below.

Numerically, the combination $\sqrt{\lambda_4}/\bar{\alpha}$ that controls the coupling $\coup$ is small,
\begin{align}
 \frac{\sqrt{\lambda_4}}{\baralpha} = \frac{1}{10^{7/2}} \left(\frac{\bar{a}}{a_0}\right)^{1/4},
\end{align}
where $\bar{a} \ll a_0$ is the minimum baryonic acceleration below which the equilibrium on galactic scales typically becomes unstable \cite{Mistele2020}.
Thus, this non-relativistic mode does not have a standard gravitational coupling constant of order $\sqrt{G}$ to normal matter.
This is because this mode couples to normal matter only indirectly through a mixing between $\theta_+$ and $\theta_-$.
This enhances the timescale $\tau_E$.

In addition to the cutoffs discussed in Sec.~\ref{sec:general}, this model requires another model-dependent cutoff, see Appendix~\ref{sec:twofieldsfdmpert:eom} or Ref.~\cite{Berezhiani2020}.
Namely, the dispersion relation $\omega = c_s k$ of the non-relativistic mode receives corrections from higher orders of $k$ for $k \gtrsim m c_s$.
Thus, our standard calculation applies only if we introduce a cutoff of order $m c_s$.
However, in practice this is not relevant since the other cutoffs introduced in Sec.~\ref{sec:general} are much more restrictive.
Roughly, the $m c_s$ cutoff becomes relevant only if $m c_s \lesssim 10^{-22}\,\mathrm{eV}$.
Typical values of $c_s$ are of order $100\,\mathrm{km}/\mathrm{s}$.
Thus, the $m c_s$ cutoff is relevant only if $m \lesssim 10^{-18}\,\mathrm{eV}$ which is much smaller than typical masses in two-field SFDM \cite{Mistele2020}.

\begin{figure}
 \centering
 \includegraphics[width=.7\textwidth]{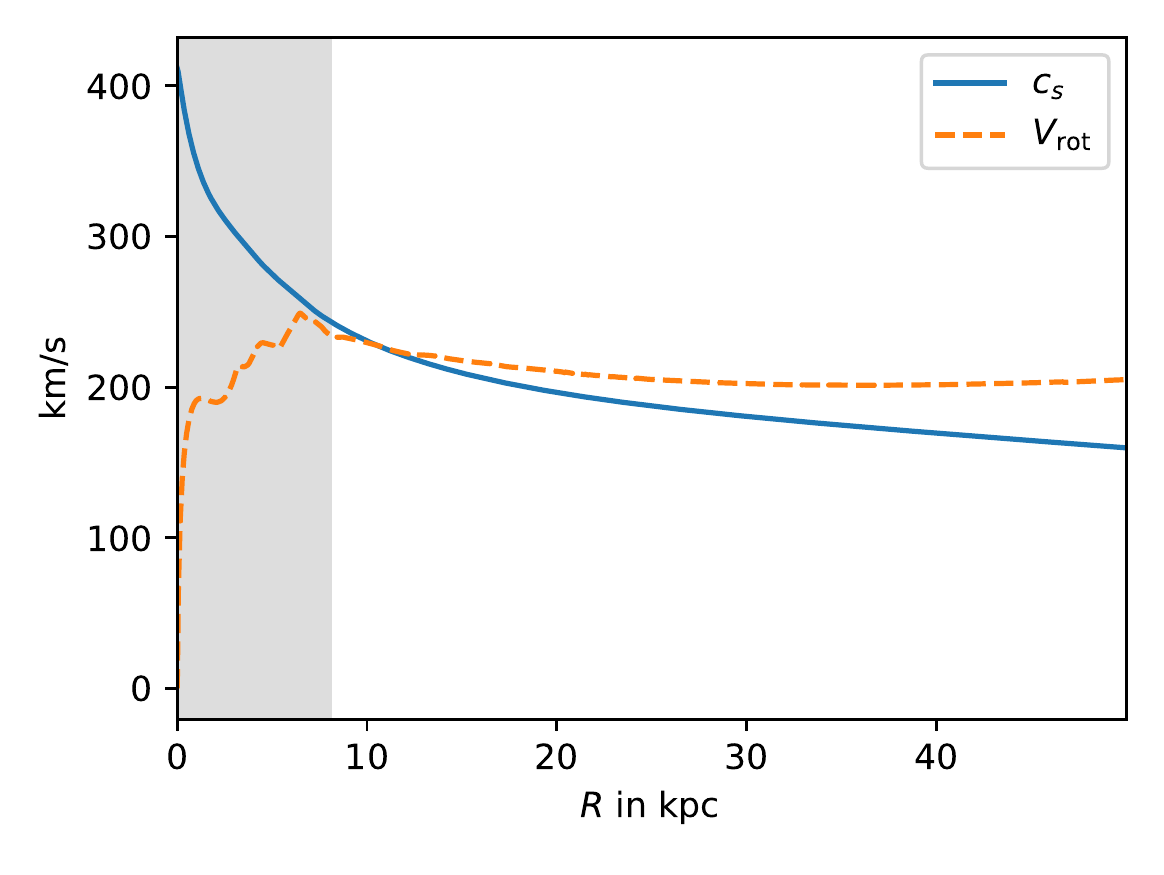}
 \caption{
     Sound speed $c_s$ and rotation curve $v_{\mathrm{rot}}$ for the two-field SFDM Milky Way model from Ref.~\cite{Mistele2020}.
     The shaded region is not in the MOND regime since $a_b^{\mathrm{gal}} > a_0$.
    }
 \label{fig:two-field-sfdm-vrot-vs-soundspeed}
\end{figure}

In many galaxies,
   at least some stars will be supersonic and lose energy through Cherenkov radiation,
    since typically $c_s = \sqrt{\hat{\mu}/m} \sim \mathcal{O}(100\,\mathrm{km}/\mathrm{s})$,
    as mentioned above.
The precise value of $c_s$ depends on the boundary condition $\mu_\infty$ of $\hat{\mu}$ which controls how much dark matter a galaxy contains \cite{Mistele2020}.
Typically, $c_s$ falls off as a function of galactocentric radius since $\hat{\mu}/m$ does.
Indeed, in spherical symmetry $\hat{\mu}'(r)/m = - G(M_b + M_{\mathrm{SF}})/r^2 < 0$.
In contrast, rotation curves become constant at large radii.
Thus, as in standard SFDM, there is typically a critical radius $R_c$ beyond which the rotational velocity is supersonic and stars emit Cherenkov radiation.
This is illustrated in Fig.~\ref{fig:two-field-sfdm-vrot-vs-soundspeed}, which shows the sound speed and rotation curve of the Milky Way model discussed in Ref.~\cite{Mistele2020}.

However, two-field SFDM actually avoids our Cherenkov radiation constraints since $\tau_E$ is larger than galactic timescales, as we will now explain.
As mentioned above, we need to revisit our calculation to include the factor of $\gamma/(1+\gamma^2)$ in $\coup$.
If, for simplicity, we first set $\gamma = 1$ in $\coup$, we can use our standard calculation and find
\begin{align}
\tau_E^{\gamma=1}
    = \frac{V^3}{f_p^2 \coup^2 \bar{c}^2 a_0} \frac{a_0}{a_b^{\mathrm{gal}}}
    = 4 \frac{V^3}{f_p^2 \bar{c}^2 a_0} \frac{a_0}{a_b^{\mathrm{gal}}} \frac{\baralpha^2}{\lambda_4} \left( \frac{|\vec{a}_{\theta_+^0}|}{a_0}\right)^2
    = 4 \cdot 10^7 \frac{V^3}{f_p^2 \bar{c}^2 a_0}  \sqrt{\frac{a_0}{\bar{a}}} \,.
\end{align}
Here, we assumed the MOND limit value $a_{\theta_+^0} = \sqrt{a_0 a_b^{\mathrm{gal}}}$ so that the factor $a_0/a_b^{\mathrm{gal}}$ cancels.
Assuming the MOND limit is usually justified in two-field SFDM \cite{Mistele2020}.
In the special cases $\vec{V} \perp \hat{a}$ and $\vec{V} \parallel \hat{a}$, the corrections due to the factor $\gamma/(1+\gamma^2)$ in $\coup$ are calculated in Appendix~\ref{sec:Edotcalculation:twofieldsfdm},
\begin{align}
 \tau_E^\parallel &= \frac{\tau_E^{\gamma=1}}{4} \cdot \left(\frac{c_s}{V} + \frac{V}{c_s}\right)^2 \,, \\
 \tau_E^\perp &= \frac{\tau_E^{\gamma=1}}{4} \cdot \frac{2^{5/2} \left(1 - \frac12 \left(\frac{c_s}{V}\right)^2 \right)^{3/2}}{1 - \left(\frac{c_s}{V}\right)^2} \,.
\end{align}
Thus, the timescale $\tau_E$ is significantly enhanced for fast stars with $\vec{V} \parallel \hat{a}$ and for barely supersonic stars with $\vec{V} \perp \hat{a}$.
Otherwise, the order of magnitude is that of $\tau_E^{\gamma=1}$,
\begin{align}
 \tau_E^{\gamma=1} = \frac{8 \cdot 10^{16}\,\mathrm{yr}}{f_p^2} \cdot \left(\frac{V/\bar{c}}{2}\right)^2 \cdot \left(\frac{V}{200\,\mathrm{km}/\mathrm{s}}\right) \cdot \left(\frac{1.2\cdot10^{-10}\,\mathrm{m}/\mathrm{s}^2}{a_0}\right) \cdot \sqrt{\frac{10^{-2} a_0}{\bar{a}}} \,.
\end{align}
As discussed in Ref.~\cite{Mistele2020},
    $\bar{a}$ must be much smaller than $a_0$ for a useful MOND limit.
Thus,
    for reasonable values of $\bar{a}$,
    the timescale $\tau_E$ is much larger than galactic timescales.
That is, our method does not constrain two-field SFDM.
The reason is that the non-relativistic gapless mode couples to normal matter only indirectly through a mixing.

\section{Application to the SZ model}
\label{sec:relmond}

Recently, Skordis and Z\l o\'{s}nik proposed a novel type of hybrid MOND dark matter model \cite{Skordis2020}.
This model has a cosmological limit that reproduces the CMB, a static limit that reproduces MOND, and has gravitational tensor modes that propagate at the speed of light.
In the cosmological limit, a scalar field $\phi(t)$ is responsible for a dark fluid that plays the role of dark matter.
In the late-time static limit relevant for galaxies, this model has $\phi = Q_0 \cdot t + \varphi$ where $\varphi$ carries the MOND-like force and $Q_0$ is a constant.
That is, the field $\phi$ plays a double role, analogously to the phonon field $\theta$ in standard SFDM.
Cosmological and galactic phenomena share a common origin.
Therefore, our method is, in principle, able to constrain this model.

This SZ model is complex and we will not attempt a full calculation of perturbations on top of a galactic background.
Instead, we follow Refs.~\cite{Skordis2020,Skordis2021} and consider perturbations on top of the late-time Minkowski background with $\phi = Q_0 t$.
We assume that this gives results that are at least qualitatively valid in galaxies.
Concretely, we consider $g^{\alpha \beta} = \eta^{\alpha \beta} - h^{\alpha \beta}$ with the Minkowski metric $\eta_{\alpha \beta}$, the vector field $A_\alpha = (-1 + \frac12 h_{00}, \vec{A})$, and the scalar field $\phi = Q_0 \cdot t + \varphi$.
Then, the second-order Lagrangian reads \cite{Skordis2020, Skordis2021}
\begin{align}
\mathcal{L} =&
\KB |\dot{\vec{A}} - \frac{1}{2} \vec{\nabla} h_{00}|^2
- 2\KB \vec{\nabla}_{[i} A_{j]} \vec{\nabla}^{[i} A^{j]}
\nonumber
\\
&
+ 2 \left(2 - \KB\right)  \left(  \dot{\vec{A}}  -  \frac{1}{2}  \vec{\nabla} h_{00} \right) \cdot \vec{\nabla} \varphi
+ 2 (2 - \KB)Q_0 \vec{A}_i \left( - \frac12 \partial_i h_{00}\right)
\nonumber
\\
&-(2-\KB) (1 + \lambda_s) \left( \vec{A}^2 Q_0^2  + (\vec{\nabla}\varphi)^2 + 2 Q_0  A^i(\partial_i \varphi) \right)
\nonumber
\\
& +2 \Ktwo \left(\dot{\varphi} + \frac12 h_{00} Q_0  \right)^2
+ \frac{1}{\tilde{M}_{\mathrm{Pl}}^2} T_{\alpha \beta} h^{\alpha \beta}
 \,,
\end{align}
where we left out the standard metric perturbations from the Ricci scalar $R$ for simplicity.
Here, $\lambda_s = \mathcal{J}'(0)$ where the function $\mathcal{J}$ determines the MOND interpolation function \cite{Skordis2020}.
The quantities $\KB$, $Q_0$, $\Ktwo$, and $\tilde{M}_{\mathrm{Pl}}$ are constants.

There is a peculiar term proportional to $\Ktwo Q_0^2 h_{00}$ in this Lagrangian.
This acts as a kind of mass term for $h_{00}$ in the static limit \cite{Skordis2020, Skordis2021}.
On galactic scales, a useful static limit with a long-range gravitational force then requires \cite{Skordis2020}
\begin{align}
 \label{eq:relmond:Ktwocond}
 m_{\mathrm{SZ}} \equiv \sqrt{\frac{2 \Ktwo}{2-\KB}} Q_0 \lesssim 1/\mathrm{Mpc} \,.
\end{align}

There is one scalar mode relevant at low energies \cite{Skordis2020, Skordis2021} with sound speed
\begin{align}
 c_s = \frac{c'}{\sqrt{\Ktwo \KB}} \,, \qquad  c'^2 = \left(2 - \KB\right) \left(1 + \frac12 \lambda_s \KB\right) \,.
\end{align}
Since we did not include the galaxy's background field,
    there is no term $(\hat{a} \vec{\nabla} \varphi)^2$ in the Lagrangian and the sound speed is $c_s = \bar{c}$ without a factor $\sqrt{1+\gamma^2}$.
In a full calculation including the galaxy's background field, we expect such a term to be present.
Here, we assume that such corrections do not change the order of magnitude of our result.

Strictly speaking, this scalar mode is a massive mode, i.e. it has dispersion relation $\omega^2 = c_s^2 k^2 + \mathcal{M}^2$ with nonzero $\mathcal{M}$ \cite{Skordis2020, Skordis2021},
\begin{align}
 \mathcal{M}^2 = \frac{2-\KB}{\KB} (1+\lambda_s) Q_0^2 \,.
\end{align}
Thus, in general, our Cherenkov radiation calculation does not apply.
However, the mass $\mathcal{M}$ is negligible for the wavevectors $\vec{k}$ we consider here,
\begin{align}
 \label{eq:relmond:masscond}
 \frac{\mathcal{M}^2}{c_s^2 k^2} = \frac{m_{\mathrm{SZ}}^2}{k^2} \frac{2-\KB}{2} \frac{1 + \lambda_s}{1 + \frac12 \lambda_s \KB} \ll 1 \,.
\end{align}
To see this, first use the condition Eq.~\eqref{eq:relmond:Ktwocond} and the fact that we only consider wavevectors $k > k_{\mathrm{min}} \sim 1/\mathrm{kpc}$.
This implies that the first factor in Eq.~\eqref{eq:relmond:masscond} is small.
The other factors cannot counter this because $0 < \KB < 2$ for stability \cite{Skordis2020} and because $\lambda_s$ is small in the MOND limit in which we are interested here \cite{Skordis2020}.

Depending on the numerical values of the parameters $\Ktwo$, $\KB$, and $\lambda_s$, the sound speed $c_s$ may be nonrelativistic.
In this case, stars in galaxies may be supersonic and therefore lose energy to Cherenkov radiation,
    if the scalar mode is coupled to matter.
One possibility to avoid constraints is to choose $c_s$ large enough to make stars subsonic.
But it turns out that this is not necessary.
The reason is that the coupling to matter is sufficiently suppressed to make the energy-loss time scale $\tau_E$ much larger than galactic timescales,
    as we will now explain.

The field $\varphi$ mediates a MOND-like force in galaxies.
Therefore, in the static limit, this field is coupled to normal matter by a standard gravitational coupling.
One might expect that this coupling survives also in dynamical situations with non-zero time derivatives.
In particular, since the low-energy scalar mode discussed above contains the field $\varphi$ \cite{Skordis2020}, one might expect that this mode has a standard gravitational coupling to matter.
However, it turns out that this coupling is suppressed -- at least on-shell and in the regime we are interested in.
That is, the energy loss through Cherenkov radiation is reduced and the timescale $\tau_E$ is enhanced
    so that our constraints are weakened.
We will now discuss this in more detail.

\subsection{Suppressed matter coupling}
\label{sec:relmond_coupling}

The coupling of $\varphi$ to matter in the static limit comes from the $(\dot{\vec{A}} - \frac12 \vec{\nabla} h_{00}) \vec{\nabla} \varphi$ term in the Lagrangian.
Specifically, in the static limit, the $\varphi$ equation of motion is
\begin{align}
 0 = \dots - \frac12 \vec{\nabla}^2 h_{00} \,,
\end{align}
where we only show the contributions from the $(\dot{\vec{A}} - \frac12 \vec{\nabla} h_{00}) \vec{\nabla} \varphi$ term.
Here, $\vec{\nabla}^2 h_{00}$ contains a term that is proportional to the matter density,
    just as in standard general relativity.
That is, the matter coupling of $\varphi$ is due to a mixing of $\varphi$ and $h_{00}$.
Consider now a dynamical situation where time derivatives may be important.
Then, in the $\varphi$ equation of motion, we have
\begin{align}
 \label{eq:sz:phieom}
 0 = \dots + \vec{\nabla} \left(\dot{\vec{A}} - \frac12 \vec{\nabla} h_{00}\right) \,.
\end{align}
The combination $\dot{\vec{A}} - \frac12 \vec{\nabla} h_{00}$ also occurs in the $\vec{A}$ equation of motion,
    from the kinetic term of $\vec{A}$ proportional to $(\dot{\vec{A}} - \frac12 \vec{\nabla} h_{00})^2$.
Roughly,
\begin{align}
 \label{eq:sz:Aeom}
 0 = \dots + \partial_t \left(\dot{\vec{A}} - \frac12 \vec{\nabla} h_{00}\right) \,.
\end{align}
For scalar perturbations, we may choose $\vec{A}$ as the gradient of a scalar field \cite{Skordis2020}.
As a result, spatial derivatives from the kinetic term of $\vec{A}$ drop out.
Then, the $\partial_t^2 \vec{A}$ term may dominate over other terms proportional to $\vec{A}$ in Eq.~\eqref{eq:sz:Aeom},
    even though the dispersion relation $\omega \approx c_s k$ may force time derivatives to be much smaller than spatial derivatives if the sound speed $c_s$ is non-relativistic.
In this case,
    the solution of Eq.~\eqref{eq:sz:Aeom} becomes $\partial_t \vec{A} = \frac12 \vec{\nabla} h_{00} + \dots$.
This cancels the $\vec{\nabla}^2 h_{00}$ term in the $\varphi$ equation of motion Eq.~\eqref{eq:sz:phieom}.
Thus, for sufficiently large $\omega$, the standard coupling of $\varphi$ to matter is absent.

We expect this to hold also for perturbations on top of a galaxy instead of empty Minkowski space.
Roughly, the galaxy introduces a new scale of order $a_0 \sim 10^{-34}\,\mathrm{eV}$.
This is much smaller than $\omega \approx c_s k$ since we typically have a lower cutoff of order $k_{\mathrm{min}} \sim 10^{-26}\,\mathrm{eV}$.
Except possibly for an extremely low sound speed, i.e. $c_s \lesssim 10^{-8} \sim \mathrm{m}/\mathrm{s}$.
We leave a detailed study of this regime for future work.

Thus, in the static limit, the field $\varphi$ has a standard gravitational coupling to matter which is crucial for the MOND force in the static limit.
However, for dynamical modes, this coupling may vanish due to the specific form of the $\vec{A}$ equation of motion.
A more detailed calculation shows that on-shell, i.e. when evaluated for $\omega^2 = c_s^2 k^2 + \mathcal{M}^2$,
    the matter coupling vanishes.
Close to but not exactly on-shell, the matter coupling is suppressed both by being close to on-shell and by an additional factor of $m_{\mathrm{SZ}}/k \ll 1$ which is small for the wavevectors under consideration here.
This is calculated explicitly in Appendix~\ref{sec:relmondpert}.

In the language of our prototype Lagrangian from Eq.~\eqref{eq:prototype},
    the matter coupling of the non-relativistic mode of the SZ model is suppressed by
\begin{align}
 g_m \propto \left(\frac{m_{\mathrm{SZ}}}{k}\right) \cdot \left(\frac{\omega}{\sqrt{c_s^2 k^2 + \mathcal{M}^2}} - 1 \right) \,.
\end{align}
This is to lowest order in both $m_{\mathrm{SZ}}/k$ and $\omega/\sqrt{c_s^2 k^2 + \mathcal{M}^2} - 1$.
As already discussed above, the factor $m_{\mathrm{SZ}}/k$ is small because $m_{\mathrm{SZ}} \lesssim 1/\mathrm{Mpc}$ and we consider only $k \gtrsim 1/\mathrm{kpc}$.

The standard leading order Cherenkov radiation calculation evaluates the coupling on-shell (see Fig.~\ref{fig:cherenkov-feynman} and Appendix~\ref{sec:Edotcalculation}).
Thus, for the SZ model, this leading order vanishes.
Therefore, we expect that the energy loss timescale $\tau_E$ is much larger than galactic timescales and does not constrain the model.

\section{Discussion}
\label{sec:discussion}

Usually, Cherenkov radiation is discussed for highly relativistic objects like cosmic rays.
Above, we have considered non-relativistic perturbers like stars.
That these can emit Cherenkov radiation is a novel feature typical of certain hybrid MOND dark matter models.
Of course, relativistic perturbers can also emit Cherenkov radiation in these models.
We have not discussed them here since our strict cutoffs do not allow useful constraints in this case.
For example, for a proton with mass $m_p$, the upper cutoff $k_{\mathrm{max}}$ that we used in Sec.~\ref{sec:general:validity} would be
\begin{align}
 k_{\mathrm{max}} \sim \sqrt{\frac{M_\odot}{m_p}} 10^{-22}\,\mathrm{eV} \sim \mathrm{MeV} \,.
\end{align}
This is very high compared to the scales where the models discussed above are usually probed, see e.g. Ref.~\cite{Berezhiani2015}.
Thus, the models we discuss may not be valid at such scales.
But even if we ignore this, we do not find a useful constraint.
Concretely, the timescale $\tau_E$ would scale as\footnote{
    Our calculation in Appendix~\ref{sec:Edotcalculation} assumes a non-relativistic perturber.
    Very roughly, we expect $d\dot{E} \sim \omega d\Gamma \to (\omega \gamma) (d\Gamma / \gamma)$  in the relativistic case, where the Lorentz factor $\gamma$ cancels.
    So we can reuse our formulas for $\dot{E}$ to get an order-of-magnitude estimate even for the relativistic case.
}
\begin{align}
 \tau_E \sim \frac{E M_{\mathrm{Pl}}^2}{c_s^2 m_p^2 k_{\mathrm{max}}^2} \sim 10^{29}\,\mathrm{yr} \,,
\end{align}
where we used $c_s = 100\,\mathrm{km}/\mathrm{s}$ and assumed a proton energy $E \sim 10^{11}\,\mathrm{GeV}$.
This is much larger than the age of the universe even for much smaller proton energies.\footnote{
    This estimate assumes that the Cherenkov radiation mode couples to the proton's rest mass.
    Depending on the details of the model under consideration in the relativistic limit,
        the proton rest mass might have to be replaced by its energy, see for example Ref.~\cite{Moore2001}.
    In this case, the cutoff would be smaller but the energy loss rate might be enhanced due to the larger coupling.
    Which effect wins out needs to be checked separately for each specific model.
    In some models, the coupling might even be suppressed compared to the non-relativistic limit.
    For example for scalar fields that are coupled to the trace of the energy-momentum tensor, see Ref.~\cite{Dalang2021}.
}

Above, we have used the Milky Way stellar rotation curve to constrain standard SFDM.
This analysis can be extended straightforwardly to a larger sample of stellar rotation curves such as those from the DiskMass survey \cite{Martinsson2013}.
An even larger sample of rotation curves is provided by the SPARC database \cite{Lelli2016}.
However, we cannot readily use these rotation curves from Ref.~\cite{Lelli2016} since these are not obtained from stars but from gas.

The problem is that gas clouds cannot easily be treated as point particles in the same way we treated stars as point particles in our analysis.
Alternatively, we could try to apply our results to individual neutral hydrogen atoms.
But this also does not work:
Naively, our minimum radius $r_{\mathrm{min}} = 1/k_{\mathrm{max}}$ is roughly $10\,\mathrm{fm}$ in this case.
This corresponds to $k_{\mathrm{max}} \sim \mathrm{MeV}$ which may be too large for the models discussed here,
    as already mentioned above.
But even if we ignore this, our naive $r_{\mathrm{min}}$ of $\sim10\,\mathrm{fm}$ is much smaller than the Bohr radius $\sim 10^5\,\mathrm{fm}$.
Thus, we must anyway decrease our usual cutoff by a factor of about $10^4$ if we want to treat the hydrogen atoms as point particles.
This enhances $\tau_E$ by a factor of about $10^8$
    so that we do not get useful constraints.

A way around these issues would be to assume that stellar and gas rotation curves are the same up to the effect of asymmetric drift which can be on the order of $10\%$ \cite{Martinsson2013}.
Then, we can redo our analysis with the gas rotation curves from the SPARC sample by adding an additional uncertainty to account for asymmetric drift.
Investigating this in more detail is left for future work.

A possible limitation of our results is that our point-particle approximation for stars breaks down if stars are closer to each other than their effective size $r_{\mathrm{min}} = 1/k_{\mathrm{max}}$.
Numerically,
\begin{align}
 r_{\mathrm{min}} = 3 \cdot 10^{-5}\,\mathrm{kpc} \cdot \sqrt{\frac{M}{M_\odot}} \cdot \frac{1}{f_p} \cdot \sqrt{\frac{a_0}{a_b^{\mathrm{gal}}}} \,,
\end{align}
for $a_0 = 1.2 \cdot 10^{-10}\,\mathrm{m}/\mathrm{s}^2$.
This is much smaller than the typical distance between stars of about $10^{-3}\,\mathrm{kpc}$ in the Milky Way.
Thus, this does not affect our results.
Another possible effect is that stars could absorb some of the Cherenkov radiation emitted by other stars.
In this way, stars may regain some of the energy lost to Cherenkov radiation.
Of course, purely geometrically, some of the Cherenkov radiation should be able to escape the galaxy,
    since Cherenkov radiation is dominated by wavelengths of order $1/k_{\mathrm{max}}$,
    while stars are typically much further away from each other than $1/k_{\mathrm{max}}$.
In addition, the coupling of Cherenkov radiation to matter is suppressed by $1/M_{\mathrm{Pl}}$ so that reabsorption happens only rarely.
Thus, we expect this to be only a small effect.
We leave a more detailed investigation for future work.

Another interesting effect was discovered by Ref.~\cite{Berezhiani2019b}.
Namely, when $r_{\mathrm{max}}$ is comparable to the Jeans length $r_J$ of a superfluid, the dispersion relation may be modified
    so that Cherenkov radiation is possible for arbitrarily low velocities $V$.
This concerns perturbations with large wavelengths.
Thus, this would not affect our result for the energy loss which is dominated by short wavelengths.
However, it means there may not be a hard cutoff of the energy loss at a critical velocity.
That is, evading our constraints by increasing the sound speed, as we have done in standard SFDM, would only decrease the energy loss but not shut it down entirely.
Of course, this applies only if our cutoff $r_{\mathrm{max}} \sim \mathrm{kpc}$ is larger than the Jeans length $r_J$.
So possibly our strict cutoffs do not allow to directly see this effect in our case.

\section{Conclusion}
\label{sec:conclusion}

In modified gravity theories, one usually tries to avoid superluminal sound speeds for theoretical reasons.
But, for empirical reasons, one must also be careful with subluminal sound speeds.
Specifically, a subluminal sound speed often leads to gravitational Cherenkov radiation which allows various astrophysical objects to lose energy.

In this paper, we have discussed a new type of Cherenkov radiation that is often possible in hybrid MOND dark matter models with a common origin for the cosmological and galactic scale phenomena.
Such models typically contain a massless mode that couples directly to normal matter (for the MOND phenomenology).
This same mode often has a non-relativistic sound speed (to account for the CDM phenomenology).
This allows even non-relativistic objects like stars to lose energy.

In our calculation, we use a controlled approximation that relies only on the MOND regime of each model,
    not on the Newtonian, high-acceleration regime.
This avoids technical issues with the non-linearities of MOND
    and the possibly non-trivial high-acceleration behavior of each model.
The price to pay is that we obtain only a conservative lower bound on the energy loss through Cherenkov radiation.
The actual energy loss may be much larger but is also much more difficult to calculate.

We have first discussed the idea behind this new type of Cherenkov radiation for a prototype model and then applied the results to the original SFDM model, two-field SFDM, and the SZ model.
For standard SFDM, we could rule out a MOND limit in the Milky Way for part of the parameter space.
Two-field SFDM avoids these constraints by weakening the link between the cosmological and galactic phenomena.
The relevant massless mode is coupled to normal matter only indirectly through a mixing.
The SZ model avoids our constraints by having a matter coupling that is a standard gravitational coupling only in the static limit.
In dynamical situations, the matter coupling is suppressed.

These results do not completely rule out any of the models considered.
Still, we have demonstrated that our method is powerful and can severely constrain hybrid MOND dark matter models.
This method can be improved upon in the future and it can be applied to more models.
Any hybrid MOND dark matter model must satisfy our constraints.

\section*{Acknowledgements}
\label{sec:acknowledgements}

I am grateful for financial support from FIAS.
I thank Sabine Hossenfelder, Stacy McGaugh, and Luciano Rezzolla for discussions.

\begin{appendices}

\section{Calculation of energy loss through Cherenkov radiation}
\label{sec:Edotcalculation}

Consider the prototype model from Sec.~\ref{sec:general}.
Consider further an incoming, non-relativistic matter particle with four-momentum $p_i^\alpha = (M + \frac12 M V^2, M \vec{V})$.
For concreteness, we assume $\vec{V} = (0, 0, V)$.
When this matter particle emits Cherenkov radiation,
    the Cherenkov radiation carries away four-momentum $k^\alpha = (\omega, \vec{k})$ which leaves the matter particle with a final four-momentum $p_f^\alpha = (E_f, \vec{p}_f) = (M + \frac12 M V'^2, M \vec{V}') $.
We have \cite{Peskin1995}
\begin{align}
d\Gamma = \frac{1}{2M} \frac{d^3\vec{p}_f}{(2\pi)^3 \, 2E_f} \frac{d^3\vec{k}}{(2\pi)^3 \, 2\omega} |\mathcal{M}|^2 (2\pi)^4 \delta^{(4)}(p_i - p_f - k) \,.
\end{align}
And we can calculate the energy loss per time from this,
\begin{align}
 \frac{dE}{dt} = -\int \omega d\Gamma \,.
\end{align}
The coupling term in the Lagrangian Eq.~\eqref{eq:prototype} is $-M^2 \chi^2 \, \coup \delta/(\sqrt{2} M_{\mathrm{Pl}})$.
The Feynman rules give a factor of $2!$ for the two factors of $\chi$.
Thus, we have
\begin{align}
 |\mathcal{M}| = \frac{\sqrt{2} \bar{c} \coup}{M_{\mathrm{Pl}}} M^2 \,.
\end{align}
The factor $\bar{c}$ is because we need to canonically normalize our field, $\delta \to \bar{c} \cdot \delta$,
    so that we can apply the standard Feynman diagram formalism.
Here, we assume that $\coup$ does not depend on $\vec{k}$.
This is not true in two-field SFDM and in the SZ model.
Corrections due to this are discussed in Appendix~\ref{sec:Edotcalculation:twofieldsfdm} and in Sec.~\ref{sec:relmond_coupling}.

Then, using $\omega = c_s |\vec{k}|$,
\begin{align}
\begin{split}
 d\Gamma &= \frac{1}{(2M)^2} \frac{1}{(2\pi)^2} \frac{d^3\vec{k}}{2 \omega} |\mathcal{M}|^2 \delta\left(\frac12 M V^2 - \frac12 M (\vec{V} - \vec{k}/M)^2 - c_s |\vec{k}|\right) \\
         &= \frac{1}{(2M)^2} \frac{1}{(2\pi)^2} \frac{d^3\vec{k}}{2 \omega} |\mathcal{M}|^2 \delta\left(V |\vec{k}| \cos \theta - \left(\frac{|\vec{k}|^2}{2M} + c_s |\vec{k}|\right)\right) \\
         &= \frac{1}{(2M)^2} \frac{1}{(2\pi)^2} \frac{d|\vec{k}| |\vec{k}| \, d\varphi \, d \cos\theta}{2 \omega V} |\mathcal{M}|^2 \delta\left(\cos \theta - \left(\frac{|\vec{k}|}{2MV} + \frac{c_s}{V} \right)\right) \,,
\end{split}
\end{align}
where $\theta$ is the angle between $\vec{k}$ and $\vec{V}$.
We see that the delta function requires $|\vec{k}|$ to be not too large. In particular, we need to cut off the $|\vec{k}|$ integral at
\begin{align}
 |\vec{k}| = 2M(V - c_s) = 2M \left( V - \bar{c} \sqrt{1 + \gamma^2}\right) \,,
\end{align}
where $\gamma$ is the cosine of the angle between $\vec{k}$ and $\hat{a}$.
This condition implies that there will be energy loss due to Cherenkov radiation only for supersonic velocities $V > c_s$.

Numerically, we are interested in $M \sim M_\odot$ and we impose an additional cutoff $k_{\mathrm{max}} \sim 10^{-22}\,\mathrm{eV}$ to stay in the regime of validity of our calculation,
    see Sec.~\ref{sec:general:validity}.
In this case, the $c_s/V$ term dominates over the $k/(2MV)$ term in the delta function.
Thus, this delta function becomes
\begin{align}
 \label{eq:Edotcalculation:delta}
 \delta\left(\cos \theta - \frac{\bar{c}}{V} \sqrt{1 + \gamma^2} \right)\,.
\end{align}
In general, $\gamma = \vec{k} \cdot \hat{a} / k$ depends on $\cos \theta$ in a complicated way.
Thus, analytically evaluating this delta function is nontrivial.
Here, we first give the result for the simpler case $\hat{a} = 0$ where the $\sqrt{1+\gamma^2}$ factor in $c_s$ is absent.
In the following sections, we then give the result including $\hat{a}$ for the special cases $\vec{V} \perp \hat{a}$ and $\vec{V} \parallel \hat{a}$.
For $\hat{a} = 0$, we have
\begin{align}
\begin{split}
|\dot{E}|^{\hat{a}=0} &= \int_{k_{\mathrm{min}}}^{k_{\mathrm{max}}} d|\vec{k}|\frac{1}{(2M)^2} \frac{1}{(2\pi)} \frac{|\vec{k}|}{2 V} |\mathcal{M}|^2 \\
          &= \frac{1}{(2M)^2} \frac{1}{(2\pi)} \frac{1}{4 V} (k_{\mathrm{max}}^2 - k_{\mathrm{min}}^2) \frac{2 \bar{c}^2 \coup^2 M^4}{M_{\mathrm{Pl}}^2} \\
          &= \frac{\bar{c}^2}{16 \pi V} \frac{\coup^2 M^2}{M_{\mathrm{Pl}}^2} (k_{\mathrm{max}}^2 - k_{\mathrm{min}}^2)\,.
\end{split}
\end{align}

\subsection{Special case: $\vec{V} \parallel \hat{a}$}

In this case, we have
\begin{align}
 \gamma^2 = \frac{(\vec{k} \hat{a})^2}{|\vec{k}|^2} = \cos^2 \theta \,,
\end{align}
since $\theta$ is the angle between $\vec{V}$ and $\vec{k}$ and $\vec{V} \parallel \hat{a}$.
Then, the delta function Eq.~\eqref{eq:Edotcalculation:delta} in $d\Gamma$ can be written as
\begin{align}
 \delta \left( \cos \theta - \frac{\bar{c}}{V} \sqrt{1 + \cos^2 \theta} \right) \,.
\end{align}
This is solved by
\begin{align}
 \cos \theta = \frac{1}{\sqrt{(V/\bar{c})^2-1}}\,.
\end{align}
Thus, Cherenkov radiation requires
\begin{align}
 V > \sqrt{2} \bar{c} \,,
\end{align}
otherwise $\cos \theta$ would be larger than $1$.
Evaluated at this solution, the delta function becomes
\begin{align}
 \frac{1}{1 - (\bar{c}/V)^2} \, \delta\left(\cos \theta - \frac{1}{\sqrt{(V/\bar{c})^2 - 1}}  \right) \,,
\end{align}
which implies
\begin{align}
 |\dot{E}| = \frac{1}{1 - (\bar{c}/V)^2} \cdot |\dot{E}|^{\hat{a} = 0} \,.
\end{align}

\subsection{Special case: $\vec{V} \perp \hat{a}$}
\label{sec:Edotcalculation:perp}

For simplicity, we choose our coordinate system such that $\hat{a}$ points into the positive $x$ direction.
This is perpendicular to $\vec{V}$ which points into the positive $z$ direction.
Then,
\begin{align}
 \gamma^2 = \frac{(\vec{k} \hat{a})^2}{|\vec{k}|^2} = \sin^2 \theta \, \cos^2 \varphi \,.
\end{align}
That is, the delta function Eq.~\eqref{eq:Edotcalculation:delta} in $d\Gamma$ becomes
\begin{align}
 \delta\left(\cos \theta - \frac{\bar{c}}{V} \sqrt{1 + (1-\cos^2 \theta) \cos^2 \varphi} \right)\,.
\end{align}
This is solved by
\begin{align}
 \cos \theta = \frac{\bar{c}}{V} \sqrt{\frac{1 + \cos^2 \varphi}{1 +  (\bar{c}/V)^2 \,\cos^2 \varphi}} \,.
\end{align}
As a result, Cherenkov radiation requires
\begin{align}
 V > \bar{c}\,,
\end{align}
otherwise $\cos \theta$ would be larger than 1.
Evaluated at this solution, the delta function becomes
\begin{align}
 \frac{1}{1 + (\bar{c}/V)^2 \cos^2 \varphi} \, \delta\left(\cos \theta - \frac{\bar{c}}{V} \sqrt{\frac{1 + \cos^2 \varphi}{1 + (\bar{c}/V)^2 \cos^2 \varphi}} \right) \,.
\end{align}
After doing the $\varphi$ integral, this implies
\begin{align}
 |\dot{E}| = \frac{1}{\sqrt{1 + (\bar{c}/V)^2}} \cdot |\dot{E}|^{\hat{a}=0} \,.
\end{align}

\subsection{Direction of friction force}
\label{sec:Edotcalculation:direction}

Above, we calculated the total energy loss, but not the direction in which the Cherenkov emission happens, i.e. the direction in which the resulting effective friction force acts.
This direction is proportional to the direction of the following integral
\begin{align}
 \label{eq:Edotcalculation:direction}
 \int \vec{k} \, d\Gamma \propto
 \int_{-1}^{+1} d \cos \theta \int_{0}^{2\pi} d\varphi \frac{1}{\sqrt{1+ \gamma^2}} \begin{pmatrix}
                                   \sin \theta \cos \varphi \\
                                   \sin \theta \sin \varphi \\
                                   \cos \theta
                                  \end{pmatrix}
                        \delta \left(\cos \theta - \bar{c}/V \sqrt{1 + \gamma^2} \right) \,.
\end{align}
Here, as in the previous section, we choose $y$ to be the direction that is perpendicular to both $\vec{V}$ and $\hat{a}$.
For the two special cases $\vec{V} \parallel \hat{a}$ and $\vec{V} \perp \hat{a}$,
    the friction force points exactly in the $z$ direction, i.e. in the direction of $\vec{V}$.
This follows by using the delta function to do the $\cos \theta$ integral and then noticing that the integrand depends on $\varphi$ only through $\cos^2 \varphi$,
    except for the factors $\cos \varphi$ and $\sin \varphi$ in the $x$ and $y$ components of $\vec{k}$.
That is, the $x$ and $y$ components vanish for these two special cases.

For a general orientation of $\vec{V}$ relative to $\hat{a}$,
    it is still true that the force in the $y$ direction vanishes.
That is, the force stays in the $x$-$z$ plane spanned by $\vec{V}$ and $\hat{a}$.
This can be seen analytically by using the delta function to do the $\varphi$ integral.
This gives two solutions for $\cos \varphi$,
\begin{align}
 \cos \varphi = \frac{1}{a_\perp \sin \theta} \left(\pm \sqrt{\left(\frac{V}{\bar{c}}\right)^2 \cos^2 \theta - 1} - a_\parallel \cos \theta \right) \,,
\end{align}
where $a_\perp$ is the component of $\hat{a}$ in the $x$ direction and $a_\parallel$ that in the $z$ direction.
Each of these solutions for $\cos \varphi$ corresponds to two different values of $\varphi$ in the interval $(0, 2\pi)$.
These two values of $\varphi$ give the same $\cos \varphi$, but $\pm \sin \varphi$.
The integrand in Eq.~\eqref{eq:Edotcalculation:direction} depends on the sign of $\sin \varphi$ only through the factor $\sin \varphi$ in the $y$ component.
Thus, the $y$ component of the force vanishes,
    since the contributions from the two different values of $\varphi$ cancel each other.

In contrast, the $x$ component does not vanish in general.
The integrals are hard to evaluate analytically, but we have numerically verified that the $x$ component can be as large as a few $10\%$ of the $z$ component.
Thus, in general,
    the friction force does not act in the direction of $\vec{V}$.

\subsection{Two-field SFDM corrections}
\label{sec:Edotcalculation:twofieldsfdm}

As discussed in Sec.~\ref{sec:twofieldsfdm} (see also Appendix~\ref{sec:twofieldsfdmpert} below),
    two-field SFDM differs from our prototype model in two ways.
First, the sound speed $c_s$ does not have the $\sqrt{1+\gamma^2}$ factor,
    so that $c_s = \bar{c} = \sqrt{\hat{\mu}/m}$ is independent of the direction of $\vec{k}$.
Second, the matter coupling $\coup$ depends on this direction through the $\gamma$-dependent factor $\gamma/(1+\gamma^2)$.

That $c_s$ has no factor $\sqrt{1+\gamma^2}$ means that the energy-conserving delta function gives $\cos \theta = c_s/V$.
Thus, there is Cherenkov radiation as long as $V > c_s = \sqrt{\hat{\mu}/m}$ and the correction factor $f_a$ is absent.

The additional factor $\gamma^2/(1+\gamma^2)^2$ in $|\mathcal{M}|^2$ modifies the result for $|\dot{E}|$.
We again orient our coordinate system such that $\vec{V}$ points in the $z$ direction and  $\hat{a}$ has zero component in the $y$ direction.
Then
\begin{align}
 \gamma^2 = \frac{(\vec{k} \hat{a})^2}{|\vec{k}|^2} = (a_z \cos \theta + a_x \sin \theta \, \cos \varphi)^2 \,.
\end{align}

If, for simplicity, we set $\gamma = 1$, we get from the $\varphi$ integration in $|\dot{E}|$
\begin{align}
 \int d\varphi \left.\frac{\gamma^2}{(1+\gamma^2)^2}\right|_{\gamma=1} = \frac{2 \pi}{4} \,.
\end{align}
For general $\gamma$, the $\varphi$ integral can be done analytically for the special cases $\vec{V} \parallel \hat{a}$ and $\vec{V} \perp \hat{a}$.
For $\hat{a} \parallel \vec{V}$, we have $a_x = 0$ and $a_z^2 = 1$, so that
\begin{align}
 \int d\varphi \left. \frac{\gamma^2}{(1+\gamma^2)^2} \right|_{\vec{V} \parallel \hat{a}}
    = 2 \pi \frac{\cos^2 \theta}{(1+\cos^2\theta)^2}
    = 2 \pi \left(\frac{c_s}{V} + \frac{V}{c_s}\right)^{-2} \,.
\end{align}
For $\hat{a} \perp \vec{V}$, we have $a_z = 0$ and $a_x^2 = 1$, so that
\begin{align}
\begin{split}
 \int d\varphi \left. \frac{\gamma^2}{(1+\gamma^2)^2} \right|_{\vec{V} \perp \hat{a}}
    &= \int d\varphi \frac{\sin^2 \theta \cos^2 \varphi}{(1 + \sin^2 \theta \cos^2 \varphi)^2}\\
    &= 2 \pi \, \frac12 \frac{\sin^2 \theta}{(1  + \sin^2\theta)^{3/2}}\\
    &= 2 \pi \, \frac12 \frac{1 - \left(\frac{c_s}{V}\right)^2}{\left(2 - \left(\frac{c_s}{V}\right)^2 \right)^{3/2}} \,.
\end{split}
\end{align}
Thus,
\begin{align}
 |\dot{E}| = 4 |\dot{E}|^{\gamma=1} \cdot \begin{cases}
                                        \left(\frac{c_s}{V} + \frac{V}{c_s}\right)^{-2} \,, & \vec{V} \perp \hat{a} \\
                                        \frac12 \frac{1 - \left(\frac{c_s}{V}\right)^2}{\left(2 - \left(\frac{c_s}{V}\right)^2 \right)^{3/2}}  \,, & \vec{V} \parallel \hat{a}
                                        \end{cases}\,.
\end{align}

For $\vec{V} \parallel \hat{a}$ and $\vec{V} \perp \hat{a}$, the direction of the effective friction force stays the same as in the case of our prototype Lagrangian.
However, for a general orientation of $\vec{V}$,
    the additional $\gamma$-dependence may give corrections also to the direction of the force.
Here, we do not calculate these since they are not important for our results.

\section{Standard SFDM perturbations}
\label{sec:standardsfdmpert}

Consider the SFDM Lagrangian Eq.~\eqref{eq:sfdm}.
If we define
\begin{align}
\begin{split}
 \bar{f}(K) &\equiv \frac{2 \Lambda}{3} (2m)^{3/2} \sqrt{|K|} \,, \\
       f(K) &\equiv \bar{f}(K) K \,,
\end{split}
\end{align}
the kinetic term in this Lagrangian can be written as $\bar{f}(X - \barbeta Y) X$.
If $X - \barbeta Y \approx X$, this becomes approximately $f(X)$.
We now consider a background equilibrium solution $\theta_0$ in the MOND limit, i.e. $(\vec{\nabla} \theta_0)^2 \gg 2 m \hat{\mu}$.
We expand $X$ and $Y$ in perturbations $\delta$ of the phonon field $\theta$ and introduce a formal expansion parameter $\epsilon$,
\begin{align}
\begin{split}
 X &= X_0 + \epsilon X_1 + \epsilon^2 X_2 + \dots \,, \\
 Y &= Y_0 + \epsilon Y_1 \,,
\end{split}
\end{align}
with
\begin{align}
\begin{split}
 X_1 &= \dot{\delta} - \frac{\vec{\nabla} \theta_0 \vec{\nabla} \delta}{m} \,, \\
 X_2 &= - \frac{(\vec{\nabla} \delta)^2}{2m} \,, \\
 Y_1 &= \dot{\delta} \,,
\end{split}
\end{align}
where we left out a possible $\epsilon^2$ term in $Y$ since $Y$ is linear in the fields.
We further define
\begin{align}
\begin{split}
 f_0' &= f'(X_0) = (2m) \Lambda |\vec{\nabla} \theta_0| \,, \\
 f_0'' &= f''(X_0) = - (2m)^2 \frac{\Lambda}{2} \frac{1}{|\vec{\nabla} \theta_0|} \,, \\
 \bar{f}_0' &= \bar{f}'(X_0) = - (2m)^2 \frac{\Lambda}{3} \frac{1}{|\vec{\nabla} \theta_0|} \,, \\
 \bar{f}_0'' &= \bar{f}''(X_0) = - (2m)^3 \frac{\Lambda}{6} \frac{1}{|\vec{\nabla} \theta_0|^3} \,.
\end{split}
\end{align}
For simplicity, we also write
\begin{align}
 \lambda \equiv \frac{\baralpha \Lambda}{M_{\mathrm{Pl}}} \,.
\end{align}
For a background galaxy in the MOND limit, spatial derivatives dominate so that $X_0 - \barbeta Y_0 \approx X_0$.
Then, the perturbed Lagrangian is

\begin{align}
\begin{split}
 \mathcal{L}
     &= X_2 f'_0 + \frac12 X_1^2 f''_0 - \barbeta X_1 Y_1 (\bar{f}_0' + X_0 \bar{f}_0'') + \frac12 \barbeta^2 Y_1^2 X_0 \bar{f}_0'' - \lambda \delta \, \delta_b \\
    &=
    \begin{multlined}[t][.77\textwidth]
    \Lambda |\vec{\nabla} \theta_0| \left(- (\vec{\nabla} \delta)^2 - \frac{m^2}{|\vec{\nabla} \theta_0|^2} \left(\dot{\delta} - \frac{\vec{\nabla} \theta_0 \vec{\nabla} \delta}{m}\right)^2 \right. \\
    \left. + \frac23 \frac{m^2}{|\vec{\nabla} \theta|^2} \barbeta \dot{\delta} \left(\dot{\delta} - \frac{\vec{\nabla} \theta \vec{\nabla} \delta}{m} \right) + \frac13 \barbeta^2 \dot{\delta}^2 \frac{m^2}{|\vec{\nabla} \theta_0|^2} \right) - \lambda \delta \, \delta_b
    \end{multlined} \\
    &=
    \begin{multlined}[t]
    2 \Lambda |\vec{\nabla} \theta_0| \left(\dot{\delta}^2 \frac{m^2}{|\vec{\nabla} \theta_0|^2} \frac16 (\barbeta-1) (\barbeta + 3) - \frac12 (\vec{\nabla} \delta)^2 - \frac12 (\hat{a} \vec{\nabla} \delta)^2 \right. \\
    \left. - \frac13 \frac{m}{|\vec{\nabla} \theta_0|} \hat{a} \vec{\nabla} \delta \dot{\delta} (\barbeta - 3)\right)- \lambda \delta \, \delta_b \,,
    \end{multlined}
\end{split}
\end{align}
where we set $\hat{a} = \vec{\nabla} \theta_0/|\vec{\nabla} \theta_0|$.
With the definition
\begin{align}
 \bar{c}^{-1} = \frac{m}{|\vec{\nabla} \theta_0|} \sqrt{\frac{(\barbeta-1)(\barbeta+3)}{3}} \,,
\end{align}
this can be written as
\begin{align}
 \mathcal{L} = 2 \Lambda |\vec{\nabla} \theta_0| \left(\frac12 \dot{\delta}^2 \bar{c}^{-2} - \frac12 (\vec{\nabla} \delta)^2 - \frac12 (\hat{a} \vec{\nabla} \delta)^2 + \bar{c}^{-1} f_\barbeta \hat{a} \vec{\nabla} \delta \dot{\delta} \right) - \lambda \delta \, \delta_b\,,
\end{align}
where, following Ref.~\cite{Berezhiani2015, Berezhiani2018}, we assumed $\barbeta < 3$ and defined
\begin{align}
 f_\barbeta =  \sqrt{\frac{(3 - \barbeta)^2}{3 (\barbeta-1)(\barbeta+3)}} \,.
\end{align}
With an appropriate rescaling we finally get
\begin{align}
 \mathcal{L} = \frac12 \dot{\delta}^2 \bar{c}^{-2} - \frac12 (\vec{\nabla} \delta)^2 - \frac12 (\hat{a} \vec{\nabla} \delta)^2 + \bar{c}^{-1} f_\barbeta \hat{a} \vec{\nabla} \delta \dot{\delta} - \frac{\coup}{\sqrt{2} M_{\mathrm{Pl}}} \delta \, \delta_b\,,
\end{align}
with
\begin{align}
 \coup = \sqrt{\frac{a_0}{|a_{\theta_0}|}} \,,
\end{align}
where we used $a_0 = \baralpha^3 \Lambda^2/M_{\mathrm{Pl}}$ and $\vec{a}_\theta = - \lambda \vec{\nabla} \theta$ \cite{Berezhiani2015}.

\section{Standard SFDM Cherenkov radiation}
\label{sec:Edotcalculation:standardsfdm}

For standard SFDM, our simple QFT calculation from Appendix~\ref{sec:Edotcalculation} does not apply
    because spatial and time derivatives are mixed in the Lagrangian for the perturbations.
Thus, we do a classical calculation instead,
    along the lines of the standard calculation of electromagnetic Cherenkov radiation \cite{Jackson1998}.

\subsection{Reference calculation}

For reference,
    we first consider a field $\bar{\phi}(\bar{t}, \vec{\bar{x}})$ with Lagrangian
    \begin{align}
    \bar{\mathcal{L}} = \frac12 \frac{1}{c_{\mathrm{eff}}^2} (\partial_{\bar{t}} \bar{\phi})^2 - \frac12  (\vec{\bar{\nabla}} \bar{\phi})^2 - \frac{\coup}{\sqrt{2} M_{\mathrm{Pl}}} \bar{\phi} \, \rho_{\mathrm{b},\mathrm{eff}} \,,
    \end{align}
with some constant $c_{\mathrm{eff}}$.
This is our prototype Lagrangian from Sec.~\ref{sec:general}
    but in the simpler case with $\hat{a} = 0$ and with $\bar{c}$ replaced by $c_{\mathrm{eff}}$.
We also introduced a modified baryonic density
    \begin{align}
     \rho_{\mathrm{b},\mathrm{eff}} = M_{\mathrm{eff}} \delta(\bar{z} - V_{\mathrm{eff}} \bar{t}) \delta(\bar{x}) \delta(\bar{y}) \,,
    \end{align}
    with constants $M_{\mathrm{eff}}$ and $V_{\mathrm{eff}}$.
This is the usual form for the perturber's density used in classical Cherenkov radiation calculations \cite{Jackson1998, Ostriker1999, Berezhiani2019b}.
Below, we will express solutions of the equation of motion of the standard SFDM perturbation field $\delta$ in terms of solutions of the equation of motion of $\bar{\phi}$.

We use the following convention for the Fourier transform and inverse Fourier transform of a function $f$
    \begin{subequations}
    \begin{align}
        f(\omega, \vec{k}) &= \frac{1}{(2\pi)^2} \int d^4x \, e^{- i \omega t + i \vec{k} \vec{x}} f(t, \vec{x}) \,, \\
             f(t, \vec{x}) &= \frac{1}{(2\pi)^2} \int d^4k \, e^{+ i \omega t - i \vec{k} \vec{x}} f(\omega, \vec{k}) \,.
    \end{align}
    \end{subequations}
The equation of motion of $\bar{\phi}$ is
    \begin{align}
    \label{eq:phibareom}
     0 = - \frac{1}{c_{\mathrm{eff}}^2} \partial_{\bar{t}}^2 \bar{\phi} + \vec{\bar{\nabla}}^2 \bar{\phi} - \frac{g_m}{\sqrt{2} M_{\mathrm{Pl}}} M_{\mathrm{eff}} \delta(\bar{z} - V_{\mathrm{eff}} \bar{t}) \delta(\bar{x}) \delta(\bar{y}) \,.
    \end{align}
In Fourier space
    \begin{align}
     0 = \left(\frac{\bar{\omega}^2}{c_{\mathrm{eff}}^2} - \vec{\bar{k}}^2\right) \bar{\phi}(\bar{\omega}, \vec{\bar{k}}) - \frac{g_m}{\sqrt{2} M_{\mathrm{Pl}}} \frac{M_{\mathrm{eff}}}{2 \pi} \delta(\bar{\omega} - V_{\mathrm{eff}} \bar{k}_{\bar{z}}) \,,
    \end{align}
where we used $\int dk e^{ikx} = (2\pi) \delta(x)$
    and introduced the notation $\bar{\omega}$ and $\vec{\bar{k}}$ to indicate that this is a Fourier transform with respect to $\bar{t}$ and $\vec{\bar{x}}$.
This gives
    \begin{align}
    \begin{split}
     \bar{\phi}(\bar{\omega}, \vec{\bar{k}})
        &= \frac{g_m}{\sqrt{2} M_{\mathrm{Pl}}} \frac{M_{\mathrm{eff}}}{2\pi} \frac{\delta(\bar{\omega} - V_{\mathrm{eff}} \bar{k}_{\bar{z}})}{\frac{\bar{\omega}^2}{c_{\mathrm{eff}}^2} - \bar{k}^2} \\
        &= \frac{1}{|V_{\mathrm{eff}}|} \frac{g_m}{\sqrt{2} M_{\mathrm{Pl}}} \frac{M_{\mathrm{eff}}}{2\pi} \frac{\delta\left(\frac{\bar{\omega}}{V_{\mathrm{eff}}} - \bar{k}_{\bar{z}}\right)}{\bar{\lambda}^2 - \bar{k}_\perp^2}  \,,
    \end{split}
    \end{align}
    where
    \begin{align}
     \bar{\lambda}^2 = \bar{\omega}^2 \left(\frac{1}{c_{\mathrm{eff}}^2} - \frac{1}{V_{\mathrm{eff}}^2}\right) \,,
    \end{align}
and $\vec{\bar{k}}_\perp$ denotes the components of $\vec{\bar{k}}$ in the $\bar{x}$ and $\bar{y}$ directions.
On-shell radiation (that is Cherenkov radiation) is possible only for $\bar{\lambda}^2 > 0$.
    Otherwise, the propagator has no pole that could be picked up.
    This gives the standard condition for Cherenkov radiation
    \begin{align}
     |V_{\mathrm{eff}}| > c_{\mathrm{eff}} \,.
    \end{align}

In this case, we have \cite{Jackson1998}
    \begin{align}
    \begin{split}
     \bar{\phi}(\bar{t}, \vec{\bar{x}})
      &= \frac{1}{(2\pi)^2} \int d^4 \bar{k} e^{i \bar{\omega} \bar{t} - i \vec{\bar{k}} \vec{\bar{x}}}
        \frac{1}{|V_{\mathrm{eff}}|} \frac{g_m}{\sqrt{2} M_{\mathrm{Pl}}} \frac{M_{\mathrm{eff}}}{2\pi} \frac{\delta\left(\frac{\bar{\omega}}{V_{\mathrm{eff}}} - \bar{k}_{\bar{z}}\right)}{\bar{\lambda}^2 - \bar{k}_\perp^2}  \\
      &=
      \begin{multlined}[t][.7\textwidth]
        \frac{1}{(2\pi)^2} \frac{1}{|V_{\mathrm{eff}}|} \frac{g_m}{\sqrt{2} M_{\mathrm{Pl}}} \frac{M_{\mathrm{eff}}}{2\pi}
            \int d\bar{\omega} e^{i\bar{\omega}\left( \bar{t} - \frac{1}{V_{\mathrm{eff}}} \bar{z}\right)} \\
            \times \int d^2\vec{\bar{k}}_\perp e^{-i \vec{\bar{k}}_\perp \vec{\bar{x}}_\perp} \frac{1}{\bar{\lambda}^2 - \bar{k}_\perp^2}
       \end{multlined} \\
     &= -\frac{M_{\mathrm{eff}}}{(2\pi)^2} \frac{1}{|V_{\mathrm{eff}}|} \frac{g_m}{\sqrt{2} M_{\mathrm{Pl}}}
        \int d\bar{\omega} e^{i\bar{\omega}\left( \bar{t} - \frac{1}{V_{\mathrm{eff}}} \bar{z}\right)} K_0(i\bar{\lambda} |\vec{\bar{x}}_\perp|) \,,
    \end{split}
    \end{align}
    where $K_0$ is the zeroth modified Bessel function of the second kind
        and $\vec{\bar{x}}_\perp$ denotes components of $\vec{\bar{x}}$ in the $\bar{x}$ and $\bar{y}$ directions.
Here, $\bar{\lambda}$ is given by
    \begin{align}
    \bar{\lambda} = \bar{\omega} \sqrt{\frac{1}{c_{\mathrm{eff}}^2} - \frac{1}{V_{\mathrm{eff}}^2}}
                  \equiv \frac{\bar{\omega}}{c_{\mathrm{eff}}} f_{\mathrm{crit}} \,.
    \end{align}
When doing the $\vec{\bar{k}}_\perp$ integral,
    we implicitly assumed the retarded propagator.
For our convention of the Fourier transform, this gives the $K_0(i \bar{\lambda} |\vec{\bar{x}}_\perp|)$.
Ref.~\cite{Jackson1998} uses the opposite sign in the $e^{\pm i \omega t}$ factors of the Fourier transform and therefore gets $\bar{\lambda} \to - \bar{\lambda}$ in the Bessel function's argument.
We see that $\bar{\phi}$ depends on $\vec{\bar{x}}$ only through the combinations
    \begin{align}
     \bar{z} - V_{\mathrm{eff}} \bar{t} \quad \mathrm{and} \quad |\vec{\bar{x}}_\perp| \,.
    \end{align}
Below, we are interested in the limit $|\bar{\lambda} \vec{\bar{x}}_\perp| \gg 1$.
In this limit,
    \begin{align}
     \bar{\phi}(\bar{t}, \vec{\bar{x}})
       = - \sqrt{\frac{\pi}{2}} \frac{M_{\mathrm{eff}}}{(2\pi)^2} \frac{1}{|V_{\mathrm{eff}}|} \frac{g_m}{\sqrt{2} M_{\mathrm{Pl}}}
        \int d\bar{\omega} e^{i\bar{\omega}\left( \bar{t} - \frac{1}{V_{\mathrm{eff}}} \bar{z}\right)} \frac{e^{-i \bar{\lambda} |\vec{\bar{x}}_\perp|}}{\sqrt{i \bar{\lambda} |\vec{\bar{x}}_\perp|}} \,.
    \end{align}

\subsection{Mapping to the reference system}

To calculate Cherenkov radiation for standard SFDM, we use the standard perturber density
    \begin{align}
     \delta_b(t, \vec{x}) = M \delta(z - V t) \delta(x) \delta(y) \,.
    \end{align}
    This assumes that the emitted Cherenkov radiation is soft.
    Otherwise, we could not assume the perturber to travel on a straight line with constant velocity.
    This is justified because we set very conservative limits on the integration boundaries.
    This is discussed in Appendix~\ref{sec:Edotcalculation} and Sec.~\ref{sec:Edotcalculation:orbits}.
The equation of motion for the perturbation $\delta$ is then
    \begin{multline}
     0 = - \frac{1}{\bar{c}^2} \partial_t^2 \delta + \vec{\nabla}^2 \delta + \vec{\nabla} \left( (\hat{a} \cdot \vec{\nabla} \delta) \hat{a}\right) - \frac{2 f_\barbeta}{\bar{c}} \hat{a} \vec{\nabla} \dot{\delta} \\
     - \frac{g_m}{\sqrt{2} M_{\mathrm{Pl}}} M \delta(z - Vt)\delta(x) \delta(y) \,.
    \end{multline}
To facilitate an analytical treatment, we discuss only the special cases $\vec{V} \parallel \hat{a}$ and $\vec{V} \perp \hat{a}$.
Here, $\vec{V}$ points in the positive $z$ direction.

For $\vec{V} \parallel \hat{a}$,
    we have $\hat{a} = a_\parallel \hat{e}_z$ where $\hat{e}_z$ is the unit vector in the $z$ direction and $a_\parallel = \pm 1$.
The solutions for $\delta(t, \vec{x})$ can be obtained from those for $\bar{\phi}(\bar{t}, \vec{\bar{x}})$ by a coordinate transformation.
Specifically, we take
    \begin{align}
    \begin{split}
     \bar{t} &= t + a_\parallel \frac{f_\barbeta}{2 \bar{c}} z \,, \\
     \bar{z} &= \frac{z}{\sqrt{2}} \,, \\
     \vec{\bar{x}}_\perp &= \vec{x}_\perp \,.
    \end{split}
    \end{align}
Then, $\delta$ defined as
    \begin{align}
     \delta(t, \vec{x}) = \bar{\phi}(\bar{t}, \vec{\bar{x}}) \,,
    \end{align}
    solves the $\delta$ equation of motion if we take
    \begin{subequations}
    \begin{align}
        c_{\mathrm{eff}} &= \bar{c} \frac{1}{\sqrt{1 + \frac12 f_\barbeta^2}} \,, \\
        V_{\mathrm{eff}} &= V \frac{\sqrt{2}}{2 + a_\parallel f_\barbeta (V/\bar{c})} \,, \\
        M_{\mathrm{eff}} &= M \frac{\sqrt{2}}{|2 + a_\parallel f_\barbeta (V/\bar{c})|} \,,
    \end{align}
    \end{subequations}
in the $\bar{\phi}$ solution.
Note that $c_{\mathrm{eff}}$, $V_{\mathrm{eff}}$, and $M_{\mathrm{eff}}$ are constants.
They do not depend on spacetime coordinates or fields.
This immediately allows to find the condition for Cherenkov radiation.
As discussed above, $\delta(t, \vec{x})$ contains a pole only if $|V_{\mathrm{eff}}| > c_{\mathrm{eff}}$.
This gives the critical velocity
    \begin{align}
     V_{\mathrm{crit}}^\parallel = \bar{c} \left(\sqrt{2 + f_\barbeta^2} + a_\parallel f_\barbeta\right) \,.
    \end{align}

For $\vec{V} \perp \hat{a}$,
    we choose our coordinates such that $\hat{a}$ points in the positive $x$ direction, $\hat{a} = \hat{e}_x$.
We further choose
    \begin{align}
    \begin{split}
     \bar{t} &= t + \frac{f_\barbeta}{2\bar{c}} x \,, \\
     \bar{x} &= \frac{x}{\sqrt{2}}\,, \\
     \bar{y} &= y \,, \\
     \bar{z} &= z \,.
    \end{split}
    \end{align}
Then, $\delta$ defined as
    \begin{align}
     \delta(t, \vec{x}) = \bar{\phi}(\bar{t}, \vec{\bar{x}}) \,,
    \end{align}
solves the $\delta$ equation of motion if we take
    \begin{align}
        c_{\mathrm{eff}} &= \bar{c} \frac{1}{\sqrt{1 + \frac12 f_\barbeta^2}} \,,\\
        V_{\mathrm{eff}} &= V \,, \\
        M_{\mathrm{eff}} &= \frac{M}{\sqrt{2}} \,,
    \end{align}
in the $\bar{\phi}$ solution.
This gives the critical velocity
    \begin{align}
     V_{\mathrm{crit}}^\perp = \bar{c} \frac{1}{\sqrt{1 + \frac12 f_\barbeta^2}} \,.
    \end{align}

\subsection{Energy loss without cutoffs}
\label{sec:Edotcalculation:standardsfdm:nocutoffs}

To calculate the energy loss through Cherenkov radiation following Ref.~\cite{Jackson1998},
    we first need to know the energy flux in different spatial directions.
The canonical energy density of the standard SFDM perturbations is
    \begin{align}
    \mathcal{H} = \frac12 \frac{1}{\bar{c}^2} \dot{\delta}^2 + \frac12 (\vec{\nabla} \delta)^2 + \frac12 (\vec{\nabla} \delta \hat{a})^2\,.
    \end{align}
    The $f_\barbeta$ term that mixes spatial and time derivatives does not contribute
        since it is linear in time derivatives.
From this we find, using the equation of motion
    \begin{align}
    \begin{split}
      \partial_t \mathcal{H}
        &= \frac{\dot{\delta} \ddot{\delta}}{\bar{c}^2} + \vec{\nabla} \delta \vec{\nabla} \dot{\delta} + (\vec{\nabla} \delta \hat{a})(\vec{\nabla} \dot{\delta} \hat{a}) \\
        &= \dot{\delta} \left[ \vec{\nabla}^2 \delta + \vec{\nabla} \left( (\vec{\nabla} \delta \cdot \hat{a}) \hat{a} \right) - \frac{2 f_\barbeta}{\bar{c}} \hat{a} \vec{\nabla} \dot{\delta} \right] + \vec{\nabla} \delta \vec{\nabla} \dot{\delta} + (\vec{\nabla} \delta \hat{a})(\vec{\nabla} \dot{\delta} \hat{a}) \\
        &= \begin{multlined}[t][.77\textwidth]
                \left[\dot{\delta}  \vec{\nabla}^2 \delta  + \vec{\nabla} \delta \vec{\nabla} \dot{\delta} \right] + \left[ \dot{\delta} \vec{\nabla} \left( (\vec{\nabla} \delta \cdot \hat{a}) \hat{a} \right)+ (\vec{\nabla} \delta \hat{a})(\vec{\nabla} \dot{\delta} \hat{a}) \right] \\
                 - \vec{\nabla} \left( \frac{f_\barbeta}{\bar{c}} \hat{a} \dot{\delta}^2 \right)
           \end{multlined} \\
        &= \vec{\nabla} \left[\dot{\delta} \vec{\nabla} \delta + \dot{\delta} \hat{a} \, (\hat{a} \cdot \vec{\nabla} \delta) - \hat{a} \frac{f_\barbeta}{\bar{c}} \dot{\delta}^2 \right] \,.
    \end{split}
    \end{align}
That is, we have $\partial_t \mathcal{H} - \vec{\nabla} \vec{J} = 0$ with
    \begin{align}
     \vec{J} = \dot{\delta} \left( \vec{\nabla} \delta + \hat{a} (\vec{\nabla} \delta \cdot \hat{a}) - \hat{a} \frac{f_\barbeta}{\bar{c}} \dot{\delta} \right) \,.
    \end{align}
To derive this, we neglected the coupling of $\delta$ to the baryonic density perturbation $\delta_b$.
Corrections from this coupling can only be non-zero at the position of the perturber,
    as long as we model the perturber as a point particle.
Below, we integrate $\vec{J}$ over a surface at a finite distance from the perturber.
For this purpose, such corrections are irrelevant.

For $\vec{V} \parallel \hat{a}$, we have
    \begin{subequations}
    \begin{align}
     J_x(t, \vec{x}) &= \dot{\delta}(t, \vec{x}) \left(\partial_x \phi(t, \vec{x}) \right) = (\partial_{\bar{t}} \bar{\phi})(\bar{t}, \vec{\bar{x}}) (\partial_{\bar{x}} \bar{\phi})(\bar{t}, \vec{\bar{x}}) \,, \\
     J_y(t, \vec{x}) &= \dot{\delta} \, \partial_y \phi = (\partial_{\bar{t}} \bar{\phi}) (\partial_{\bar{y}} \bar{\phi}) \,, \\
     J_z(t, \vec{x}) &= \dot{\delta} \left( 2 \partial_z \delta - a_\parallel \frac{f_\barbeta}{\bar{c}} \dot{\delta}\right)
                      = \sqrt{2} (\partial_{\bar{t}} \bar{\phi}) (\partial_{\bar{z}} \bar{\phi}) \,,
    \end{align}
    \end{subequations}
    where, for brevity, we explicitly show the arguments only in the first line.
And for $\vec{V} \perp \hat{a}$,
    \begin{subequations}
    \begin{align}
     J_x(t, \vec{x}) &= \dot{\delta} \left(2 \partial_x \delta -  \frac{f_\barbeta}{\bar{c}} \dot{\delta} \right) = \sqrt{2} (\partial_{\bar{x}} \bar{\phi}) (\partial_{\bar{t}} \bar{\phi}) \,, \\
     J_y(t, \vec{x}) &= \dot{\delta} (\partial_y \delta) = (\partial_{\bar{t}} \bar{\phi}) (\partial_{\bar{y}} \bar{\phi}) \,, \\
     J_z(t, \vec{x}) &=  \dot{\delta} (\partial_z \delta) = (\partial_{\bar{t}} \bar{\phi}) (\partial_{\bar{z}} \bar{\phi}) \,.
    \end{align}
    \end{subequations}
Note that the $f_\barbeta/\bar{c}$ terms drop out when written in terms of $\bar{\phi}$.

To streamline the calculation, we write derivatives with respect to $\bar{t}$ and $\vec{\bar{x}}$ of $\bar{\phi}$ as
    \begin{align}
      (\partial_{\bar{\mu}} \bar{\phi}) =- \sqrt{\frac{\pi}{2}} \frac{M_{\mathrm{eff}}}{(2\pi)^2} \frac{1}{|V_{\mathrm{eff}}|} \frac{g_m}{\sqrt{2} M_{\mathrm{Pl}}}
        \int d\bar{\omega} (i \bar{\omega}) f_{\bar{\mu}} e^{i\bar{\omega}\left( \bar{t} - \frac{1}{V_{\mathrm{eff}}} \bar{z}\right)} \frac{e^{-i \bar{\lambda} |\vec{\bar{x}}_\perp|}}{\sqrt{i \bar{\lambda} |\vec{\bar{x}}_\perp|}} \,,
    \end{align}
    with
    \begin{align}
     \label{eq:fbarmu}
     f_{\bar{\mu}} = \begin{cases}
                      1  \,,                                                                         &\bar{\mu} = \bar{t} \\
                      -\frac{1}{V_{\mathrm{eff}}}                                  \,,                   &\bar{\mu} = \bar{z} \\
                      -\frac{f_{\mathrm{crit}}}{c_{\mathrm{eff}}} \frac{\bar{x}}{|\vec{\bar{x}}_\perp|} \,,  &\bar{\mu} = \bar{x} \\
                      -\frac{f_{\mathrm{crit}}}{c_{\mathrm{eff}}} \frac{\bar{y}}{|\vec{\bar{x}}_\perp|} \,,  &\bar{\mu} = \bar{y}
                     \end{cases}\,,
    \end{align}
    where we assumed the $|\bar{\lambda} \vec{\bar{x}}_\perp| \gg 1$ limit.
Consider then the integral over the surface of a cylinder oriented along the $z$ axis with radius $a \to \infty$
    \begin{align}
     I^g_{\bar{\mu} \bar{\nu}} \equiv a \int d\varphi \int dz \, g(\varphi) \, (\partial_{\bar{\mu}} \bar{\phi})(\bar{t}, \vec{\bar{x}}) (\partial_{\bar{\nu}} \bar{\phi})(\bar{t}, \vec{\bar{x}})\,,
    \end{align}
    with $x = a \cos \varphi$ and $y = a \sin \varphi$.
    Here, $g$ is a function of $\varphi$.
    Later, we will consider $g = \cos \varphi$ and $g = \sin \varphi$.
We can now calculate (assuming $\bar{z}/z$ to be constant)
    \begin{align}
    \label{eq:Edotcalculation:standardsfdm:Imunu}
    \begin{split}
     I^g_{\bar{\mu} \bar{\nu}}
        &=\begin{multlined}[t][.7\textwidth]
            a \frac{1}{64 \pi^3} \frac{M_{\mathrm{eff}}^2}{V_{\mathrm{eff}}^2} \frac{g_m^2}{M_{\mathrm{Pl}}^2}
            \int d\varphi g(\varphi) \int dz \int d \bar{\omega} (i\bar{\omega})
             f_{\bar{\mu}} \int d\bar{\omega}' (-i \bar{\omega}') f_{\bar{\nu}}'^\dagger \\
             \times  e^{i \bar{t}(\bar{\omega} - \bar{\omega}')} e^{-i \bar{z} \frac{1}{V_{\mathrm{eff}}}(\bar{\omega} - \bar{\omega}')} \frac{e^{-i |\vec{\bar{x}}_\perp| (\bar{\lambda} - \bar{\lambda}')}}{\sqrt{\bar{\lambda} \bar{\lambda}'} |\vec{\bar{x}}_\perp|}
          \end{multlined} \\
        &=\begin{multlined}[t][.75\textwidth]
            a \frac{1}{64 \pi^3} \frac{M_{\mathrm{eff}}^2}{V_{\mathrm{eff}}^2} \frac{g_m^2}{M_{\mathrm{Pl}}^2} \left(2\pi \frac{1}{\left|\frac{\bar{z}}{z} \frac{1}{V_{\mathrm{eff}}} - a_\parallel \frac{f_\barbeta}{2 \bar{c}}\right|} \right) \\
            \times \int d\varphi g(\varphi) \frac{1}{|\vec{\bar{x}}_\perp|}  \int d \bar{\omega} \frac{\bar{\omega}^2}{\bar{\lambda}} f_{\bar{\mu}} f_{\bar{\nu}}^\dagger
          \end{multlined} \\
        &=\begin{multlined}[t][.75\textwidth]
            \frac{1}{16 \pi^2} \frac{M_{\mathrm{eff}}^2}{|V_{\mathrm{eff}}|} \frac{g_m^2}{M_{\mathrm{Pl}}^2} \left(\frac{c_{\mathrm{eff}}}{f_{\mathrm{crit}}} \frac{1}{|V_{\mathrm{eff}}| \left|\frac{\bar{z}}{z} \frac{1}{V_{\mathrm{eff}}} - a_\parallel \frac{f_\barbeta}{2 \bar{c}}\right|} \right) \\
            \times \int d\varphi g(\varphi) \frac{a}{|\vec{\bar{x}}_\perp|}  \int_+ d \bar{\omega} \bar{\omega} \, f_{\bar{\mu}} f_{\bar{\nu}}^\dagger \,.
          \end{multlined}
    \end{split}
    \end{align}
Here, $\int_+$ means we integrate only over positive values.
This result holds for both $\vec{V} \parallel \hat{a}$ and $\vec{V} \perp \hat{a}$.
For $\vec{V} \perp \hat{a}$, we have $a_\parallel = 0$ in Eq.~\eqref{eq:Edotcalculation:standardsfdm:Imunu}.
For $\vec{V} \parallel \hat{a}$, we have $a_\parallel = \pm 1$.

The energy loss due to Cherenkov radiation is an integral over such a cylinder surface \cite{Jackson1998}.
We take this cylinder to be infinitely long along the $z$ direction with a large but finite radius $a$.
We consider only the side of the cylinder, since the top and bottom bases do not contribute, as we show below,
    \begin{align}
     \dot{E} = \int d\vec{S} \vec{J} =  a \int d\varphi \int dz (\cos \varphi J_x + \sin \varphi J_y) \,.
    \end{align}
For $\vec{V} \parallel \hat{a}$, we have $\vec{\bar{x}}_\perp = \vec{x}_\perp$ such that
    \begin{align}
    \begin{split}
     \dot{E}
        &= I^{\cos \varphi}_{\bar{t} \bar{x}} + I^{\sin \varphi}_{\bar{t} \bar{y}} \\
        &=\begin{multlined}[t][.7\textwidth]
            -\frac{1}{16 \pi^2} \frac{M_{\mathrm{eff}}^2}{|V_{\mathrm{eff}}|} \frac{g_m^2}{M_{\mathrm{Pl}}^2} \frac{1}{|V_{\mathrm{eff}}|\left|\frac{1}{\sqrt{2}} \frac{1}{V_{\mathrm{eff}}} - a_\parallel \frac{f_\barbeta}{2 \bar{c}}\right|} \\
            \times \int d\varphi (\cos^2 \varphi + \sin^2 \varphi)
            \int_+ d \bar{\omega} \bar{\omega}
          \end{multlined} \\
        &= -\frac1{8 \pi} \frac{M^2}{V} \frac{g_m^2}{M_{\mathrm{Pl}}^2} \int_+ d\bar{\omega} \bar{\omega} \,.
    \end{split}
    \end{align}
For $\vec{V} \perp \hat{a}$, we have
    \begin{align}
     |\vec{\bar{x}}_\perp| = a \sqrt{\frac{1 + \sin^2 \varphi}{2}} \,.
    \end{align}
This gives
    \begin{align}
    \begin{split}
     \dot{E}
        &= \sqrt{2} I_{\bar{t} \bar{x}}^{\cos \varphi} + I_{\bar{t} \bar{y}}^{\sin \varphi} \\
        &=\begin{multlined}[t][.7\textwidth]
            - \frac{1}{16 \pi^2} \frac{M_{\mathrm{eff}}^2}{|V_{\mathrm{eff}}|} \frac{g_m^2}{M_{\mathrm{Pl}}^2} \\
            \times \int d\varphi \sqrt{\frac{2}{1 + \sin^2 \varphi}}^2 \left(\sqrt{2} \frac{1}{\sqrt{2}} \cos^2 \varphi + \sin^2 \varphi \right) \int_+ d\bar{\omega} \bar{\omega}
          \end{multlined}\\
        &= - \frac1{\sqrt{2} \, 8\pi} \frac{M^2}{V} \frac{g_m^2}{M_{\mathrm{Pl}}^2} \int_+ d\bar{\omega} \bar{\omega} \,.
    \end{split}
    \end{align}

The contributions of the top and bottom cylinder bases are proportional to
\begin{align}
 \label{eq:Edotcalculation:standardsfdm:cylinderbases}
 \left. \int_{|\vec{x}_\perp| < a} d^2\vec{x}_\perp (\partial_{\bar{t}} \bar{\phi}) (\partial_{\bar{z}} \bar{\phi})\right|_{z \to \pm \infty} \,.
\end{align}
This holds for both $\vec{V} \parallel \hat{a}$ and $\vec{V} \perp \hat{a}$.
In position space \cite{Jackson1998}, we have
\begin{align}
 \label{eq:Edotcalculation:standardsfdm:positionspace}
 \bar{\phi} \propto \frac{1}{\sqrt{(\bar{z} - V_{\mathrm{eff}} \bar{t})^2 - \frac{f_{\mathrm{crit}}^2}{1 - f_{\mathrm{crit}}^2} |\vec{\bar{x}}_\perp|^2}} \,,
\end{align}
whenever the argument of the square root is positive.
Otherwise, $\bar{\phi}$ vanishes,
\begin{align}
 \label{eq:Edotcalculation:standardsfdm:insideshockwave}
 \bar{\phi} = 0
    \quad \mathrm{for} \quad (\bar{z} - V_{\mathrm{eff}} \bar{t})^2 - \frac{f_{\mathrm{crit}}^2}{1 - f_{\mathrm{crit}}^2} |\vec{\bar{x}}_\perp|^2 < 0 \,.
\end{align}
For both $\vec{V} \parallel \hat{a}$ and $\vec{V} \perp \hat{a}$, the limit $z \to \pm \infty$ at fixed $t$ and $\vec{x}_\perp$ implies $|\bar{z} - V_{\mathrm{eff}} \bar{t}| \to \infty$.
Thus, the integrand in Eq.~\eqref{eq:Edotcalculation:standardsfdm:cylinderbases} vanishes at least as $1/z^4$.
Therefore, the cylinder bases do not contribute to $\dot{E}$.

\subsection{Cutoffs}
\label{sec:Edotcalculation:standardsfdm:cutoffs}

The above calculations were without any kind of cutoff.
But, as discussed in Sec.~\ref{sec:general},
    we need to impose cutoffs.
In our QFT calculation in Appendix~\ref{sec:Edotcalculation},
    we imposed cutoffs $k_{\mathrm{min}}$ and $k_{\mathrm{max}}$ on the spatial wavevector $\vec{k}$ of the emitted radiation.
In our classical computation, $\vec{k}$ corresponds to the prefactor of $i \vec{x}$ in the $e^{i \vec{k} \vec{x}}$ factor in the Fourier transform of $\delta$.
We express $\delta(t, \vec{x})$ through $\bar{\phi}(\bar{t}, \vec{\bar{x}})$
    which depends on the coordinates as $e^{i (\bar{\omega} \bar{t} - \vec{\bar{k}} \vec{\bar{x}})}$.
We can write the dependence of this exponential on $\vec{x}$ as $e^{-i \vec{k} \vec{x}}$ for some $\vec{k}$ using the transformation between $\bar{t}$, $\vec{\bar{x}}$ and $t$, $\vec{x}$.
Then, we can try to impose our standard cutoffs on the absolute value of $\vec{k}$.
The calculation in the previous section naturally leaves an $\bar{\omega}$ integral in the end.
Thus, we will try to express $|\vec{k}|$ through $\bar{\omega}$.
Then, our standard cutoffs on $|\vec{k}|$ translate to cutoffs on $\bar{\omega}$.

Consider first $\vec{V} \parallel \hat{a}$.
Then,
    \begin{align}
    \begin{split}
    &\bar{\omega} \bar{t} - \vec{\bar{k}} \vec{\bar{x}} = \bar{\omega} \bar{t} - \vec{\bar{k}}_\perp \vec{\bar{x}}_\perp - \bar{k}_{\bar{z}} \bar{z} \\
        &= \bar{\omega} t - \vec{\bar{k}}_\perp \vec{x}_\perp - z \left( \frac{1}{\sqrt{2}} \bar{k}_{\bar{z}} - a_\parallel\frac{f_\barbeta}{2\bar{c}} \bar{\omega} \right) \,.
    \end{split}
    \end{align}
Thus
    \begin{align}
    \label{eq:k2parallel}
    |\vec{k}|^2 = \bar{k}_\perp^2 + \left( \frac{1}{\sqrt{2}} \bar{k}_{\bar{z}} - a_\parallel\frac{f_\barbeta}{2\bar{c}} \bar{\omega} \right)^2 \,.
    \end{align}
The source term $\delta_{b,\mathrm{eff}} \propto \delta(\bar{z} - V_{\mathrm{eff}} \bar{t})$ gives in Fourier space $\bar{\omega} = V_{\mathrm{eff}} \bar{k}_{\bar{z}}$.
Further, we are interested only in the on-shell contribution from the pole $\bar{\omega} = c_{\mathrm{eff}} \bar{k}$.
Thus, using $\bar{k}_\perp^2 = \bar{k}^2  - \bar{k}_{\bar{z}}^2$, we have
    \begin{align}
    \begin{split}
    |\vec{k}|^2
        &= \bar{\omega}^2 \left[\left(\frac{1}{c_{\mathrm{eff}}^2} - \frac{1}{V_{\mathrm{eff}}^2} \right) + \left(\frac{1}{\sqrt{2}} \frac{1}{V_{\mathrm{eff}}} - a_\parallel \frac{f}{2\bar{c}} \right)^2 \right] \\
        &= \frac{\bar{\omega}^2}{\bar{c}^2} \left(\frac{1}{1 - (\bar{c}/V)^2 - 2 a_\parallel f_\barbeta (\bar{c}/V) }\right)^{-1}
        \equiv \frac{\bar{\omega}^2}{\bar{c}^2} \frac{1}{\left(f^\parallel_{\mathrm{max}}\right)^2}
        \,.
    \end{split}
    \end{align}
This is a relation between $k$ and $\bar{\omega}$.
We can use this to impose our standard cutoffs on $k$, i.e. $k < k_{\mathrm{max}}$ and $k > k_{\mathrm{min}}$, in the $\bar{\omega}$ integrals in the result for $\dot{E}$ found in the previous section.
This gives the final result for the energy loss
    \begin{align}
        \dot{E} = -\frac{\bar{c}^2}{16 \pi} \frac{M^2}{V} \frac{g_m^2}{M_{\mathrm{Pl}}^2} \frac{1}{1 - (\bar{c}/V^2) - 2 a_\parallel f_\barbeta (\bar{c}/V)} (k_{\mathrm{max}}^2 - k_{\mathrm{min}}^2)  \,.
    \end{align}

Consider next $\vec{V} \perp \hat{a}$.
Then,
    \begin{align}
    \begin{split}
      &\bar{\omega} \bar{t} - \vec{\bar{k}} \vec{\bar{x}} = \bar{\omega} \bar{t} - \bar{k}_{\bar{x}} \bar{x} - \bar{k}_{\bar{y}} \bar{y} - \bar{k}_{\bar{z}} \bar{z} \\
        &= \bar{\omega} t - x \left(\frac{1}{\sqrt{2}} \bar{k}_{\bar{x}} - \frac{f_\barbeta}{2\bar{c}} \bar{\omega} \right) - \bar{k}_{\bar{y}} y - \bar{k}_{\bar{z}} z \,.
    \end{split}
    \end{align}
That is
    \begin{align}
    |\vec{k}|^2 =\left(\frac{1}{\sqrt{2}} \bar{k}_{\bar{x}} - \frac{f_\barbeta}{2 \bar{c}} \bar{\omega} \right)^2+ \bar{k}_{\bar{y}}^2 + \bar{k}_{\bar{z}}^2 \,.
    \end{align}
We can again use the two relations $\bar{\omega} = V_{\mathrm{eff}} \bar{k}_{\bar{z}}$ and $\bar{\omega} = c_{\mathrm{eff}} \bar{k}$.
But, in contrast to the $\vec{V} \parallel \hat{a}$ case,
    this does not uniquely specify $|\vec{k}|$ in terms of $\bar{\omega}$.
Instead, $|\vec{k}|$ depends on both $\bar{\omega}$ and the orientation of $\vec{\bar{k}}$, i.e. the spherical angles $\bar{\theta}$ and $\bar{\varphi}$.
Concretely
    \begin{align}
    \begin{split}
    |\vec{k}|^2
        &= \frac12 \bar{k}_{\bar{x}}^2 + \bar{k}_{\bar{y}}^2 + \bar{k}_{\bar{z}}^2+ \left(\frac{f_\barbeta}{2 \bar{c}}\right)^2 \bar{\omega}^2 - \frac1{\sqrt{2}} \frac{f_\barbeta}{\bar{c}} \bar{k}_{\bar{x}} \bar{\omega}\\
        &= \frac12 \bar{k}_{\bar{x}}^2 + (\bar{k}^2 - \bar{k}_{\bar{x}}^2 - \bar{k}_{\bar{z}}^2) + \bar{k}_{\bar{z}}^2 + \left(\frac{f_\barbeta}{2 \bar{c}}\right)^2 \bar{\omega}^2 - \frac1{\sqrt{2}} \frac{f_\barbeta}{\bar{c}} \bar{k}_{\bar{x}} \bar{\omega} \\
        &= \bar{k}^2 - \frac12 \bar{k}^2 s_{\bar{\theta}}^2 c_{\bar{\varphi}}^2 - \frac{1}{\sqrt{2}} \frac{f_\barbeta}{\bar{c}} \bar{k} \bar{\omega} s_{\bar{\theta}} c_{\bar{\varphi}} + \left(\frac{f_\barbeta}{2 \bar{c}}\right)^2 \bar{\omega}^2 \\
        &= \bar{k}^2 + \left(\frac{f_\barbeta}{2\bar{c}}\right)^2 \bar{\omega}^2 - \frac12 \left(\bar{k} s_{\bar{\theta}} c_{\bar{\varphi}} + \frac{1}{\sqrt{2}} \frac{f_\barbeta}{\bar{c}} \bar{\omega} \right)^2 + \frac14 \frac{f_\barbeta^2}{\bar{c}^2} \bar{\omega}^2 \,,
    \end{split}
    \end{align}
where $s_{\bar{\theta}} = \sin \bar{\theta}$ and $c_{\bar{\varphi}} = \cos \bar{\varphi}$.
To get a relation between $k$ and $\bar{\omega}$,
    we could use the relations $\bar{\omega} = V_{\mathrm{eff}} \bar{k}_{\bar{z}}$ and $\bar{\omega} = c_{\mathrm{eff}} \bar{k}$ to fix $\cos \bar{\theta} = c_{\mathrm{eff}}/V_{\mathrm{eff}}$.
But this still leaves $\bar{\varphi}$ undetermined.

To avoid complications from this, we do something simpler.
We set $\bar{\varphi}$ to the value that gives the largest value $(|\vec{k}|^2)_{\mathrm{largest}}$ of $|\vec{k}|$.
Using this largest value of $|\vec{k}|$ to set the cutoff on $\bar{\omega}$ gives a more conservative cutoff than the cutoffs we used previously.
This largest possible value of $|\vec{k}|$ is
    \begin{align}
     \label{eq:kperplargestgeneral}
     \left(|\vec{k}|^2\right)_{\mathrm{largest}} = \bar{k}^2 + \frac{f_\barbeta^2}{2} \left(\frac{\bar{\omega}}{\bar{c}}\right)^2 \,.
    \end{align}
With $\bar{\omega} = c_{\mathrm{eff}} \bar{k}$, this gives
    \begin{align}
    \begin{split}
      \left(|\vec{k}|^2\right)_{\mathrm{largest}}
        &= \left(\frac{\bar{\omega}}{\bar{c}}\right)^2 \left[\left(\frac{\bar{c}}{c_{\mathrm{eff}}}\right)^2 +\frac12 f_\barbeta^2 \right] \\
        &= \left(\frac{\bar{\omega}}{\bar{c}}\right)^2  (1 + f_\barbeta^2)
        \equiv \left(\frac{\bar{\omega}}{\bar{c}}\right)^2 \frac{1}{ (f^\perp_{\mathrm{max}})^2 }\,.
    \end{split}
    \end{align}
We will use this relation between $k$ and $\bar{\omega}$ to set the upper cutoff in the $\bar{\omega}$ integral.
For the lower cutoff,
    we should in principle use the value of $\bar{\varphi}$ that gives the smallest possible $|\vec{k}|^2$.
However, our result is dominated by the upper cutoff.
Thus,
    for simplicity,
    we take the same value of $\bar{\varphi}$ for both the upper and the lower cutoff.
Imposing our standard cutoffs on $k$ then gives the final result for the energy loss
    \begin{align}
        \dot{E} = - \frac{\bar{c}^2}{\sqrt{2} \, 16\pi} \frac{M^2}{V} \frac{g_m^2}{M_{\mathrm{Pl}}^2} \frac{1}{1 + f_\barbeta^2} (k_{\mathrm{max}}^2 - k_{\mathrm{min}}^2) \,.
    \end{align}

Above, we have ignored three subtleties regarding cutoffs that we discuss now.
First, we have calculated the energy loss through Cherenkov radiation by
    integrating the energy flux over the surface of an infinitely large cylinder.
But we justified our cutoffs by arguing that the perturbation's Lagrangian is valid only between a distance $r_{\mathrm{min}}$ and $r_{\mathrm{max}}$ away from the perturber.
So maybe we should use a cylinder with radius and length at most $r_{\mathrm{max}}$ instead of an infinitely large cylinder.
Here, we do not do this.
Instead, we keep the infinitely large cylinder but use the field $\bar{\phi}$ as calculated with a momentum-space cutoff.
One reason is that energy conservation guarantees that the integral over an infinitely large cylinder gives the same result as that over a finite cylinder.
Of course, the infinitely large cylinder is unphysical.
But it suffices that, mathematically, the result will be the same in both cases.
More precisely, this holds only up to small corrections because, with the cutoffs, the perturber is no longer a strictly localized point particle.
Instead, the perturber is smeared out over a distance $r_{\mathrm{min}}$ so that some of the Cherenkov radiation is produced outside any finite volume.
Corrections from this are negligible as long as $r_{\mathrm{min}} \ll r_{\mathrm{max}}$, which is true in our case.
Another reason why we keep the infinitely large cylinder is that this is what we already did implicitly when we employed the standard QFT formalism with momentum-space cutoffs in our standard calculation in Appendix~\ref{sec:Edotcalculation}.

The second subtlety is that, in our calculation of $\bar{\phi}(\bar{t}, \vec{\bar{x}})$, we used the following result from Ref.~\cite{Jackson1998}
\begin{align}
 \label{eq:standardsfdm:I}
 I \equiv \int d^2\bar{k}_\perp e^{-i \vec{\bar{k}}_\perp \vec{\bar{x}}_\perp} \frac{1}{\bar{\lambda}^2 - \bar{k}_\perp^2}
    = -2\pi K_0(i \bar{\lambda} |\vec{\bar{x}}_\perp|) \,.
\end{align}
This, in principle, includes wavevectors $\bar{k}_\perp$ that are arbitrarily large.
Instead we should use a version of this integral with appropriate cuts on $\bar{k}_\perp$.
However, the $e^{-i\vec{\bar{k}}_\perp \vec{\bar{x}}_\perp}$ factor already effectively cuts off the integral at
    \begin{align}
     \bar{k}_{\perp} \sim 1/|\vec{\bar{x}}_\perp| \,.
    \end{align}
Thus, our previous calculation is valid as long as $|\vec{\bar{x}}_\perp|$ is larger than $1/\bar{k}_{\perp,\mathrm{max}}$
    where $\bar{k}_{\perp,\mathrm{max}}$ is the cutoff we should impose on $\bar{k}_\perp$.
For smaller values of $|\vec{\bar{x}}_\perp|$ we should use a version of the integral with cutoffs.

Specifically, we have for the upper cutoff on $\bar{k}_\perp$,
\begin{subequations}
\begin{align}
 \left(\bar{k}^\parallel_{\perp,\mathrm{max}}\right)^2
    &= k_{\mathrm{max}}^2 - \left(\frac{\bar{\omega}}{\bar{c}}\right)^2 \left(\frac{1}{\sqrt{2}} \frac{\bar{c}}{V_{\mathrm{eff}}} - a_\parallel \frac{f_\barbeta}{2} \right)^2 \,, \\
  \left(\bar{k}^\perp_{\perp,\mathrm{max}}\right)^2
    &= k_{\mathrm{max}}^2 - \left(\frac{\bar{\omega}}{\bar{c}}\right)^2 \left( \left(\frac{\bar{c}}{V_{\mathrm{eff}}}\right)^2+ \frac12 f_\barbeta^2\right)
    \,,
\end{align}
\end{subequations}
where $\bar{k}_{\perp,\mathrm{max}}^\parallel$ applies for $\vec{V} \parallel \hat{a}$ and $\bar{k}_{\perp,\mathrm{max}}^\perp$ applies for $\vec{V} \perp \hat{a}$.
This follows from Eq.~\eqref{eq:k2parallel} and Eq.~\eqref{eq:kperplargestgeneral} using $\bar{k}_{\bar{z}} = \bar{\omega}/V_{\mathrm{eff}}$.
Both cases can be written as
\begin{align}
\begin{split}
 (\bar{k}_{\perp,\mathrm{max}})^2
 &= k_{\mathrm{max}}^2 - \bar{\omega}^2 \left( \left(\frac{1}{\bar{c} f_{\mathrm{max}}}\right)^2 - \left(\frac{\bar{\lambda}}{\bar{\omega}}\right)^2 \right) \\
 &= k_{\mathrm{max}}^2 \left(1 - \left(\frac{\bar{\omega}}{\bar{\omega}_{\mathrm{max}}}\right)^2 \left(1 - f_{\mathrm{crit}}^2 f_{\mathrm{max}}^2 \left(\frac{\bar{c}}{c_{\mathrm{eff}}}\right)^2 \right) \right) \,.
\end{split}
\end{align}
The smallest possible value of $\bar{k}_{\perp,\mathrm{max}}^2$ is exactly $\bar{\lambda}^2$ evaluated at $\bar{\omega} = \bar{\omega}_{\mathrm{max}}$.
Thus, our previous calculation is valid if
\begin{align}
 |\vec{\bar{x}}_\perp|
    \gtrsim \frac{1}{\bar{\lambda}|_{\bar{\omega} = \bar{\omega}_{\mathrm{max}}}}
    = r_{\mathrm{min}} \cdot \frac{c_{\mathrm{eff}}}{\bar{c}} \frac{1}{f_{\mathrm{max}} f_{\mathrm{crit}} } \,.
\end{align}
The right-hand side is of the same of order of magnitude as $r_{\mathrm{min}}$
    except if $c_{\mathrm{eff}}/V_{\mathrm{eff}}$ is close to $1$,
    i.e. for barely supersonic perturbers.
In this case, $f_{\mathrm{crit}}$ becomes small.
But, as we will discuss below, we consider only perturbers that are at most $1\%$ away from the critical velocity.
So this is never much larger than $r_{\mathrm{min}}$ for our purposes.
Thus, for our energy loss calculation we can keep our previous result.

In principle, there is also a lower cutoff on $\bar{k}_\perp$.
However, this cutoff is 0 except for $\bar{\omega}$ close to its lower cutoff.
As already mentioned above,
    our results are dominated by the upper cutoff of $\bar{\omega}$.
Thus, we leave out this lower cutoff on $\bar{k}_\perp$ for simplicity.

The third subtlety is regarding stars that are barely supersonic, $V \to V_{\mathrm{crit}}$.
One may suspect a logarithmic divergence in $\bar{\phi}$ at $V \to V_{\mathrm{crit}}$, i.e. at $V_{\mathrm{eff}} \to c_{\mathrm{eff}}$.
The reason is that $\bar{\lambda}^2 \propto f_{\mathrm{crit}}^2 = 1 - (c_{\mathrm{eff}}/V_{\mathrm{eff}})^2$ and $K_0(i \bar{\lambda} |\vec{\bar{x}}_\perp|)$ diverges logarithmically for small arguments.
Indeed, such a divergence exists in our case and we will restrict ourselves to perturbers at least $1\%$ away from $V_{\mathrm{crit}}$ in order to stay away from this divergence.

This logarithmic divergence does not affect the calculated energy loss.
The energy loss is calculated from an expansion of $K_0(i \bar{\lambda} |\vec{\bar{x}}_\perp|)$ for large arguments, while the divergence comes from an expansion of this function for small arguments.
However, the logarithmic divergence is relevant when judging whether or not perturbations are small at various locations, for example at $\vec{\bar{x}}_\perp = 0$ where the argument of $K_0$ is definitely small.
This is discussed in more detail in Appendix~\ref{sec:standasfdm:perturbationssmall}.
In electrodynamics, such logarithms occur as well (see for example Eq.~(13.39) in Ref.~\cite{Jackson1998} and the discussion below this equation).

For later reference, we explicitly give the result of the $\bar{k}_\perp$ integral at $|\vec{\bar{x}}_\perp| = 0 $ including cutoffs,
\begin{align}
\begin{split}
 \label{eq:standardsfdm:kbarintegral-cutoff}
  &\frac{1}{2\pi} \int d^2\bar{k}_\perp e^{-i \vec{\bar{k}}_\perp \vec{\bar{x}}_\perp} \frac{\Theta(\bar{k}_{\perp,\mathrm{max}} - \bar{k}_\perp)}{\bar{\lambda}^2 - k_\perp^2} \\
    &= - K_0(i\bar{\lambda} |\vec{\bar{x}}_\perp|)
        -  \int^{\infty}_{\bar{k}_{\perp,\mathrm{max}}} d\bar{k}_\perp \frac{\bar{k}_\perp J_0(\bar{k}_\perp |\bar{\vec{x}}_\perp|)}{\bar{\lambda}^2 - \bar{k}_\perp^2} \\
    &\approx - K_0(i\bar{\lambda} |\vec{\bar{x}}_\perp|)
        - \int_{\bar{k}_{\perp,\mathrm{max}}}^{N |\bar{\lambda}|} d\bar{k}_\perp \frac{\bar{k}_\perp}{\bar{\lambda^2} - \bar{k}_\perp^2}
        - \int_{N |\bar{\lambda}|}^{\infty} d\bar{k}_\perp \frac{J_0(\bar{k}_\perp |\vec{\bar{x}}_\perp|)}{-\bar{k}_\perp}
        \\
    &\approx \left(\gamma_E + \ln\left(\frac{|\bar{\lambda} \vec{\bar{x}}_\perp|}{2}\right) + i\sigma(\bar{\lambda}) \frac{\pi}{2} \right) \\
    & \qquad    - \left(
            -\frac12 \left[\ln\left(\bar{k}_\perp^2 - \bar{\lambda}^2\right) \right]^{N|\bar{\lambda}|}_{\bar{k}_{\perp,\mathrm{max}}}
            +\gamma_E + \ln\left(\frac{N |\vec{\bar{x}}_\perp| |\bar{\lambda}|}{2} \right)
            \right) \\
    &\approx - \frac12 \ln\left(\frac{\bar{k}_{\perp,\mathrm{max}}^2}{\bar{\lambda}^2} - 1\right) + i \sigma(\bar{\lambda}) \frac{\pi}{2} \,,
\end{split}
\end{align}
where we expanded in $|\vec{\bar{x}}_\perp|$,
$N$ is a large integer,
and we used the fact that $\bar{k}_{\perp,\mathrm{max}}^2$ is always larger than $\bar{\lambda}^2$.
If we define $\bar{f}_{\mathrm{crit}} \equiv f_{\mathrm{crit}} \cdot f_{\mathrm{max}} \cdot (\bar{c} / c_{\mathrm{eff}})$,
    we can write the remaining logarithm as
\begin{align}
 \ln\left(\frac{1 - (1 - \bar{f}_{\mathrm{crit}}^2) (\bar{\omega}/\bar{\omega}_{\mathrm{max}})^2}{\bar{f}_{\mathrm{crit}}^2 (\bar{\omega}/\bar{\omega}_{\mathrm{max}})^2} -1 \right)
 = \ln\left(\frac{1 - (\bar{\omega}/\bar{\omega}_{\mathrm{max}})^2}{(\bar{\omega}/\bar{\omega}_{\mathrm{max}})^2}\right)
  - \ln\left(\bar{f}_{\mathrm{crit}}^2\right)\,.
\end{align}

\subsection{Direction of friction force}
\label{sec:Edotcalculation:standardsfdm:direction}

Above, we calculated the total energy loss of a perturber due to Cherenkov radiation.
But in which direction does this energy loss push the perturber?
To answer this question
    we can calculate the linear momentum loss in different directions.
Or,
    a bit simpler,
    we can calculate the relative amount of the momentum loss in different directions.

The canonical linear momentum of perturbations is
    \begin{align}
     \vec{P}
        = \frac{\partial \mathcal{L}}{\partial \dot{\delta}} (\vec{\nabla} \delta)
        = \left(\frac{\dot{\delta}}{\bar{c}^2} + \frac{f_\barbeta}{\bar{c}} \hat{a} \vec{\nabla} \delta\right) (\vec{\nabla} \delta) \,.
    \end{align}
Each component $P_i$ satisfies a continuity equation $\partial_t P_i - \partial_j (J^P_i)^j = 0$.
If we know $(J^P_i)^j$
    we can calculate the momentum loss in a way similar to how we calculated the energy loss,
    i.e. by evaluating a surface integral.

We have, using the $\delta$ equation of motion
    \begin{align}
    \begin{split}
    \partial_t P_i
        &= \left(\frac{\ddot{\delta}}{\bar{c}^2} + \frac{f_\barbeta}{\bar{c}} \hat{a} \vec{\nabla}  \dot{\delta} \right) (\partial_i \delta) + \left(\frac{\dot{\delta}}{\bar{c}^2} + \frac{f_\barbeta}{\bar{c}} \hat{a} \vec{\nabla} \delta \right) (\partial_i \dot{\delta}) \\
        &=\begin{multlined}[t][.7\textwidth]
            \left(\vec{\nabla}^2 \delta + \vec{\nabla} \left( (\vec{\nabla} \delta \hat{a}) \hat{a} \right) - 2 \hat{a} \frac{f_\barbeta}{\bar{c}} \vec{\nabla} \dot{\delta}+ \frac{f_\barbeta}{\bar{c}} \hat{a} \vec{\nabla} \dot{\delta} \right) (\partial_i \delta) \\
            + \left(\frac{\dot{\delta}}{\bar{c}^2} + \frac{f_\barbeta}{\bar{c}} \hat{a} \vec{\nabla} \delta \right) (\partial_i \dot{\delta})
          \end{multlined} \\
        &=\begin{multlined}[t][.7\textwidth]
            \partial_j \left(
                \partial_i \delta \partial_j  \delta +
                a_j (a_k \partial_k \delta) (\partial_i \delta) -
                \frac{f_\barbeta}{\bar{c}} a_j \dot{\delta} (\partial_i \delta)
                \right) \\
            + \partial_i \left(
                \frac12 \frac{\dot{\delta}^2}{\bar{c}^2} -
                \frac12 (\partial_j \delta)^2 -
                \frac12 (a_j \partial_j \delta)^2 +
                \frac{f_\barbeta}{\bar{c}} (a_j \partial_j \delta) \dot{\delta}
                \right) \,.
          \end{multlined}
    \end{split}
    \end{align}
Thus
    \begin{multline}
        (J^P_i)^j =
            \partial_i \delta \partial_j  \delta +
            a_j (a_k \partial_k \delta) (\partial_i \delta) -
            \frac{f_\barbeta}{\bar{c}} a_j \dot{\delta} (\partial_i \delta) \\
            + \delta_i^j \left(
                \frac12 \frac{\dot{\delta}^2}{\bar{c}^2} -
                \frac12 (\partial_k \delta)^2 -
                \frac12 (a_k \partial_k \delta)^2 +
                \frac{f_\barbeta}{\bar{c}} (a_k \partial_k \delta) \dot{\delta}
                \right) \,.
    \end{multline}

The linear momentum loss of the $i$-th component through the surface of a cylinder with radius $a \to \infty$ is then
    \begin{align}
     \dot{P}_i = a \int dz \int d\varphi \left(\cos \varphi (J^P_i)^x + \sin \varphi (J^P_i)^y\right) \,.
    \end{align}
As in the case of the energy loss,
    we can express $(J^P_i)^j$ as a linear combination of $(\partial_{\bar{\mu}} \bar{\phi})(\partial_{\bar{\nu}} \bar{\phi})$
    using the transformation between $\bar{t}$, $\vec{\bar{x}}$ and $t$, $\vec{x}$.
Then, $\dot{P}_i$ is just a linear combination of the integrals $I^{\cos \varphi}_{\bar{\mu} \bar{\nu}}$ and $I^{\sin \varphi}_{\bar{\mu} \bar{\nu}}$ defined in Appendix~\ref{sec:Edotcalculation:standardsfdm:nocutoffs}.
Many of these $I^g_{\bar{\mu} \bar{\nu}}$ terms vanish after doing the $\varphi$ integral.
The only non-zero integrals for $g = \cos \varphi$ are
    \begin{align}
        I^{\cos \varphi}_{\bar{x} \bar{t}} \,, \quad  I^{\cos \varphi}_{\bar{x} \bar{z}} \,,
    \end{align}
and the only non-zero integrals for $g = \sin \varphi$ are
    \begin{align}
        I^{\sin \varphi}_{\bar{y} \bar{t}} \,, \quad I^{\sin \varphi}_{\bar{x} \bar{z}} \,.
    \end{align}
This can be seen from the fact that the $\varphi$ integral is of the form
    \begin{align}
     \int d\varphi \, \tilde{g}(\sin^2 \varphi) \, g(\varphi) \, f_{\bar{\mu}} f_{\bar{\nu}} \,,
    \end{align}
where $\tilde{g}$ is some function of $\sin^2 \varphi$ and then using the specific form of $f_{\bar{\mu}}$ and $f_{\bar{\nu}}$ given by Eq.~\eqref{eq:fbarmu}.

For $\vec{V} \parallel \hat{a}$, we have
    \begin{align}
    \begin{split}
     (J_x^P)^x
        = (\partial_x \delta)^2 + \frac12 \frac{\dot{\delta}^2}{\bar{c}^2}
            - \frac12 (\vec{\nabla} \delta)^2 - \frac12 (\partial_z \delta)^2 + \frac{f_\barbeta}{\bar{c}} (a_\parallel \partial_z \delta) \dot{\delta} \,.
    \end{split}
    \end{align}
    When written in terms of $\bar{\phi}$,
        this contains no terms proportional to $(\partial_{\bar{x}} \bar{\phi})(\partial_{\bar{t}} \bar{\phi})$ or $(\partial_{\bar{x}} \bar{\phi})(\partial_{\bar{z}} \bar{\phi})$.
    The same is true for $(J_x^P)^y$.
    Thus, $\dot{P}_x = 0$.
    An analogous argument shows that $\dot{P}_y = 0$.
    Thus, the friction force points in the $z$ direction.

For $\vec{V} \perp \hat{a}$, we have for $J_x^P$
    \begin{align}
    \begin{split}
     (J_x^P)^x
        &=\begin{multlined}[t][.7\textwidth]
            (\partial_x \delta)^2 + (\partial_x \delta)^2 - \frac{f_\barbeta}{\bar{c}} \dot{\delta} (\partial_x \delta)
            + \frac12 \frac{\dot{\delta}}{\bar{c}^2} - \frac12 (\vec{\nabla} \delta)^2 \\
            - \frac12 (\partial_x \delta)^2
            + \frac{f_\barbeta}{\bar{c}} (\partial_x \delta) \dot{\delta}
          \end{multlined} \\
        &= (\partial_x \delta)^2 + \dots
        = \frac{1}{\sqrt{2}} \frac{f_\barbeta}{\bar{c}} (\partial_{\bar{t}} \bar{\phi}) (\partial_{\bar{x}}) \bar{\phi} + \dots \,,
    \end{split}
    \end{align}
    where ``$\dots$'' denotes terms that cannot contribute to $\dot{P}_x$.
    Similarly,
    \begin{align}
    \begin{split}
     (J_x^P)^y
        &= (\partial_x \delta)(\partial_y \delta) = \left(\frac{1}{\sqrt{2}} (\partial_{\bar{x}} \bar{\phi}) + \frac{f_\barbeta}{2\bar{c}} (\partial_{\bar{t}} \bar{\phi}) \right) (\partial_{\bar{y}} \bar{\phi}) \\
        &= \frac{f_\barbeta}{2 \bar{c}} (\partial_{\bar{t}} \bar{\phi})(\partial_{\bar{y}} \bar{\phi}) + \dots \,.
    \end{split}
    \end{align}
For $(J_y^P)$, we find
    \begin{align}
    \begin{split}
     (J_y^P)^x
        &= (\partial_y \delta)(\partial_x \delta) + (\partial_x \delta)(\partial_y \delta) - \frac{f_\barbeta}{\bar{c}} \dot{\delta} (\partial_y \delta) \\
        &= (\partial_y \delta) \left(2 \partial_x \delta - \frac{f_\barbeta}{\bar{c}} \dot{\delta}\right) \\
        &= (\partial_{\bar{y}} \bar{\phi}) \left(2 \frac{1}{\sqrt{2}} (\partial_{\bar{x}} \bar{\phi}) + 2 \frac{f_\barbeta}{2 \bar{c}} (\partial_{\bar{t}} \bar{\phi}) - \frac{f_\barbeta}{\bar{c}} (\partial_{\bar{t}} \bar{\phi}) \right)
         = 0 + \dots \,,
    \end{split}
    \end{align}
and
    \begin{align}
    \begin{split}
     (J_y^P)^y
       &= (\partial_y \delta)^2 + \frac12 \frac{\dot{\delta}^2}{\bar{c}^2} - \frac12 (\vec{\nabla} \delta)^2 - \frac12 (\partial_x \delta)^2 + \frac{f_\barbeta}{\bar{c}} (\partial_x \delta) \dot{\delta} \\
       &= (\partial_x \delta) \left( -(\partial_x \delta) + \frac{f_\barbeta}{\bar{c}} \dot{\delta} \right) \\
       &= \left(\frac{1}{\sqrt{2}} (\partial_{\bar{x}} \bar{\phi}) + \frac{f_\barbeta}{2\bar{c}} (\partial_{\bar{t}} \bar{\phi}) \right) \left(-\frac{1}{\sqrt{2}} (\partial_{\bar{x}} \bar{\phi}) + \frac{f_\barbeta}{2 \bar{c}} (\partial_{\bar{t}} \bar{\phi}) \right) = 0 + \dots \,.
    \end{split}
    \end{align}
Thus, $\dot{P}_y = 0$, but $\dot{P}_x \neq 0$.
That is, the direction of the friction force lies in the $x$-$z$ plane, i.e. the plane spanned by $\vec{V}$ and $\hat{a}$.

To find the direction within this plane, we can compare $\dot{P}_x$ to $\dot{P}_z$.
We have
    \begin{align}
    \begin{split}
     (J^P_z)^x
      &= (\partial_z \delta) (\partial_x \delta) + (\partial_x \delta) (\partial_z \delta) - \frac{f_\barbeta}{\bar{c}} \dot{\delta} (\partial_z \delta) \\
      &= 2 (\partial_z \delta) (\partial_x \delta) + \dots = \sqrt{2} (\partial_{\bar{z}} \bar{\phi}) (\partial_{\bar{x}} \bar{\phi}) + \dots \,,
    \end{split}
    \end{align}
and
    \begin{align}
     (J^P_z)^y = (\partial_z \delta) (\partial_y \delta) = (\partial_{\bar{z}} \bar{\phi})(\partial_{\bar{y}} \bar{\delta}) \,.
    \end{align}
Using $(\partial_{\bar{z}} \bar{\phi}) = (-1/V_{\mathrm{eff}}) (\partial_{\bar{t}} \bar{\phi})$, we then have
    \begin{align}
     \frac{\dot{P}_x}{\dot{P}_z}
     = - V_{\mathrm{eff}} \frac{f_\barbeta}{\bar{c}} \frac{\frac{1}{\sqrt{2}} I^{\cos \varphi}_{\bar{t} \bar{x}} + \frac12 I^{\sin \varphi}_{\bar{t} \bar{y}}}{\sqrt{2} I^{\cos \varphi}_{\bar{t} \bar{x}} + I^{\sin \varphi}_{\bar{t} \bar{y}}} \,.
    \end{align}
The relevant $\varphi$ integral in $I^{\cos \varphi}_{\bar{t} \bar{x}}$ is
    \begin{align}
    \begin{split}
     \int d\varphi \frac{a \cos \varphi}{|\vec{\bar{x}}_\perp|} f_{\bar{x}}
     &= \int d\varphi \frac{a \cos \varphi}{|\vec{\bar{x}}_\perp|} \left(-\frac{\bar{x}}{|\vec{\bar{x}}_\perp|}\right) \\
     &= -\sqrt{2} \int d\varphi \frac{\cos^2 \varphi}{1 + \sin^2 \varphi} = 2 \pi (\sqrt{2} - 2) \,.
    \end{split}
    \end{align}
The corresponding $\varphi$ integral in $I^{\sin \varphi}_{\bar{t} \bar{y}}$ is
    \begin{align}
    \begin{split}
     \int d\varphi \frac{a \sin \varphi}{|\vec{\bar{x}}_\perp|} f_{\bar{y}}
     &= \int d\varphi \frac{a \sin \varphi}{|\vec{\bar{y}}_\perp|} \left(-\frac{\bar{y}}{|\vec{\bar{x}}_\perp|}\right) \\
     &= -2 \int d\varphi \frac{\sin^2 \varphi}{1 + \sin^2 \varphi} = 2 \pi \cdot 2(\sqrt{2} - 2) \,.
    \end{split}
    \end{align}
Thus,
    \begin{align}
    \begin{split}
     \frac{\dot{P}_x}{\dot{P}_z}
        &= - \frac{f_\barbeta V}{\bar{c}} \frac{\frac{1}{\sqrt{2}} + \frac12 \cdot 2}{\sqrt{2} + 2} \\
        &= - \frac12 \frac{f_\barbeta V}{\bar{c}} \,,
    \end{split}
    \end{align}
where we used that $V_{\mathrm{eff}} = V$ for $\vec{V} \perp \hat{a}$.
Usually, a friction force points into the negative $\vec{V}$ direction, i.e. in the negative $z$ direction.
Our result shows that there is also a component into the positive $x$ direction in our case.
For standard SFDM, $\hat{a} = \hat{e}_x$ points into the opposite direction of the background MOND force.
Thus, this additional component of the friction force pushes the perturber away from the galactic center.

\subsection{When are perturbations small}
\label{sec:standasfdm:perturbationssmall}

Let us now check explicitly when perturbations are smaller than the background fields.
This determines when we can expand in them so that the perturbations' Lagrangian we derived above is valid.
In terms of our prototype calculation in Sec.~\ref{sec:general:validity}, this determines the factor $f_p$.
The standard SFDM Lagrangian has a coupling to matter that is linear in the phonon field $\theta$.
Thus, the coupling term can always be expanded, no matter how large the perturbations are.
The kinetic term is more complicated.
In the MOND limit, it is a function of
    \begin{align}
     \label{eq:standardsfdm:XYfull}
     X  - \barbeta Y = (1 - \barbeta) \dot{\delta}_\theta + \frac{(\vec{\nabla} \theta_0 + \vec{\nabla} \delta_\theta)^2}{2m} \,,
    \end{align}
where $\theta = \theta_0 + \delta_\theta$ is the full phonon field with background field $\theta_0$ and perturbation $\delta_\theta$.
The kinetic term also depends on $X$ without the $-\barbeta Y$ term.
This has the same form as Eq.~\eqref{eq:standardsfdm:XYfull} but with $1$ instead of $1 - \barbeta$.
We can expand in $\delta_\theta$ as long as
    $\vec{\nabla} \delta_\theta$ is smaller than $\vec{\nabla} \theta_0$
    and both $(1-\barbeta) \dot{\delta}_\theta$  and $\dot{\delta}_\theta$ are smaller than $(\vec{\nabla} \theta_0)^2/2m$.

We can estimate in the MOND limit
\begin{align}
 |\vec{\nabla}\theta_0| \approx a_\theta/\lambda \approx \sqrt{a_0 a_b^{\mathrm{gal}}}/\lambda  = \bar{\alpha}^2 \Lambda \sqrt{\frac{a_b^{\mathrm{gal}}}{a_0}} \,,
\end{align}
where $\lambda = \bar{\alpha} \Lambda/M_{\mathrm{Pl}}$.
The variables $\delta$ and $\delta_\theta$ are related by the rescaling $\delta^2 = 2 \Lambda |\vec{\nabla} \theta_0| \delta_\theta^2$,
    see Appendix~\ref{sec:standardsfdmpert}.
This gives
\begin{align}
    \delta_\theta = \frac{1}{\sqrt{2\Lambda |\vec{\nabla} \theta_0|}} \delta \approx \frac{1}{\sqrt{2} \bar{\alpha} \Lambda} \left(\frac{a_0}{a_b^{\mathrm{gal}}}\right)^{1/4} \delta \,.
\end{align}
Our goal is to check if $\delta_\theta$ is sufficiently small.
For simplicity, here we check explicitly at a few specific points:
First, at the position of the perturber.
Second, at a distance $r_{\mathrm{min}}$ purely in the direction of the perturber's trajectory,
    purely in the perpendicular direction,
    and on the shockwave front.
That is, we consider $(z - Vt)^2 + |\vec{x}_\perp|^2 = r_{\mathrm{min}}^2$ together with one of the following conditions
\begin{subequations}
\begin{align}
 |\vec{x}_\perp| &= 0 \,, \\
 z - Vt &= 0 \,, \\
 \label{eq:standardsfdm:shockwave}
 \bar{t} - \bar{z}/V_{\mathrm{eff}} &= |\vec{\bar{x}}_\perp| \frac{f_{\mathrm{crit}}}{c_{\mathrm{eff}}} \,.
\end{align}
\end{subequations}
Without our cutoffs,
    we could skip some of these cases.
For example, without cutoffs, the field at $|\vec{\bar{x}}_\perp| = r_{\mathrm{min}}$ vanishes for $\bar{z} - V_{\mathrm{eff}} \bar{t} = 0$ (corresponding to $|\vec{x}_\perp| = r_{\mathrm{min}}$ and $z- Vt = 0$ for $\vec{V} \parallel \hat{a}$).
However, with our cutoffs, everything is smeared out and the field can be nonzero there.

To check the smallness of perturbations, we will again express $\delta$ in terms of $\bar{\phi}$.
For $\vec{V} \parallel \hat{a}$ we have
\begin{align}
 \bar{t} - \frac{\bar{z}}{V_{\mathrm{eff}}} = t - \frac{z}{V} \,, \quad |\vec{\bar{x}}_\perp| = |\vec{x}_\perp| \,.
\end{align}
And for $\vec{V} \perp \hat{a}$,
\begin{align}
 \bar{t} - \frac{\bar{z}}{V_{\mathrm{eff}}} = t - \frac{z}{V} + \frac{f_\barbeta}{2\bar{c}} |\vec{x}_\perp| c_\varphi \,, \quad |\vec{\bar{x}}_\perp| = |\vec{x}_\perp| \left(1 - \frac12 c_\varphi^2\right) \,,
\end{align}
where $\varphi$ is the azimuthal angle in the $\vec{x}_\perp$ plane.
Including our cutoffs, the coordinate-dependence of the field $\bar{\phi}$ is given by the integral
\begin{align}
 \label{eq:standardsfdm:Ibar}
 \bar{I} =2 \Re \left[ \frac{1}{2\pi} \int_{\bar{\omega}_{\mathrm{min}}}^{\bar{\omega}_{\mathrm{max}}} d\bar{\omega} e^{i \bar{\omega} \left(\bar{t} - \frac{\bar{z}}{V_{\mathrm{eff}}}\right)} \int d^2\bar{k}_\perp e^{-i \vec{\bar{k}}_\perp \vec{\bar{x}}_\perp} \frac{\Theta(\bar{k}_{\perp,\mathrm{max}} - \bar{k}_\perp)}{\bar{\lambda}^2 - \bar{k}_\perp^2} \right] \,,
\end{align}
where $\Theta$ denotes the Heaviside theta function.
For large $|\vec{\bar{x}}_\perp|$, the $\bar{k}_{\perp,\mathrm{max}}$ cutoff can be neglected,
    as discussed in Appendix~\ref{sec:Edotcalculation:standardsfdm:cutoffs}.
For $|\vec{\bar{x}}_\perp| = 0$, it is required.
For $|\vec{\bar{x}}_\perp| \sim r_{\mathrm{min}}$, it may or may not be necessary depending on the precise numerics.
As a conservative estimate, we impose the cutoff for $|\vec{\bar{x}}_\perp| = 0$,
    but leave it out at $|\vec{\bar{x}}_\perp| \gtrsim r_{\mathrm{min}}$.
The $\bar{k}_\perp$ integral then gives the standard $-2\pi K_0(i\bar{\lambda} |\vec{\bar{x}}_\perp|)$ for $|\vec{\bar{x}}_\perp| \gtrsim r_{\mathrm{min}}$.
For $|\vec{\bar{x}}_\perp| = 0$, we can use Eq.~\eqref{eq:standardsfdm:kbarintegral-cutoff}.
That is, we have
\begin{multline}
 \bar{I} = 2 \Re \left[\int_{\bar{\omega}_{\mathrm{min}}}^{\bar{\omega}_{\mathrm{max}}} d\bar{\omega} e^{i \bar{\omega} \left(\bar{t} - \frac{\bar{z}}{V_{\mathrm{eff}}}\right)} \right. \\
 \times \left. \begin{cases}
        - K_0(i (\bar{\lambda}/\bar{\omega}) \bar{\omega} |\vec{\bar{x}}_\perp|) \,, &|\vec{\bar{x}}_\perp| \gtrsim r_{\mathrm{min}} \,, \\
        - \frac12 \ln\left(\frac{1 - (\bar{\omega}/\bar{\omega}_{\mathrm{max}})^2}{\bar{f}_{\mathrm{crit}}^2 (\bar{\omega}/\bar{\omega}_{\mathrm{max}})^2 }\right) + i \frac{\pi}{2} \,, &|\vec{\bar{x}}_\perp| = 0
    \end{cases}
 \right] \,.
\end{multline}
At the shockwave front, Eq.~\eqref{eq:standardsfdm:shockwave},
    both $|\vec{\bar{x}}_\perp| \sim r_{\mathrm{min}}$ and $|\vec{\bar{x}}_\perp| \ll r_{\mathrm{min}}$ are possible,
    depending on whether or not $V$ is close to $V_{\mathrm{crit}}$.
For $V$ close to $V_{\mathrm{crit}}$, we have $|\vec{\bar{x}}_\perp| \sim r_{\mathrm{min}}$.
For $V \gg V_{\mathrm{crit}}$, we have $|\vec{\bar{x}}_\perp| \ll r_{\mathrm{min}}$.
Thus, we should leave out the cutoff only for $V \sim V_{\mathrm{crit}}$.
Here, we leave out the cutoff in both cases as a conservative estimate.

Below, we are interested in derivatives of the integral $\bar{I}$.
After a rescaling
    \begin{align}
     \bar{\omega} = \bar{\omega}_{\mathrm{max}} \hat{\omega} = \bar{c} f_{\mathrm{max}} k_{\mathrm{max}} \hat{\omega} = \frac{\bar{c} f_{\mathrm{max}}}{r_{\mathrm{min}}} \hat{\omega} \,,
    \end{align}
we have
\begin{align}
\begin{split}
 \label{eq:standardsfdm:dIbar}
 \partial_{\bar{\mu}} \bar{I}
    &=\begin{multlined}[t][.7\textwidth]
        \hat{f}_{\bar{\mu}} \left(\frac{\bar{c} f_{\mathrm{max}}}{r_{\mathrm{min}}}\right)^2 2 \Re \left[
        \int_{\frac{\bar{\omega}_{\mathrm{min}}}{\bar{\omega}_{\mathrm{max}}}}^{1} d\hat{\omega} (i\hat{\omega}) e^{i \hat{\omega} f_{\mathrm{exp}} } \right. \\
        \left. \times \begin{cases}
        - K_{\hat{n}}(i \hat{\omega} f_K) \,, &|\vec{\bar{x}}_\perp| \gtrsim r_{\mathrm{min}} \,, \\
        - \frac12 \ln\left(\frac{1 - \hat{\omega}^2}{f_K'^2 \hat{\omega}^2}\right) + i \frac{\pi}{2} \,, &|\vec{\bar{x}}_\perp| = 0
        \end{cases}
        \right]
      \end{multlined}\\
 &\equiv \hat{f}_{\bar{\mu}} \left(\frac{\bar{c} f_{\mathrm{max}}}{r_{\mathrm{min}}}\right)^2 \hat{I}_{\bar{\mu}} \,,
\end{split}
\end{align}
where
\begin{subequations}
\begin{align}
 f_{\mathrm{exp}} &= f_{\mathrm{max}} \left(\frac{\bar{c} \bar{t}}{r_{\mathrm{min}}} - \frac{\bar{c}}{V_{\mathrm{eff}}}\frac{\bar{z}}{r_{\mathrm{min}}} \right) \,, \\
          f_K &= f_{\mathrm{max}} \frac{\bar{c} \bar{\lambda}}{\bar{\omega}} \frac{|\vec{\bar{x}}_\perp|}{r_{\mathrm{min}}} \,, \\
         f_K' &= \bar{f}_{\mathrm{crit}} = f_{\mathrm{max}} \frac{\bar{c} \bar{\lambda}}{\bar{\omega}}\,,
\end{align}
\end{subequations}
and
\begin{align}
 \hat{f}_{\bar{\mu}} = \begin{cases}
                        1\,, & \bar{\mu} = \bar{t} \,, \\
                        -\frac{1}{V_{\mathrm{eff}}} \,, & \bar{\mu} = \bar{z} \,, \\
                        - \frac{\bar{\mu}}{|\vec{\bar{x}}_\perp|} \frac{\bar{\lambda}}{\bar{\omega}} \,, & \bar{\mu} = \bar{x}, \bar{y} \,, |\vec{\bar{x}}_\perp| \gtrsim r_{\mathrm{min}} \,, \\
                        0 \,, & \bar{\mu} = \bar{x}, \bar{y} \,, |\vec{\bar{x}}_\perp| =0 \,,
                       \end{cases}
\end{align}
and
\begin{align}
 \hat{n} = \begin{cases}
            0 \,, &\bar{\mu} = \bar{t}, \bar{z} \,, \\
            1 \,, &\bar{\mu} = \bar{x}, \bar{y} \,.
           \end{cases}
\end{align}
Note that $f_{\mathrm{exp}}$, $f_K$, $f_K'$, and $\hat{f}_{\bar{\mu}}$ are independent of $\hat{\omega}$.

We can now calculate various ratios of $\partial_{\bar{\mu}} \bar{\phi}$ and $\vec{\nabla} \theta_0$.
For example,
\begin{align}
\begin{split}
 \bar{r}_{\bar{\mu}}
   &\equiv \frac{\frac{1}{\sqrt{2 \Lambda |\vec{\nabla} \theta_0|}} \partial_{\bar{\mu}} \bar{\phi}}{|\vec{\nabla} \theta_0|}
   = \frac{1}{\sqrt{2} a_0 M_{\mathrm{Pl}}} \left(\frac{a_0}{a_b^{\mathrm{gal}}}\right)^{3/4} \partial_{\bar{\mu}} \bar{\phi} \\
   &= \frac{1}{\sqrt{2} a_0 M_{\mathrm{Pl}}} \left(\frac{a_0}{a_b^{\mathrm{gal}}}\right)^{3/4} \frac{M_{\mathrm{eff}}}{(2\pi)^2} \frac{1}{|V_{\mathrm{eff}}|} \frac{g_m}{\sqrt{2} M_{\mathrm{Pl}}} \hat{f}_{\bar{\mu}} \left(\frac{\bar{c} f_{\mathrm{max}}}{r_{\mathrm{min}}}\right)^2 \hat{I}_{\bar{\mu}} \,.
\end{split}
\end{align}
We can simplify this using
\begin{align}
 g_m = \left(\frac{a_0}{a_b^{\mathrm{gal}}}\right)^{1/4} \,, \quad r_{\mathrm{min}}^{-2} = f_p^2 \frac{8 \pi M_{\mathrm{Pl}}^2 a_0}{M} \frac{a_b^{\mathrm{gal}}}{a_0} \,.
\end{align}
We find
\begin{align}
  \bar{r}_{\bar{\mu}}
  = \frac{f_p^2 f_{\mathrm{max}}^2}{\pi} \frac{\hat{f}_{\bar{\mu}} \bar{c}^2}{|V_{\mathrm{eff}}|} \frac{M_{\mathrm{eff}}}{M} \hat{I}_{\bar{\mu}} \,.
\end{align}
We also need
\begin{align}
 \bar{r}_{\bar{\mu}}'
   &\equiv \frac{\frac{1}{\sqrt{2 \Lambda |\vec{\nabla} \theta_0|}} \partial_{\bar{\mu}} \bar{\phi}}{|\vec{\nabla} \theta_0|^2 / (2m)}
   = \frac{\bar{r}_{\bar{\mu}}}{\bar{c}} \frac{2 m \bar{c}}{|\vec{\nabla} \theta_0|} \,.
\end{align}
We can simplify this using
\begin{align}
 \bar{c} = 3 \bar{f}_\barbeta \sqrt{\frac{a_b^{\mathrm{gal}}}{a_0}} \frac{\bar{\alpha}^2 \Lambda}{m} \,.
\end{align}
We find
\begin{align}
 \bar{r}_{\bar{\mu}}' = 6 \bar{f}_\barbeta \frac{\bar{r}_{\bar{\mu}}}{\bar{c}} \,.
\end{align}

As discussed above, we are interested in the following ratios that tell us whether or not we can expand in the perturbation $\delta_\theta$,
\begin{align}
 r_t \equiv \frac{\max(|1-\barbeta|, 1) \dot{\delta}_\theta}{(\vec{\nabla}\theta_0)^2/(2m)} \,, \quad
 r_i \equiv \frac{\partial_i \delta_\theta}{|\vec{\nabla} \theta_0|} \,,
\end{align}
where $i = x,y,z$.
We can express these ratios in terms of the ratios $\bar{r}_{\bar{\mu}}$ and $\bar{r}'_{\bar{\mu}}$ by expressing $\delta_\theta$ through $\bar{\phi}$.
We find
\begin{align}
 r_t = \max(|1-\barbeta|, 1) \bar{r}_{\bar{t}}' \,, \quad
 r_y = \bar{r}_{\bar{y}} \,.
\end{align}
Further, for $\vec{V} \parallel \hat{a}$
\begin{align}
r_x = \bar{r}_{\bar{x}}\,, \quad
r_z = \frac{1}{\sqrt{2}} \bar{r}_{\bar{z}} + a_\parallel \frac{f_\barbeta}{2 \bar{c}} \bar{r}_{\bar{t}} \,.
\end{align}
And for $\vec{V} \perp \hat{a}$
\begin{align}
r_x = \frac{1}{\sqrt{2}} \bar{r}_{\bar{x}} + \frac{f_\barbeta}{2 \bar{c}} \bar{r}_{\bar{t}} \,, \quad
r_z = \bar{r}_{\bar{z}}\,.
\end{align}
These should all be smaller than $1$ in magnitude.
Otherwise, we must adjust our choice of $r_{\mathrm{min}}$, i.e. the factor $f_p$.

It can easily be seen that $\bar{r}_{\vec{\bar{x}}}$, $\bar{r}'_{\bar{t}}$, and $(f_{\barbeta}/2\bar{c}) \bar{r}_{\bar{t}}$ cannot be much larger than $1$
    for supersonic velocities $V_{\mathrm{eff}} > c_{\mathrm{eff}}$,
    if the integral $\hat{I}_{\bar{\mu}}$ is not much larger than $1$.
Indeed, this integral cannot be much larger than $1$ (except for $V \to V_{\mathrm{crit}}$ which we discuss below).
In particular, at a distance $r_{\mathrm{min}}$ from the perturber, the background field and the perturbation are of the same order of magnitude.
This roughly confirms our choice of $r_{\mathrm{min}}$ from Sec.~\ref{sec:general:validity}.
As discussed in Appendix~\ref{sec:Edotcalculation:standardsfdm:cutoffs},
    one caveat to this is that our integral diverges logarithmically for $V \to V_{\mathrm{crit}}$ at $|\vec{\bar{x}}_\perp| = 0$,
    corresponding to the fact that $f_K' \to 0$ and $f_K \to 0$ for $V \to V_{\mathrm{crit}}$.
Thus, we restrict ourselves to velocities at least $1\%$ above the critical velocity,
    as also discussed above.

\begin{figure}
 \centering
 \includegraphics[width=.49\textwidth]{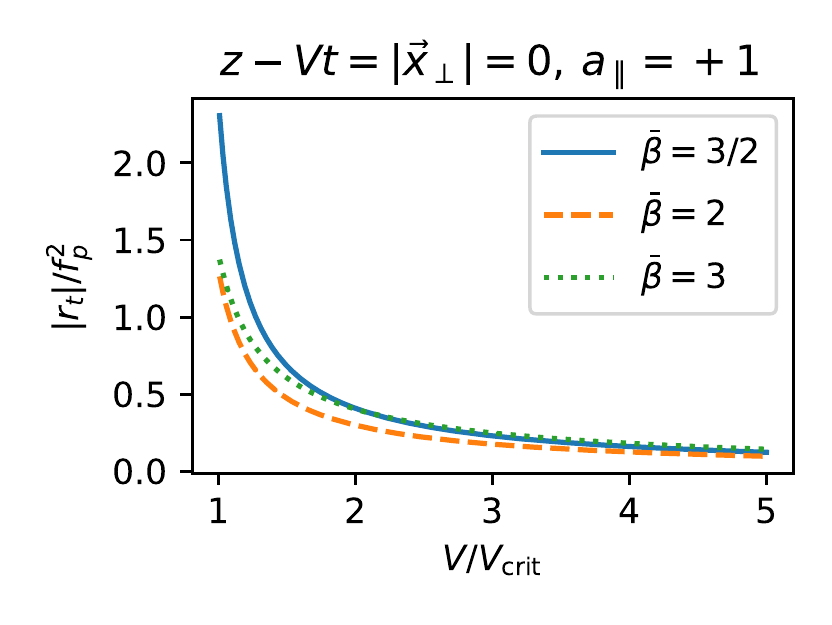}
 \includegraphics[width=.49\textwidth]{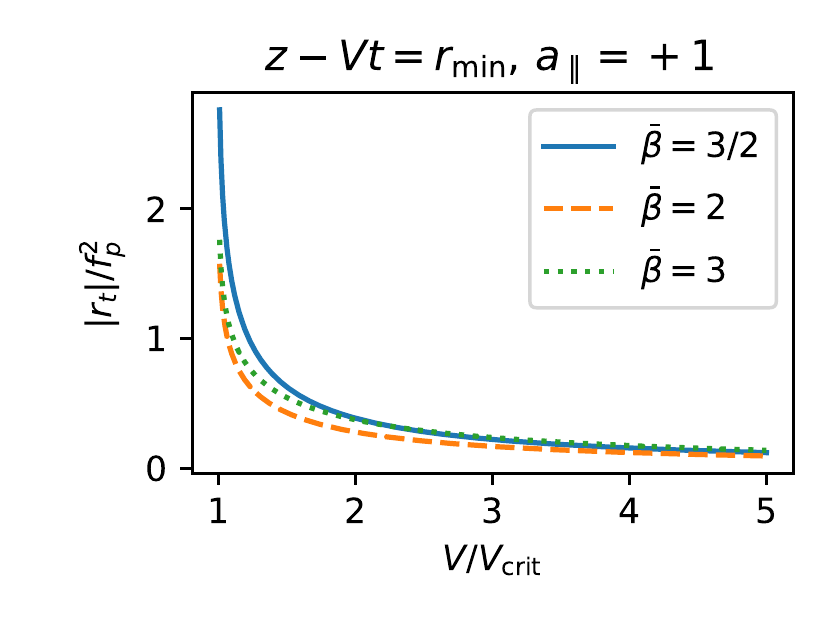}
 \caption{
    The absolute value of the ratio
        $r_t = \frac{\dot \delta_\theta \max(|1-\barbeta|,1)}{(\vec{\nabla} \theta_0)^2 / 2 m}$ as a function of the perturber's velocity $V$ relative to the critical velocity $V_{\mathrm{crit}}$.
    The ratio $r_t$ determines whether or not the perturbation's time derivative is smaller than the relevant combination of background fields.
     Left: For $\vec{V} \parallel \hat{a}$ with $a_\parallel = +1$ at $|\vec{x}_\perp| = z - V t = 0$.
     Right: For $\vec{V} \parallel \hat{a}$ at $|\vec{x}_\perp| = 0$ and $z - Vt = r_{\mathrm{min}}$.
     These are the two worst cases, i.e. the cases with the largest $|r_t|/f_p^2$.
 }
 \label{fig:perturbationsSmall}
\end{figure}

We can make this more precise by numerically evaluating these ratios in Mathematica \cite{Mathematica12}.
We find that they are smaller than $1$ for many cases.
In these cases, we can safely keep $r_{\mathrm{min}}$ as it is, i.e. set $f_p = 1$.
In some cases, $r_t/f_p^2$ is larger than $1$.
In particular, this can happen for $\vec{V} \parallel \hat{a}$ with $a_\parallel = +1$
    and at velocities that are only slightly larger than the critical velocity.
The ratio is larger for smaller values of $\barbeta$.
We show the two worst cases in Fig.~\ref{fig:perturbationsSmall}.
These occur for $\barbeta = 3/2$ and $V = 1.01 V_{\mathrm{crit}}$.
We have $|r_t| \approx 2.3 f_p^2$ for $|\vec{x}_\perp| = z - Vt = 0$.
Similarly, we have $|r_t| \approx 2.8 f_p^2$ for $z - Vt = r_{\mathrm{min}}$ and $|\vec{x}_\perp| = 0$.
In this case, we may need to make $f_p$ smaller by a factor of about $1/\sqrt{2.8}$.

\section{Two-field SFDM perturbations}
\label{sec:twofieldsfdmpert}

We consider perturbations on top of a background galaxy in equilibrium in two-field SFDM.
Ref.~\cite{Mistele2020} derived the second-order Lagrangian for such perturbations.
With the definition $\lambda \equiv \baralpha \Lambda/M_{\mathrm{Pl}}$ and including the $-\lambda \delta_+ \delta_b$ coupling term,

\begin{align}
\label{eq:Limppertfull}
\begin{split}
 \mathcal{L}
 = &+\frac12 \left(\dot{\delta}_\rho^2 - (\vec{\nabla} \delta_\rho)^2 \right) - (\mu_0^2 g^{00} -m^2) \delta_\rho^2 \\
   &+\left(f_0' \dot{\delta}_+^2 - (f_0' - 2 f_0'' \gamma^2 |\vec{\nabla} \theta_+|^2)  (\vec{\nabla} \delta_+)^2 \right) \\
   &+\left((f_0' + 2 f_0'' \mu_0^2 + A) \dot{\delta}_-^2 - (f_0' + A) (\vec{\nabla} \delta_-)^2 \right) \\
   &+\left(- 4 f_0'' \mu_0 |\vec{\nabla} \theta^0_+| \gamma\right) |\vec{\nabla} \delta_+| \dot{\delta}_- \\
   &+2\mu_0 \rho_0 \delta_\rho \dot{\delta}_- \\
   &-\lambda \delta_+ \delta_b \,.
\end{split}
\end{align}
Here, $\delta_\rho$ is the perturbation on top of the background solution $\rho_0$ of the field $\rho_-$ and $\delta_\pm$ are the perturbations on top of the background solutions $\theta_\pm^0$ of the fields $\theta_\pm$.
Further, $\mu_0$ is the background chemical potential and $A = \rho_0^2/2$.

\subsection{Equations of motion}
\label{sec:twofieldsfdmpert:eom}

The $\delta_\rho$ equation of motion in Fourier space gives
\begin{align}
 \delta_\rho = - \frac{2 i \omega \mu_0 \rho_0}{\omega^2 - \vec{k}^2 -2 (\mu^2 g^{00} - m^2)} \delta_- \,.
\end{align}
Plugging this into the $\delta_-$ equation, we find
\begin{align}
\begin{split}
 0 = & (\delta_- e^{ikx} + \delta_-^* e^{-ikx}) \begin{aligned}[t] \left[\left( 2(f_0' + 2 f_0'' \mu^2 + A) \omega^2 - 2(f_0' + A) \vec{k}^2 \right) \right. \\
                                                                   \left. - \frac{(2 \mu_0 \rho_0 \omega)^2}{\omega^2 - \vec{k}^2 -2 (\mu_0^2 g^{00} - m^2)}\right] \end{aligned} \\
 &+ \left(4 f_0'' \mu_0 |\vec{\nabla} \theta^0_+| \gamma\right) |\vec{k}| \omega (\delta_+ e^{ikx} + \delta_+^* e^{-ikx}) \,.
\end{split}
\end{align}
The $\delta_+$ equation reads
\begin{align}
\begin{split}
 0 = &\left( 2 f_0' \omega^2 - 2 (f_0' - 2 f_0'' \gamma^2 |\vec{\nabla} \theta^0_+|^2) \vec{k}^2 \right) (\delta_+ e^{ikx} + \delta_+^* e^{-ikx})  \\
 &+ \left(4 f_0'' \mu_0 |\vec{\nabla} \theta^0_+| \gamma\right) |\vec{k}| \omega (\delta_- e^{ikx} + \delta_-^* e^{-ikx}) \\
 &+ \lambda (\delta_b e^{ikx} + cc) \,.
\end{split}
\end{align}
Solving the $\delta_+$ equation for $ \delta_+ $ gives
\begin{align}
 \label{eq:deltapmmix}
 \delta_+ = -\frac{\delta_- 4 f_0'' \mu_0 |\vec{\nabla} \theta^0_+| \gamma |\vec{k}| \omega + \lambda \delta_b}{2 f_0' \omega^2 - 2(f_0' - 2f_0'' \gamma^2 |\vec{\nabla} \theta^0_+|^2)\vec{k}^2} \,.
\end{align}

Plugging this result into the $\delta_-$ equation and dividing by $ A$ yields
\begin{align}
\begin{split}
 0 = &\begin{aligned}[t]
        &\left[ \left( 2\left(\frac{f_0' + 2 f_0'' \mu_0^2}{A} + 1\right) \omega^2 - 2\left(\frac{f_0'}{A} + 1\right) \vec{k}^2 \right) \right . \\
        &\quad - \frac{8 \mu_0^2 \omega^2}{\omega^2 - \vec{k}^2 -2 (\mu_0^2 g^{00} - m^2)} \\
        &\quad \left. \, - \frac{2 f_0'' \mu_0^2}{A} \frac{4 \cdot 2 f_0'' |\vec{\nabla} \theta^0_+|^2 \gamma^2 |\vec{k}|^2 \omega^2}{2 f_0' \omega^2 - 2(f_0' - 2 f_0'' \gamma^2 |\vec{\nabla} \theta^0_+|^2) \vec{k}^2} \right] (\delta_- e^{ikx} + cc)
      \end{aligned} \\
     & - \frac{1}{A} \frac{ 4 \mu_0 f_0'' |\vec{\nabla} \theta_+^0| \gamma |\vec{k}| \omega}{2 f_0' \omega^2 - 2(f_0' - 2f_0'' \gamma^2 |\vec{\nabla} \theta^0_+|^2)\vec{k}^2} \lambda (\delta_b e^{ikx} + cc) \,.
\end{split}
\end{align}
Here,
\begin{align}
 A = \frac{\rho_0^2}{2} = \frac{m^2}{\lambda_4} \frac{\hat{\mu}_0}{m} \,,
\end{align}
where $2 m \hat{\mu}_0 \approx \mu_0^2 g^{00} - m^2$.
Numerically, $ 2 f_0'' |\vec{\nabla} \theta^0_+|^2 \approx -f_0' $.
So we can write
\begin{align}
\begin{split}
 \label{eq:dispersion}
 0 = &\begin{aligned}[t]
        &\left[ \left( 2\left(\frac{f_0' + 2 f_0'' \mu_0^2}{A} + 1\right) \omega^2 - 2\left(\frac{f_0'}{A} + 1\right) \vec{k}^2 \right) \right. \\
        &\quad - \frac{8 \mu_0^2 \omega^2}{\omega^2 - \vec{k}^2 -2 (\mu_0^2 g^{00} - m^2)} \\
        &\quad \left.-\left(- \frac{2 f_0'' \mu_0^2}{A}\right) \frac{2 \cdot \gamma^2 |\vec{k}|^2 \omega^2}{ \omega^2 - (1 + \gamma^2) \vec{k}^2} \right] (\delta_- e^{ikx} + cc) \\
      \end{aligned} \\
     & + \frac{\mu_0}{A |\vec{\nabla} \theta_+^0|} \frac{\omega}{|\vec{k}|} \frac{ \gamma}{\omega^2/\vec{k}^2 - (1 +\gamma^2 )} \lambda (\delta_b e^{ikx} + cc) \,.
\end{split}
\end{align}
As discussed in Ref.~\cite{Mistele2020}, at low energies with $\omega = c_s |\vec{k}|$ and for a non-relativistic background, this equation takes the form
\begin{align}
\begin{split}
 0 = &\left( - 2 \vec{k}^2 + \frac{2 m \omega^2}{\hat{\mu}_0} \right) (\delta_- e^{ikx} + cc) \\
     & + \frac{\mu_0}{A |\vec{\nabla} \theta_+^0|} \frac{\omega}{|\vec{k}|} \frac{ \gamma}{\omega^2/\vec{k}^2 - (1 +\gamma^2 )} \lambda (\delta_b e^{ikx} + cc) \,.
\end{split}
\end{align}
Note that this not only assumes $k \ll m$ but also $k \ll m \sqrt{\hat{\mu}_0/m}$ \cite{Berezhiani2020}.
This introduces an additional cutoff on $k$.
However, numerically this is not relevant compared to our cutoff $k_{\mathrm{max}} \sim 10^{-22}\,\mathrm{eV}$ that we impose anyway for independent reasons.

We will now write this in terms of
\begin{align}
 \bar{\delta}_- \equiv \sqrt{2 A} \delta_- = \rho_0 \delta_- \,.
\end{align}
This is a useful normalization as we will see below.
The dispersion relation is
\begin{align}
 \omega = c_s |\vec{k}| = \sqrt{\frac{\hat{\mu}_0}{m}} |\vec{k}| \,.
\end{align}
Since $\hat{\mu}_0 \ll m$, we will have $\omega \ll k$ for on-shell radiation.
We can then write
\begin{align}
\begin{split}
 0 = &\left( - \vec{k}^2 + \frac{m \omega^2}{\hat{\mu}_0} \right) (\bar{\delta}_- e^{ikx} + cc) \\
     & - \frac{m}{\sqrt{2 A} |\vec{\nabla} \theta_+^0|} \frac{\omega}{|\vec{k}|} \frac{\gamma}{1+ \gamma^2} \lambda (\delta_b e^{ikx} + cc) \,,
\end{split}
\end{align}
which gives
\begin{align}
\begin{split}
 \label{eq:twofield_eom_perturber}
 0 = &\left( - \vec{k}^2 + \frac{m \omega^2}{\hat{\mu}} \right) (\bar{\delta}_- e^{ikx} + cc) \\
     & - \frac{1}{\sqrt{2} M_{\mathrm{Pl}}} \frac{\sqrt{\lambda_4} \gamma}{\baralpha} \frac{a_0}{|\vec{a}_{\theta_+^0}|} \frac{\gamma}{1+\gamma^2} (\delta_b e^{ikx} + cc) \,.
\end{split}
\end{align}

\subsection{Effective Lagrangian}
\label{sec:twofieldsfdmpert:effective}

Our calculation so far tells us how the non-relativistic mode of two-field SFDM reacts to a given perturber.
But not how much energy such a mode carries.
For this, we calculate the associated effective Lagrangian.
I.e.,
    we simply put our results for $\delta_\rho$ and $\delta_+$ back into the original Lagrangian $\mathcal{L}$ from Eq.~\eqref{eq:Limppertfull}.
In our case,
    we have $\omega \ll k$ and $k^2 \ll m^2 c_s^2$.
This gives
    \begin{subequations}
    \begin{align}
    \delta_\rho &\approx \frac12 (i \omega) \frac{\sqrt{2A}}{\hat{\mu}} \delta_- \,, \\
    \begin{split}
    \delta_+
        &\approx \frac{2 f_0'' \mu_0 \vec{\nabla} \theta_+^0 \vec{k} \omega}{f_0' (1+\gamma^2) k^2} \delta_-
            + \frac{\lambda \delta_b}{2 f_0' k^2 (1 + \gamma^2)}\\
        &\approx - c_s \frac{\bar{\alpha} m \Lambda}{M_{\mathrm{Pl}} a_{\theta_0}} \frac{\gamma}{1+\gamma^2} \delta_-
            + \frac{\lambda \delta_b}{2 f_0' k^2 (1 + \gamma^2)}
            \equiv (\delta_+)_- + (\delta_+)_b \,.
     \end{split}
    \end{align}
    \end{subequations}
Numerically,
    \begin{align}
     (\delta_+)_- = - \frac{c_s}{10^{7/4}} \frac{\gamma}{1+\gamma^2} \frac{a_0}{a_{\theta_0}} \frac{1}{\sqrt{a_0 r_0}} \left(\frac{\bar{a}}{a_0}\right)^{1/8} \delta_- \,,
    \end{align}
    where the prefactor of $\delta_-$ is typically not much larger than $1$.

The non-negligible terms in the Lagrangian are
    \begin{align}
    \begin{split}
     \mathcal{L}
        \approx &-(\mu_0^2 g^{00} - m^2) \delta_\rho^2 - (f_0' - 2 f_0'' \gamma^2 |\vec{\nabla} \theta_+|^2) (\vec{\nabla} \delta_+)^2 \\
                  & - (f_0' + A) (\vec{\nabla} \delta_-)^2 +\left(- 4 f_0'' \mu_0 |\vec{\nabla} \theta^0_+| \gamma\right) |\vec{\nabla} \delta_+| \dot{\delta}_- \\
                &+2\mu_0 \rho_0 \delta_\rho \dot{\delta}_-
                - \lambda \delta_+ \delta_b \,.
    \end{split}
    \end{align}
A coupling of $\delta_-$ to the matter density $\delta_b$ can in principle come from three terms.
From the $\delta_+ \delta_b$ term, from the $(\vec{\nabla} \delta_+)^2$ term, and from the $\vec{\nabla} \delta_+ \dot{\delta}_-$ term.
However, the contributions from the latter two terms cancel,
\begin{align}
\begin{split}
  &- f_0' (1 + \gamma^2) 2 (-i \vec{k})^2 (\delta_+)_b (\delta_+)_-
 -4 f_0'' \mu_0 \vec{\nabla} \theta_+^0 (-i \vec{k}) (i \omega) (\delta_+)_b \delta_- \\
 =& 2 (\delta_+)_b \left(f_0' (1+\gamma^2) k^2 (\delta_+)_- - 2 f_0'' \mu_0 \vec{k} \omega \vec{\nabla} \theta_+^0 \delta_-  \right) =0 \,.
\end{split}
\end{align}
Thus, the coupling to matter comes only from the $\delta_+ \delta_b$ term.
We further have
    \begin{subequations}
    \begin{align}
        2 \mu_0 \rho_0 \delta_\rho \dot{\delta}_- &\approx \rho_0^2 \frac{1}{c_s^2} \dot{\delta}_-^2 \,, \\
        - (\mu_0^2 g^{00} - m^2) \delta_\rho^2 &\approx - \frac12 \rho_0^2 \frac{1}{c_s^2} \dot{\delta}_-^2 \,.
    \end{align}
    \end{subequations}
Both terms together give the $\dot{\delta}_-^2$ term.
The mixing term $\vec{\nabla} \delta_+ \dot{\delta}_-$ in principle also contributes to the $\dot{\delta}_-^2$ term.
However, its size relative to the other contributions is of order
\begin{align}
 \frac{m^2}{|\vec{\nabla} \theta_+^0|^2} \frac{f_0'}{A} c_s^2
 =  \frac{1}{10^7} \left(\frac{a_0}{a_{\theta_0}}\right) \left(\frac{\bar{a}}{a_0}\right)^{1/2}\,,
\end{align}
which is typically much smaller than $1$.

This finally gives, using $A \gg f_0'$,
    \begin{align}
    \begin{split}
     \mathcal{L}
        &= \frac12 \rho_0^2 \frac{1}{c_s^2} \dot{\delta}_-^2 - \frac12 \rho_0^2 (\vec{\nabla} \delta_-)^2 - \lambda \delta_+ \delta_b \\
        &= \frac12 \frac{1}{c_s^2} \dot{\bar{\delta}}_-^2 - \frac12 (\vec{\nabla} \bar{\delta}_-)^2 - \frac{\lambda}{\rho_0} \frac{\delta_+}{\delta_-} \, \bar{\delta}_- \delta_b \\
        &= \frac12 \frac{1}{c_s^2} \dot{\bar{\delta}}_-^2 - \frac12 (\vec{\nabla} \bar{\delta}_-)^2 + \frac{1}{\sqrt{2} M_{\mathrm{Pl}}} \frac{\sqrt{\lambda_4}}{\bar{\alpha}} \frac{a_0}{a_{\theta_0}} \frac{\gamma}{1+\gamma^2} \, \bar{\delta}_- \delta_b \,.
    \end{split}
    \end{align}
This is our prototype Lagrangian without the $\hat{a}$ term and with
    \begin{align}
    \bar{c} = \sqrt{\frac{\hat{\mu}}{m}} \,, \quad g_m = -\frac{\sqrt{\lambda_4}}{\bar{\alpha}} \frac{a_0}{a_{\theta_0}} \frac{\gamma}{1+\gamma^2} \,.
    \end{align}
To calculate the energy loss through Cherenkov radiation,
    we can reuse our standard calculation of $\dot{E}$ from Appendix~\ref{sec:Edotcalculation}
    after adjustments for the missing $\hat{a}$ term and the factor $\gamma/(1+\gamma^2)$ in the coupling $\coup$.
This is discussed in Sec.~\ref{sec:twofieldsfdm} and Appendix~\ref{sec:Edotcalculation:twofieldsfdm}.

\section{SZ model perturbations}
\label{sec:relmondpert}

In this Appendix, we consider perturbations on top of a Minkowski background in the SZ model.
Here, we use the metric signature $(-, +, +, +)$ to facilitate easier comparison to Ref.~\cite{Skordis2020}.
Following Ref.~\cite{Skordis2020} we multiply the Lagrangian of the SZ model by $16 \pi \tilde{G} = 2/\tilde{M}_{\mathrm{Pl}}^2$.
Then, the Lagrangian for the perturbations is \cite{Skordis2020}
\begin{align}
\begin{split}
S =&  \int d^4x  \bigg\{ (\mathrm{standard\,perturbations\,from}\,R) +  \frac{1}{\tilde{M}_{\mathrm{Pl}}^2} T_{\alpha\beta} h^{\alpha\beta}
\\
&
+ \KB |\dot{\vec{A}} - \frac{1}{2} \vec{\nabla} h_{00}|^2
- 2\KB \vec{\nabla}_{[i} A_{j]} \vec{\nabla}^{[i} A^{j]}
 \\
 &
 + 2 \left(2 - \KB\right)  \left(  \dot{\vec{A}}  -  \frac{1}{2}  \vec{\nabla} h_{00} \right) \cdot \vec{\nabla} \varphi
 + 2 (2 - \KB)Q_0 \vec{A}_i \left( - \frac12 \partial_i h_{00}\right)
 \\
 &\begin{multlined}[t][.8\textwidth]
    -(2-\KB) (1 + \lambda_s) \left( \vec{A}^2 Q_0^2  + (\vec{\nabla}\varphi)^2
      + 2 Q_0 A^i(\partial_i \varphi) \right)
  \end{multlined}
 \\
& +2 \Ktwo \left(\dot{\varphi} + \frac12 h_{00} Q_0 \right)^2
 \bigg\} \,,
\end{split}
\end{align}
where ``$(\mathrm{standard\,perturbations\,from}\,R)$'' denotes the metric perturbations from the Ricci scalar as in standard General Relativity.

The dispersion relation of the scalar mode we are interested in was already calculated in Refs.~\cite{Skordis2020, Skordis2021}.
The new result of this Appendix will be how this mode couples to matter in the regime we are interested in.

\subsection{Equations of motion}

The $h_{\alpha \beta}$ equation is
\begin{align}
\begin{split}
 (\mathrm{standard})_{\alpha \beta} &=
    \begin{multlined}[t][.7\textwidth]
         \left(
        \KB \vec{\nabla} (\dot{\vec{A}} - \frac12 \vec{\nabla} h_{00})
        + (2-\KB) (\vec{\nabla}^2 \varphi + Q_0 \vec{\nabla} \vec{A}) \right. \\
        \left. + 2 \Ktwo Q_0\left(\dot{\varphi}+ \frac12 h_{00} Q_0 \right)
        \right)\delta_\alpha^0 \delta_\beta^0
        + \frac{1}{\tilde{M}_{\mathrm{Pl}}^2} T_{\alpha \beta}
    \end{multlined}\\
 &\equiv \frac{1}{M_{\mathrm{Pl}}^2} \bar{T}_{\alpha \beta} \,,
\end{split}
\end{align}
where ``$(\mathrm{standard})_{\alpha \beta}$'' denotes terms from the metric perturbations from the Ricci scalar as in standard General Relativity.
In the harmonic gauge \cite{Weinberg1972},
\begin{align}
 \frac12 \Box h_{\alpha \beta} &= - \frac{1}{\tilde{M}_{\mathrm{Pl}}^2} \left(\bar{T}_{\alpha \beta} - \frac12 \eta_{\alpha \beta} \bar{T}^\rho_\rho \right)\,.
\end{align}
For the $00$ component, this is
\begin{multline}
 (\vec{\nabla}^2 - \partial_t^2)h_{00}=
    - \frac{2}{\tilde{M}_{\mathrm{Pl}}^2} S_{00}
    - \KB \, \vec{\nabla} (\dot{\vec{A}} - \frac12 \vec{\nabla} h_{00})
    - (2-\KB) (\vec{\nabla}^2 \varphi + Q_0 \vec{\nabla} \vec{A}) \\
    - 2 \Ktwo Q_0 \left(\dot{\varphi} + \frac12 h_{00} Q_0 \right)\,,
\end{multline}
where $S_{\alpha\beta} = T_{\alpha\beta} - \frac12 \eta_{\alpha \beta} T_\rho^\rho$.
We can write this as
\begin{multline}
 \left((1 - \KB/2)\vec{\nabla}^2 - \partial_t^2 + \Ktwo Q_0^2\right)h_{00} \\
 =
    - \frac{2}{\tilde{M}_{\mathrm{Pl}}^2} S_{00}
    - \KB \, \vec{\nabla} \dot{\vec{A}}
    - (2-\KB) (\vec{\nabla}^2 \varphi + Q_0 \vec{\nabla} \vec{A})
    - 2 \Ktwo Q_0 \dot{\varphi} \,.
\end{multline}
In Fourier space this gives
\begin{align}
 h_{00} = \frac{
     -\frac{2}{\tilde{M}_{\mathrm{Pl}}^2} S_{00}
     - \KB \omega \vec{k} \vec{A} + (2-\KB) (\vec{k}^2 \varphi + i Q_0 \vec{k} \vec{A})
     - 2 \Ktwo Q_0 i \omega \varphi
    }{\omega^2 - \vec{k}^2(1 - \KB/2) + \Ktwo Q_0^2} \,.
\end{align}

The $\vec{A}$ equation reads
\begin{align}
\begin{split}
 0 = &-2 \KB \partial_t \left(\partial_t \vec{A} - \frac12 \vec{\nabla} h_{00}\right) + 2 \KB \left(\vec{\nabla}^2 \vec{A} - \vec{\nabla} (\vec{\nabla} \cdot \vec{A})\right) \\
     & - 2 (2 - \KB) \partial_t \vec{\nabla} \varphi - 2 (2-\KB) (1+\lambda_s) \left(Q_0^2 \vec{A} + Q_0 \vec{\nabla} \varphi \right) \\
     & + 2(2-\KB) Q_0 \left( - \frac12 \vec{\nabla} h_{00}\right) \,.
\end{split}
\end{align}
In Fourier space, we can write this as
\begin{align}
\begin{split}
&\left(2 \KB \omega^2 - 2 (2 - \KB)(1+\lambda_s)Q_0^2 \right) \vec{A} - 2 \KB \left( \vec{k}^2 \vec{A} - \vec{k} (\vec{k} \cdot \vec{A}) \right)  \\
&= - \omega \vec{k} \left(\KB h_{00} - 2(2-\KB) \varphi\right) - 2(2-\KB) iQ_0 \vec{k} \left( (1+\lambda_s) \varphi + \frac12 h_{00} \right) \,.
\end{split}
\end{align}
The only direction besides $\vec{A}$ in this equation is that of $\vec{k}$, so we can assume $\vec{A} \parallel \vec{k}$.
Then,
\begin{align}
\begin{split}
 \vec{A} %
         &= -\frac{\vec{k} (2 - \KB) }{2 \KB} \frac{h_{00} \left(\omega \frac{\KB}{2 - \KB}  +  i Q_0\right) + 2\left((1+\lambda_s) i Q_0 - \omega \right) \varphi }{\omega^2 - (2-\KB)(1+\lambda_s) Q_0^2/\KB} \\
         &= -\frac{\vec{k} (2 - \KB) }{2 \KB \omega} \frac{h_{00} \left(\frac{\KB}{2 - \KB}  +  i Q_0/\omega\right) - 2\left(1 - (1+\lambda_s) i Q_0/\omega \right) \varphi}{1 - (2-\KB)(1+\lambda_s) Q_0^2/(\KB\omega^2)} \,.
\end{split}
\end{align}

The $\varphi$ equation of motion is
\begin{align}
\begin{split}
 0 = &-4 \Ktwo \ddot{\varphi} + 2(2-\KB)(1 +\lambda_s) \left(\vec{\nabla}^2 \varphi + Q_0 \vec{\nabla} \vec{A} \right) \\
     & + (2 - \KB) \vec{\nabla}^2 h_{00} - 2(2-\KB) \partial_t \vec{\nabla} \vec{A} -4 \Ktwo \left( \frac12 \dot{h}_{00} Q_0 \right) \,.
    \end{split}
\end{align}
In Fourier space, this is
\begin{align}
\begin{split}
 0 = &4 \Ktwo \omega^2 \varphi
    - 2(2-\KB)(1 +\lambda_s) \vec{k}^2 \left(\varphi + \frac{i Q_0 \vec{k}}{\vec{k}^2} \vec{A}  \right) \\
    &- (2 - \KB) \vec{k}^2 h_{00} \left(1 + \frac{1}{2 - \KB} \frac{i \omega}{Q_0} \frac{2 \Ktwo Q_0^2}{\vec{k}^2} \right)
    - 2(2-\KB) \omega \vec{k} \vec{A} \,.
\end{split}
\end{align}

We now write the $h_{00}$ and $\vec{A}$ equations in a more compact way,
\begin{align}
 (2-\KB) \vec{k}^2 h_{00} = 2 f_{00} \left(
    \frac{2}{\tilde{M}_{\mathrm{Pl}}^2} S_{00}
    + \KB f_A' \omega \vec{k} \vec{A}
    - (2-\KB) f_\varphi \vec{k}^2 \varphi
    \right)\,,
\end{align}
and
\begin{align}
 \vec{A} &= - f_A \frac{\vec{k}}{\omega} \left(\frac12 h_{00} f_{Q1} - \frac{2-\KB}{\KB} f_{Q2} \varphi\right) \,,
\end{align}
with
\begin{align}
\begin{split}
      f_\varphi &\equiv 1 - \frac{1}{2 - \KB} \frac{2 \Ktwo Q_0^2}{k^2} \frac{i \omega}{Q_0} \,, \\
    f_{00}^{-1} &\equiv 1 - \omega^2/(\vec{k}^2(1 - \KB/2)) - \Ktwo Q_0^2/(k^2(1 - \KB/2)) \,, \\
       f_A^{-1} &\equiv 1 - (2-\KB)(1+\lambda_s)Q_0^2/(\omega^2\KB) \,, \\
         f_{Q1} &\equiv 1 + (i Q_0/\omega) (2-\KB)/\KB \,, \\
         f_{Q2} &\equiv 1 - (1+\lambda_s) (iQ_0/\omega) \,, \\
         f_A'   &\equiv 1 - (2 - \KB) \frac{i Q_0}{\KB \omega}  \,,
\end{split}
\end{align}
Plugging the expression for $\vec{A}$ into the $\varphi$ equation of motion gives
\begin{align}
\label{eq:relmond:phieomforX}
 0 = 4 \Ktwo \omega^2 \varphi
    - 2(2-\KB)(1 +\lambda_s) \vec{k}^2 \varphi F_k
    - (2 - \KB) \vec{k}^2 h_{00} F_h
\end{align}
where
\begin{subequations}
\begin{align}
  F_k &\equiv 1 + \frac{2 - \KB}{\KB} f_A f_{Q2} \left(\frac{i Q_0}{\omega} + \frac{1}{1 + \lambda_s} \right)\,, \\
  F_h &\equiv (1-f_A f_{Q1}) + \frac{1}{2 - \KB} \frac{i \omega}{Q_0} \frac{2 \Ktwo Q_0^2}{\vec{k}^2} - \frac{iQ_0}{\omega} (1+\lambda_s) f_A f_{Q1} \,.
\end{align}
\end{subequations}
We can simplify $F_k$ and $F_h$ and find
\begin{align}
 2(2-\KB)(1+\lambda_s) F_k k^2 &= \omega^2 4\Ktwo \frac{c_s^2 k^2}{\omega^2 - \mathcal{M}^2} \,,
\end{align}
with
\begin{subequations}
\begin{align}
 c_s^2 &\equiv \frac{c'^2}{\Ktwo \KB} \equiv \frac{(2-\KB)\left(1 + \frac12 \lambda_s \KB\right)}{\Ktwo \KB} \,, \\
 \mathcal{M}^2 &\equiv \frac{(2-\KB)(1+\lambda_s) Q_0^2}{\KB} \,.
\end{align}
\end{subequations}
Plugging this back into the $\varphi$ equation of motion gives, after multiplying with $\omega^2 - \mathcal{M}^2$,
\begin{align}
 0 = 4 \Ktwo \omega^2 \left(
        \omega^2 \varphi - \mathcal{M}^2 \varphi
        - c_s^2 k^2 \varphi
        \right)
    - (2 - \KB) \vec{k}^2 h_{00} F_h (\omega^2 - \mathcal{M}^2) \,.
\end{align}
We can also simplify $F_h$, the prefactor of the coupling to $h_{00}$, and find
\begin{align}
 (2-\KB) F_h &= i\frac{\omega \mathcal{M}}{c_s^2 k^2} \frac{(2+\KB \lambda_s) \sqrt{2-\KB}}{\sqrt{\KB (1+ \lambda_s)}} \left(1 - \frac{c_s^2 k^2}{\omega^2 - \mathcal{M}^2}\right) \,.
\end{align}
Thus the dispersion relation is $\omega^2 = c_s^2 k^2 + \mathcal{M}^2$ in agreement with Refs.~\cite{Skordis2020, Skordis2021} and this cannot be changed by finite terms from the $h_{00}$ equation of motion because, on-shell, the prefactor $F_h$ of $h_{00}$ vanishes.
In principle, a factor $1/F_h$ from the $h_{00}$ equation might change this.
But we will see below that this is not the case.

This also shows that, on-shell, this scalar mode does not couple to matter (again, unless the $F_h$ factor is canceled which does not happen in practice).
In the static limit in galaxies, $\varphi$ is coupled to matter due to a $k^2 h_{00}$ term with constant prefactor.
But this term cancels on-shell due to the factor $F_h$.

Consider then the $h_{00}$ equation of motion.
Plugging the $\vec{A}$ solution into the $h_{00}$ equation gives
\begin{multline}
 (2-\KB) \vec{k}^2 h_{00}
 = 2 f_{00} \left(
    \frac{2}{\tilde{M}_{\mathrm{Pl}}^2} S_{00} - \KB \frac12 h_{00} f_{Q1} f_A f_A' \vec{k}^2 \right. \\
    \left. +(2-\KB) f_{Q2} f_A f_A' \vec{k}^2 \varphi -(2-\KB) f_\varphi \vec{k}^2 \varphi
    \right)\,.
\end{multline}
This gives
\begin{align}
 \vec{k}^2 h_{00} = F_{00} \left( \frac{2}{\tilde{M}_{\mathrm{Pl}}^2} S_{00}  -(2-\KB) F_\varphi \vec{k}^2 \varphi\right)\,,
\end{align}
with
\begin{subequations}
\begin{align}
F_{00} &\equiv \frac{2 f_{00}}{2-\KB + \KB f_{00} f_{Q1} f_A f_A'} \,, \\
F_\varphi &\equiv f_\varphi-f_{Q2} f_A f_A' \,.
\end{align}
\end{subequations}

In the non-relativistic limit,
\begin{align}
 S_{00} = T_{00} - \frac12 \eta_{00} T^{0}_{0} = \frac12 T_{00} = \frac12 \rho_m \,,
\end{align}
where $\rho_m$ is the matter density.
This finally gives
\begin{align}
 k^2 h_{00} = F_{00} \left( \frac{\rho_m}{\tilde{M}_{\mathrm{Pl}}^2} -(2-\KB) F_\varphi \vec{k}^2 \varphi\right)\,.
\end{align}

Next, we plug this result for $k^2 h_{00}$ back into the $\varphi$ equation to get the final result.
As already mentioned above, there are no terms canceling the factor $F_h$.
So the dispersion relation is unaffected and the coupling to matter vanishes on-shell.
Concretely, we find
\begin{align}
  0 = &\left(\omega^2 c_s^{-2} \varphi - \mathcal{M}^2 c_s^{-2} \varphi - k^2 \varphi\right) \nonumber \\
 &- (2 - \KB)\left( \frac{\rho_m}{\tilde{M}_{\mathrm{Pl}}^2} + (2-\KB) F_\varphi k^2 \varphi \right) \frac{F_h F_{00} \KB}{4 c'^2} (1 - \mathcal{M}^2/\omega^2) \,.
\end{align}
To lowest order in $\delta$ and $\bar{\epsilon}$ defined as
\begin{subequations}
\begin{align}
\delta &\equiv \frac{\omega}{\sqrt{c_s^2 k^2 + \mathcal{M}^2}} - 1 \,, \\
 \bar{\epsilon} &\equiv \frac{\sqrt{\Ktwo} Q_0}{c' k} = \frac{m_{\mathrm{SZ}}}{k} \sqrt{\frac{1}{2 +  \lambda_s \KB}} \,,
\end{align}
\end{subequations}
we then have
\begin{align}
  0 = \left(\omega^2 c_s^{-2} \bar{\varphi} - \mathcal{M}^2 c_s^{-2} \bar{\varphi} - k^2 \bar{\varphi}\right)
 -i \sqrt{2 (2-\KB)} \frac{\delta \bar{\epsilon}}{1- c_s^2} \frac{\rho_m}{\sqrt{2} \tilde{M}_{\mathrm{Pl}}}  \,,
\end{align}
where we also defined
\begin{align}
 \bar{\varphi} \equiv \tilde{M}_{\mathrm{Pl}} \sqrt{\frac{2-\KB}{\KB}} \varphi \,.
\end{align}
Here, we used the fact that $F_h = \mathcal{O}(\delta \bar{\epsilon})$, $F_\varphi = \mathcal{O}(\delta \bar{\epsilon})$ and $F_{00} = (1-c_s^2)^{-1}$ to lowest order in $\delta$ and $\bar{\epsilon}$.
The quantity $\delta$ parametrizes how close $\omega$ is to being on-shell, i.e. how well $\omega$ satisfies the full scalar mode dispersion relation $\omega^2 = c_s^2 k^2 + \mathcal{M}^2$ \cite{Skordis2020, Skordis2021}.
On-shell, we have $\delta = 0$.
Note that $\bar{\epsilon} \ll 1$ for the wavevectors that interest us since $0 < \KB < 2$ for stability \cite{Skordis2020}, $\lambda_s$ is small in the MOND limit that interests us \cite{Skordis2020}, and $m_{\mathrm{SZ}} \lesssim 1/\mathrm{Mpc}$ while we consider only $k \gtrsim 1/\mathrm{kpc}$.
We are further interested in $c_s \ll 1$ so that we don't keep the $(1-c_s^2)^{-1}$ factor in what follows.

\subsection{Effective Lagrangian}

So far, our calculation tells us how the scalar mode of the SZ model reacts to a perturber density $\delta_b$.
But,
    as in Appendix~\ref{sec:twofieldsfdmpert:effective} for two-field SFDM,
    this does not tell us how much energy this mode carries.
This requires the normalization of the effective Lagrangian.
We have $\omega \ll k$, $m_{\mathrm{SZ}} \ll k$, $0 < \KB < 2$, and $\lambda_s$ small.
This implies to lowest order in $\bar{\epsilon}$ and $\delta$,
\begin{align}
 h_{00} &= \mathcal{O}\left(\bar{\epsilon} \delta \frac{2-\KB}{\sqrt{\KB}} \varphi\right) \,, \\
 \vec{A} &= \mathcal{O}\left(\frac{1}{c_s} \frac{2-\KB}{\KB} \varphi \right) \,.
\end{align}
Then, the spatial derivative terms in the effective Lagrangian scale as
\begin{align}
 \label{eq:SZkinnormalization}
 \frac{(2-\KB) \tilde{M}_{\mathrm{Pl}}^2}{\KB} (\vec{\nabla} \varphi)^2 \sim (\vec{\nabla} \bar{\varphi})^2 \,.
\end{align}
Here, we are only interested in the order of magnitude of the terms in the effective Lagrangian since that is sufficient for our purposes, as discussed in Sec.~\ref{sec:relmond_coupling}.
Thus, Eq.~\eqref{eq:SZkinnormalization} together with the equation of motion for $\bar{\varphi}$ suffices to fix the effective Lagrangian.
For the coupling to matter, we have in terms of our prototype Lagrangian
\begin{align}
 \coup = \mathcal{O}\left(\bar{\epsilon} \delta\right) = \mathcal{O}\left(\frac{m_{\mathrm{SZ}}}{k} \cdot \left(\frac{\omega}{\sqrt{c_s^2 k^2 + \mathcal{M}^2}} - 1\right) \right)\,,
\end{align}
where we used that $\bar{\epsilon}$ and $m_{\mathrm{SZ}}/k$ are related by an $\mathcal{O}(1)$ factor.
There is no $\hat{a}$ term for the non-relativistic mode of the SZ model because $\hat{a}$ corresponds to the background galaxy, while we assume a Minkowski background for simplicity.

\end{appendices}

\appto{\bibsetup}{\sloppy}

\printbibliography[heading=bibintoc]

\end{document}